\newif\ifREVTEX
\REVTEXtrue

\ifREVTEX
	\documentclass[twocolumn,prb,superscriptaddress]{revtex4}
\else
	\documentclass{naturefigTCL}
\fi

\usepackage{times}
\usepackage{graphicx}
\usepackage{latexsym,amsmath,amssymb,bm,euscript}
\usepackage{color}
\usepackage{hyperref}
\usepackage{subfigure}
\usepackage{epstopdf}

\def\ra{\rangle}
\def\la{\langle}
\def\up{\uparrow}
\def\dn{\downarrow}
\def\Hc{{\rm H.c.}}

\def\rhotot{{\rho {\rm tot}}}

\def\LSCO{La$_{2-x}$Sr$_{x}$CuO$_{4}$~}
\def\rvec{\mathbf{r}}
\def\Rvec{\mathbf{R}}
\def\kvec{\mathbf{k}}
\def\qvec{\mathbf{q}}

\def\tJK{$t$-$J$-$K$~}
\def\Ke{{\mathsf K}_e}

\begin{document}

\ifREVTEX
	\def\convScaleFac{1.0}
\else
	\def\onlinecite{\cite}
	\def\corrScaleFac{0.65}
	\def\convScaleFac{0.55}
\fi

\title{Non-Fermi-liquid $d$-wave metal phase of strongly interacting electrons}

\ifREVTEX

	\author{Hong-Chen Jiang}
	\affiliation{Kavli Institute for Theoretical Physics, University of California, Santa Barbara, CA, 93106, USA}
	
	\author{Matthew S. Block}
	\affiliation{Department of Physics and Astronomy, University of Kentucky, Lexington, Kentucky 40506, USA}
	
	\author{Ryan V. Mishmash}
	\author{\\James R. Garrison}
	\affiliation{Department of Physics, University of California, Santa Barbara, California 93106, USA}
	
	\author{D. N. Sheng}
	\affiliation{Department of Physics and Astronomy, California State University, Northridge, California 91330, USA}
	
	\author{Olexei I. Motrunich}
	\affiliation{Department of Physics, California Institute of Technology, Pasadena, California 91125, USA}
	
	\author{Matthew P. A. Fisher}
	\affiliation{Department of Physics, University of California, Santa Barbara, California 93106, USA}
	
	\date{\today}

\else

	\author{Hong-Chen~Jiang$^1$, Matthew~S.~Block$^2$, Ryan~V.~Mishmash$^3$, James~R.~Garrison$^3$, D.~N.~Sheng$^4$, Olexei~I.~Motrunich$^5$, Matthew~P.~A.~Fisher$^3$}
	
	\maketitle
	
	\begin{affiliations}
	 \item Kavli Institute for Theoretical Physics, University of California, Santa Barbara, CA, 93106, USA
	 \item Department of Physics and Astronomy, University of Kentucky, Lexington, Kentucky 40506, USA
	 \item Department of Physics, University of California, Santa Barbara, California 93106, USA
	 \item Department of Physics and Astronomy, California State University, Northridge, California 91330, USA
	 \item Department of Physics, California Institute of Technology, Pasadena, California 91125, USA
	\end{affiliations}

\fi

\begin{abstract}
\ifREVTEX
	\bf{
\fi
Developing a theoretical framework for conducting electronic fluids qualitatively distinct from those described by Landau's Fermi-liquid theory is of central importance to many outstanding problems in condensed matter physics. One such problem is that, above the transition temperature and near optimal doping, high-transition-temperature copper-oxide superconductors exhibit `strange metal' behaviour that is inconsistent with being a traditional Landau Fermi liquid. Indeed, a microscopic theory of a strange-metal quantum phase could shed new light on the interesting low-temperature behaviour in the pseudogap regime and on the $d$-wave superconductor itself. Here we present a theory for a specific example of a strange metal---the `$d$-wave metal'. Using variational wavefunctions, gauge theoretic arguments, and ultimately large-scale density matrix renormalization group calculations, we show that this remarkable quantum phase is the ground state of a reasonable microscopic Hamiltonian---the usual $t$-$J$ model with electron kinetic energy $t$ and two-spin exchange $J$ supplemented with a frustrated electron `ring-exchange' term, which we here examine extensively on the square lattice two-leg ladder. These findings constitute an explicit theoretical example of a genuine non-Fermi-liquid metal existing as the ground state of a realistic model.
\ifREVTEX
	}
\fi
\end{abstract}

\ifREVTEX
	\maketitle
\fi

Over the past several decades, experiments on strongly correlated materials have routinely revealed, in certain parts of the phase diagram, conducting liquids with physical properties qualitatively inconsistent with Landau's Fermi liquid theory.\cite{Baym91_FermiLiquids}  Examples of these so-called non-Fermi liquid metals\cite{Schofield99_ContPhys_40_95} include the strange metal phase of the cuprate superconductors\cite{LeeNagaosaWen, Boebinger09_Science_323_590} and heavy fermion materials near a quantum critical point.\cite{Stewart01_RevModPhys_73_797,Si08_NaturePhys_4_186}  However, such non-Fermi liquid behavior has been notoriously challenging to characterize theoretically, largely owing to the failure of a weakly interacting quasiparticle description.  It is even ambiguous to define a non-Fermi liquid, although possible deviations from Fermi liquid theory include, for example, violation of Luttinger's\cite{Luttinger60_PhysRev_119_1153} famed volume theorem, vanishing quasiparticle weight, and/or anomalous thermodynamics and transport.\cite{Anderson88_PRL_60_132,Varma89_PRL_63_1996,Stewart01_RevModPhys_73_797,Senthil08_PRB_78_035103,Faulkner10_Science_329_1043,Sachdev10_PRL_105_151602}  This theoretical quandary is rather unfortunate as it is likely prohibiting a full understanding of the mechanism behind high-temperature superconductivity, as well as stymying theoretically-guided searches for new exotic materials.

Pioneering early theoretical work on the cuprates relied on two main premises,\cite{Anderson87_Science_235_1196, Baskaran87_SolidStateComm_63_973, Lee90_PRL_64_2450, Lee92_PRB_46_5621, Wen96_PRL_76_503, LeeNagaosaWen} from which we will be guided but not constrained in our pursuit and understanding of a particular non-Fermi liquid metal:  (1) that the microscopics can be described by the square lattice Hubbard model with on-site Coulomb repulsion, which at strong coupling reduces in its simplest form to the $t$-$J$ model; and (2) that the physics of the system can be faithfully represented by the ``slave-boson'' technique, wherein the physical electron operator is written as a product of a slave boson (``chargon''), which carries the electronic charge, and a spin-1/2 fermionic ``spinon,''\,\cite{Anderson87_PRL_58_2790} which carries the spin, both strongly coupled to an emergent gauge field.  However, within the slave-boson formulation, it has been difficult to access non-Fermi liquid physics at low temperatures because this requires the chargons to be in an uncondensed, yet conducting, quantum phase,\cite{Feigelman93_PRB_48_16641} i.e., some sort of the elusive ``Bose metal.''  Early attempts to describe the strange metal in this framework treated it as a strictly finite-temperature phenomenon in which the slave bosons form an uncondensed, but classical, Bose fluid,\cite{Lee90_PRL_64_2450, Lee92_PRB_46_5621} a treatment which precludes the possibility that the strange metal is a true quantum phase at all.

In our view, the strange metal should be viewed as a genuine two-dimensional (2D) quantum phase, which can perhaps be unstable to superconducting or pseudogap behavior.  Indeed, recent experimental work on \LSCO has shown that when superconductivity is stripped away by high magnetic fields, strange metal behavior persists over a wide doping range down to extremely low temperatures.\cite{Cooper09_Science_323_603}  Thus, the strange metal in the cuprates is quite possibly a true, extended, zero-temperature quantum phase.\cite{Boebinger09_Science_323_590}

Inspired by these results and building on our previous work which proposed\cite{DBL} and succeeded in realizing\cite{Sheng2008_2legDBL, Block2011_3legGMI, Mishmash2011_4legDBL} a genuine, zero-temperature Bose metal, we employ a novel variant of the slave-boson approach to construct and analyze an exotic 2D non-Fermi liquid quantum phase, which we refer to as the ``$d$-wave metal'' and abbreviate as ``$d$-metal.''  The $d$-wave metal is modeled by a variational wave function consisting of a product of a $d$-wave Bose metal wave function\cite{DBL, Sheng2008_2legDBL, Block2011_3legGMI, Mishmash2011_4legDBL} for the chargons and a usual Slater determinant for the spinon.  Importantly, placing the chargons into the $d$-wave Bose metal state imparts the many-electron wave function with a sign structure qualitatively distinct from that of a simple Slater determinant, and in particular, imprints strong singlet $d$-wave two-particle correlations.  This results in a gapless, conducting quantum fluid with an electron momentum distribution function which exhibits a critical, singular surface that violates Luttinger's volume theorem,\cite{Luttinger60_PhysRev_119_1153} as well as prominent critical Cooper pairs with $d$-wave character.  The $d$-wave nature of our phase is tantalizingly suggestive of incipient $d$-wave superconductivity and thus of possible relevance to the cuprates.

Furthermore, tying back into premise (1) above, we propose a reasonably simple model Hamiltonian to stabilize the $d$-metal by augmenting the traditional $t$-$J$ model with a four-site ring-exchange term $K$.  Then, owing to the afforded numerical and analytical tractability provided by the density matrix renormalization group (DMRG)\cite{White92_PRL_69_2863, White93_PRB_48_10345} and bosonization,\cite{Giamarchi03_1D, Shankar_Acta, Lin98, Fjaerestad02} we place the problem on a quasi-one-dimensional (quasi-1D) two-leg ladder geometry (see Fig.~\ref{fig:model}).  In this system, we establish several lines of compelling evidence that the $d$-metal phase exists as the quantum ground state of our $t$-$J$-$K$ model Hamiltonian, and we are able to characterize and understand the phase very thoroughly.  Importantly, our realized two-leg $d$-metal state is \emph{non-perturbative} in that it cannot be understood within conventional Luttinger liquid theory\cite{Giamarchi03_1D} starting from free electrons.\cite{Balents96_PRB_53_12133}  We believe this study to be one of the first unbiased numerical demonstrations of a non-Fermi liquid metal as the stable ground state of a local Hamiltonian.  Finally, in our concluding remarks, we will discuss straightforward extensions of these results to two dimensions, as well as comment on their potential relevance to the actual non-Fermi liquids observed in experiment.

\section*{\uppercase{Gauge theory and variational wave functions for the $\lowercase{d}$-wave metal}}

Our theoretical description of the non-Fermi liquid $d$-metal begins by writing the electron operator for site $\rvec$ and spin state $s=\,\up,\dn$ as the product of a bosonic chargon $b(\rvec)$ and fermionic spinon $f_s(\rvec)$; that is, $c_s(\rvec)=b(\rvec)f_s(\rvec)$.   With $b(\rvec)$ a hard-core boson operator, this construction prohibits doubly occupied sites, an assumption we make from here on.  The physical electron Hilbert space is recovered by implementing at each site the constraint $b^\dagger(\rvec)b(\rvec)=\sum_s f_s^\dagger(\rvec)f_s(\rvec)=\sum_s c_s^\dagger(\rvec)c_s(\rvec)=n_e(\rvec)$, which physically means that a given site is either empty or contains a chargon and exactly one spinon to compose an electron.  Theoretically, this is achieved by strongly coupling the $b$'s and $f$'s via an emergent gauge field.\cite{LeeNagaosaWen}

Under the natural assumption that the spinons are in a Fermi sea state, the behavior of the chargons determines the resulting electronic phase.  Condensing the bosonic chargons  so that $\langle b(\rvec)\rangle\neq0$ implies $c_s(\rvec)\sim f_s(\rvec)$; thus, in this case, the electronic phase is that of a Fermi liquid.  It then follows that in order to describe a non-Fermi liquid conducting quantum fluid within this framework, we require that the chargons not condense, $\langle b(\rvec)\rangle=0$, yet still conduct.  However, accessing such a ``Bose metal'' phase has proven extremely difficult over the years.  In recent work,\cite{DBL, Sheng2008_2legDBL, Block2011_3legGMI, Mishmash2011_4legDBL} we have indeed succeeded in realizing a concrete, genuine Bose metal phase, which we named the ``$d$-wave Bose liquid'' or, equivalently, ``$d$-wave Bose metal'' (DBM).  The DBM is central to our construction of the $d$-wave metal.  Specifically, in the DBM, we decompose the hard-core boson as $b(\rvec)=d_1(\rvec)d_2(\rvec)$ with the constraint $d_1^\dagger(\rvec)d_1(\rvec)=d_2^\dagger(\rvec)d_2(\rvec)=b^\dagger(\rvec)b(\rvec)$, where $d_1$ ($d_2$) are fermionic slave particles (``partons'') with \emph{anisotropic hopping patterns}:  $d_1$ ($d_2$) is chosen to hop preferentially in the $\hat{x}$ ($\hat{y}$) direction.  The resulting bosonic phase is a conducting, yet uncondensed, quantum fluid, which is precisely the phase into which we place the charge sector of the $d$-metal.  That is, for the $d$-metal we take a novel \emph{all fermionic} decomposition of the electron,
\begin{equation}
c_s(\rvec) = d_1(\rvec)d_2(\rvec)f_s(\rvec),
\label{d1d2fs}
\end{equation}
subject to the constraint
\begin{equation}
d_1^\dagger(\rvec)d_1(\rvec)=d_2^\dagger(\rvec)d_2(\rvec)=\sum_s f_s^\dagger(\rvec)f_s(\rvec)=n_e(\rvec).
\label{constraints}
\end{equation}
The resulting theory now includes two gauge fields:  one to glue together $d_1$ and $d_2$ to form the chargon and another to glue together $b$ and $f$ to form the electron.  In 
the Supplementary Information, we give a detailed bosonization analysis of this novel gauge theory for the two-leg ladder study on which we focus below.

\def\modelCaption
{
\textbf{Schematic of the $t$-$J$-$K$ model Hamiltonian.}
Top:  Picture of the full $t$-$J$-$K$ model, Eq.~(\ref{eqn:fullmodel}), on the two-leg ladder.  In this work, we use periodic boundary conditions in the long ($\hat{x}$) direction for all calculations.  Bottom:  Action of the ring term $H_K$, Eq.~(\ref{eqn:ringsing}), on a single plaquette, elucidating its ``singlet-rotation'' nature.
}

\ifREVTEX
	\begin{figure}[t]
	\centerline{\includegraphics[width=\columnwidth]{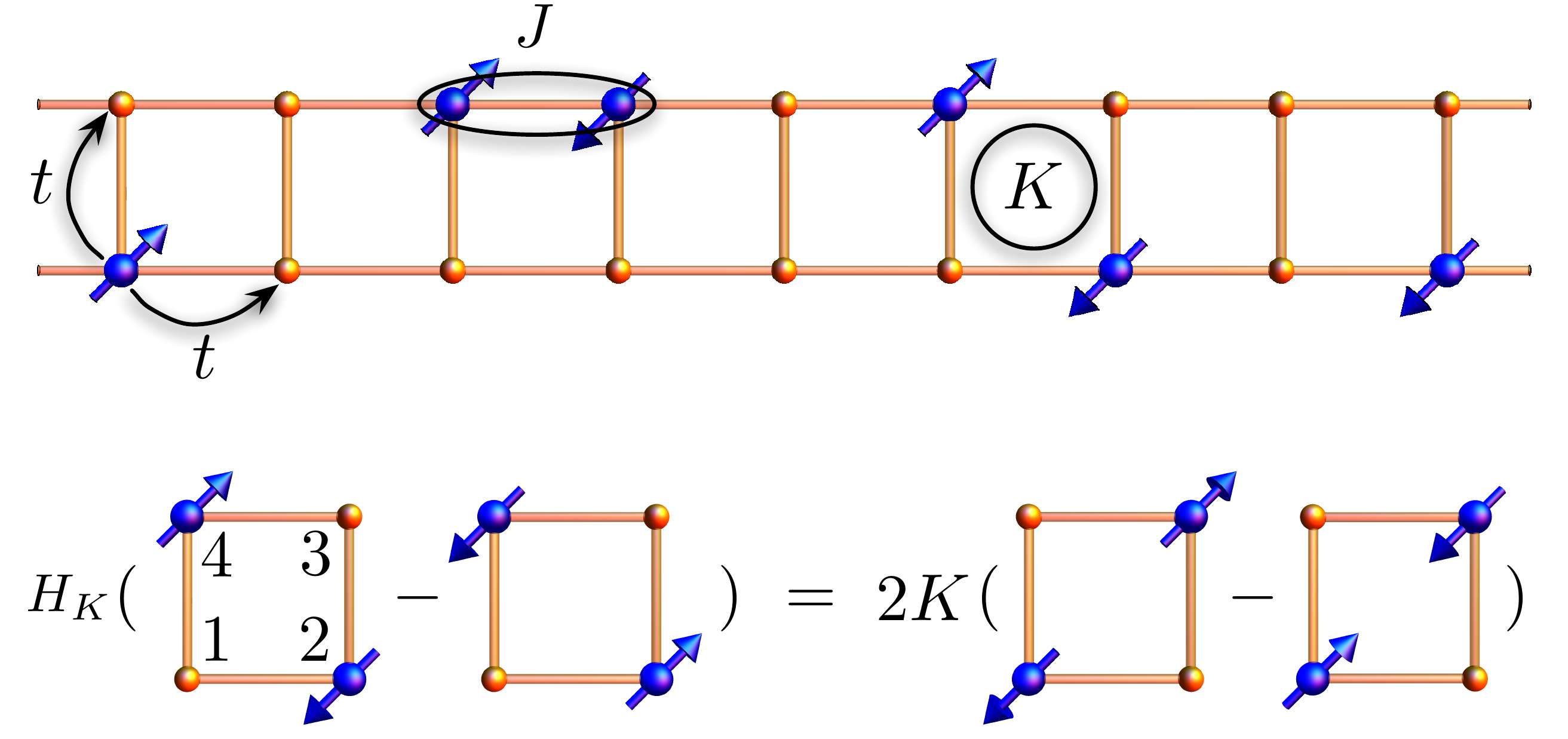}}
	\caption{\modelCaption}
	\label{fig:model}
	\end{figure}
\fi

Guided by the slave-boson construction, one can naturally construct electronic variational wave functions by taking the product of a hard-core bosonic wave function $\psi_b$ with a fermionic wave function $\psi_f$ and evaluating them at the same coordinates (Gutzwiller projection):
\begin{equation}
\label{eqn:generalwf}
\psi_c(\{\rvec_i^\up\},\{\rvec_i^\dn\})=\mathcal{P}_G\left[\psi_b(\{\Rvec_i\})\times\psi_f(\{\rvec_i^\up\},\{\rvec_i^\dn\})\right],
\end{equation}
where $\mathcal{P}_G$ performs the projection into the physical electronic Hilbert space:  $\{\Rvec_i\}=\{\rvec_i^\up\}\bigcup\{\rvec_i^\dn\}$.  If we put the $f$'s into a spin-singlet Fermi sea state with orbitals $\{\kvec_j\}$ (Slater determinant), i.e., $\psi_f(\{\rvec_i^\up\},\{\rvec_i^\dn\})=\det[e^{i\kvec_j\cdot\rvec_i^\up}]\det[e^{i\kvec_j\cdot\rvec_i^\dn}]=\psi_f^{\mathrm{FS}}$, then we can model both the Fermi liquid metal and the non-Fermi liquid $d$-metal in a unified way.  In both cases, the wave functions are straightforward to implement using variational Monte Carlo (VMC) methods.\cite{Ceperley77_PRB_16_3081, Gros89_AnnPhys_189_53, HellbergMele1991}

For the Fermi liquid, we put the $b$'s into a superfluid wave function $\psi_b^\mathrm{SF}$ via a typical Jastrow form, so that, schematically, $\psi_c^{\mathrm{FL}}=\mathcal{P}_G[\psi_b^\mathrm{SF}\times\psi_f^{\mathrm{FS}}]$.  Note that since $\psi_b^\mathrm{SF}$ is a positive wave function, the sign structure\cite{Ceperley91_JStatPhys_63_1237} of $\psi_c^{\mathrm{FL}}$ is identical to that of the noninteracting Fermi sea state.  In contrast, to model the $d$-metal, we put the $b$'s into a Bose metal wave function according to the DBM construction of Refs.~\onlinecite{DBL, Sheng2008_2legDBL, Block2011_3legGMI, Mishmash2011_4legDBL}:
\begin{equation}
\label{eqn:dblwf}
\psi_b(\{\Rvec_i\})=\psi_{d_1}(\{\Rvec_i\})\times\psi_{d_2}(\{\Rvec_i\})=\psi_b^\mathrm{DBM},
\end{equation}
where $\psi_{d_1}$ ($\psi_{d_2}$) is a Slater determinant with a Fermi sea compressed in the $\hat{x}$ ($\hat{y}$) direction.\cite{DBL}  Then, we have
\begin{equation}
\label{eqn:dmetalwf}
\psi_c^{d-\mathrm{metal}}=\mathcal{P}_G\left[\psi_b^\mathrm{DBM}\times\psi_f^{\mathrm{FS}}\right]=\mathcal{P}_G\left[\psi_{d_1}\times\psi_{d_2}\times\psi_f^{\mathrm{FS}}\right].
\end{equation}

Interestingly, this construction, Eq.~(\ref{eqn:dmetalwf}), is actually a time-reversal invariant analog of the composite Fermi liquid description of the half-filled Landau level,\cite{Halperin93_PRB_47_7312} where the $d$-wave Bose metal wave function\cite{DBL} plays the role of Laughlin's $\nu=1/2$ bosonic state.\cite{Laughlin83_PRL_50_1395}  Just as Laughlin's wave function imprints a nontrivial complex phase pattern on the Slater determinant, the DBM wave function imprints a nontrivial $d$-wave sign structure, hence our designation ``$d$-wave metal.''  As we explore in detail below, there are many physical signatures associated with putting the chargons into the DBM phase, making the $d$-wave metal dramatically distinguishable from the traditional Landau Fermi liquid.

\section*{\uppercase{Microscopic ring-exchange model}}

The $t$-$J$-$K$ model Hamiltonian which we propose to stabilize the $d$-metal phase is given by
\begin{align}
H & = H_{tJ} + H_K, \label{eqn:fullmodel} \\
H_{tJ}  & = -t\sum_{\substack{\langle i,j \rangle,\,s=\up,\dn}}\left(c_{is}^\dagger c_{js} + \Hc\right) + J\sum_{\langle i,j \rangle}\vec{S}_{i} \cdot \vec{S}_{j}, \\
H_K & = 2K \sum_{\square} (\mathcal{S}_{13}^\dagger\mathcal{S}_{24}+\Hc), \label{eqn:ringsing}
\end{align}
where $\langle i,j \rangle$ and $\square$ indicate sums over all nearest-neighbor bonds and all elementary plaquettes of the 2D square lattice, respectively.  In the spirit of the $t$-$J$ model, we choose to work in the subspace of no doubly occupied sites, but for simplicity, we do ignore the term $-\frac{J}{4}n_i n_j$ present in typical definitions of the $t$-$J$ model.\cite{LeeNagaosaWen}  In Eq.~(\ref{eqn:ringsing}), we have defined a singlet creation operator on two sites as $\mathcal{S}_{ij}^\dagger=\frac{1}{\sqrt{2}}(c_{i\up}^\dagger c_{j\dn}^\dagger-c_{i\dn}^\dagger c_{j\up}^\dagger)$, so that $H_K$ can be viewed as a four-site singlet-rotation term (see Fig.~\ref{fig:model}).  For $K>0$, the ground state of $H_K$ on a single plaquette with two electrons is a $d_{xy}$-orbital spin-singlet; thus, loosely speaking, $H_K$ has a tendency to build in $d$-wave correlations in the system and qualitatively alter the sign structure of the electronic ground state.  Further arguments for studying this model in our search for the $d$-metal can be found in 
the Supplementary Information.

While not being particularly conventional, our ring-exchange term $H_K$ (which should not be confused with four-site cyclic spin-exchange\cite{Normand04_PRB_70_134407, Sheng2009_zigzagSBM, Block11_PRL_106_157202}) is in fact present when projecting the continuum many-body Hamiltonian for screened Coulomb-interacting electrons into a narrow, tight-binding band\cite{Imada2010} (see 
Supplementary Information).  In fact, estimating the strength of $K$, or coefficients on related terms, in real materials such as \LSCO is an interesting open question.

\section*{\uppercase{Two-leg study:  DMRG and VMC}}

\def\partonbandsCaption
{
\textbf{Picture of the parton bands for the $d$-metal phase.}
We show orbitals for a $48\times2$ system, showing partially occupied bonding ($k_y=0$) and antibonding ($k_y=\pi$) bands for $d_1$ and partially occupied bonding bands for $d_2$ and $f_{\up/\dn}$; that is, each Slater determinant in Eq.~(\ref{eqn:dmetalwf}) consists of momentum-space orbitals as depicted here.  The total electron number is $N_e = N_{c\up} + N_{c\dn} = N_{d1} = N_{d2} = N_{f\up} + N_{f\dn}=32$, with $N_{f\up}=N_{f\dn}=16$ so that $S_\mathrm{tot}=0$; the longitudinal boundary conditions are periodic for $d_1$ and antiperiodic for $d_2$ and $f_{\up/\dn}$.  This is precisely the same $d$-metal configuration for which we display characteristic measurements in Fig.~\ref{fig:dmetcorrs}.
}

\ifREVTEX
	\begin{figure}[b]
	\centerline{\includegraphics[width=\columnwidth]{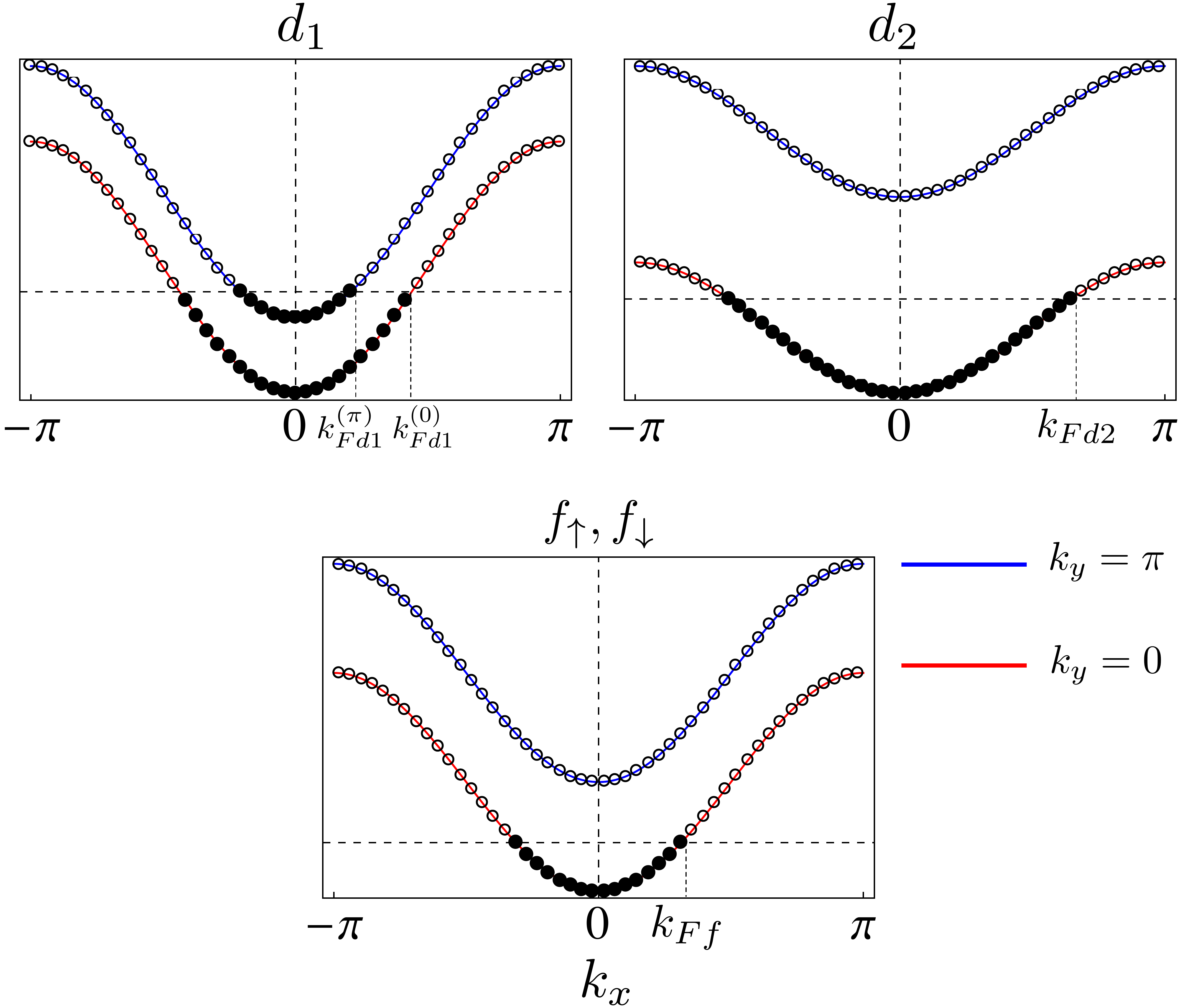}}
	\caption{\partonbandsCaption}
	\label{fig:partonbands}
	\end{figure}
\fi

Unfortunately, as with any interacting fermionic model, our $t$-$J$-$K$ Hamiltonian suffers from the so-called ``fermionic sign problem,''  rendering quantum Monte Carlo calculations inapplicable.\cite{Troyer05_PRL_94_170201}  We thus follow the heretofore successful\cite{Sheng2008_2legDBL, Sheng2009_zigzagSBM, Block2011_3legGMI, Block11_PRL_106_157202, Mishmash2011_4legDBL} approach of accessing 2D gapless phases by studying their \emph{quasi-1D descendants} on ladder geometries, relying heavily on large-scale DMRG calculations.  In fact, we have already established\cite{Sheng2008_2legDBL, Block2011_3legGMI, Mishmash2011_4legDBL} that for two, three, and four legs, the DBM phase itself is the stable ground state of a boson ring-exchange model analogous to Eq.~(\ref{eqn:fullmodel}).  Here, we take the important first step of placing the electron ring $t$-$J$-$K$ model on the two-leg ladder in search of a two-leg descendant of the $d$-metal.

For concreteness, we now consider the model, Eq.~(\ref{eqn:fullmodel}), on the two-leg ladder (see Fig.~\ref{fig:model}) at a generic electron density of $\rho=N_e/(2L_x)=1/3$, where $N_e=N_{c\up}+N_{c\dn}$ is the total number of electrons and $L_x$ is the length of our two-leg ladder (i.e., the system has $L_x\times 2$ total sites).  At this density, $\rho=1/3<1/2$, on the two-leg ladder, the noninteracting ground state is a spin-singlet wherein electrons of each spin partially fill the bonding band ($k_y=0$), leaving the antibonding band ($k_y=\pi$) empty.  Thus, for $t\gg K$, we expect the system to be in a simple one-band metallic state, which is a two-leg analog of the Fermi liquid.  Formally speaking, this phase is a conventional Luttinger liquid with two one-dimensional (1D) gapless modes (central charge $c=2$).  For moderate values of ring exchange, $K\gtrsim t$, we anticipate the unconventional non-Fermi liquid $d$-metal to be a candidate ground state.  On the two-leg ladder at this density, the $d$-metal phase has characteristic band filling configurations for the $d_1$, $d_2$, and $f_{\up/\dn}$ partons as shown in Fig.~\ref{fig:partonbands}: $d_1$ partially fills both bonding and antibonding bands, while $d_2$ and $f_{\up/\dn}$ only fill the bonding band.  (The $d_1$ and $d_2$ configurations constitute the phase denoted ``DBL[2,1]'' in Ref.~\onlinecite{Sheng2008_2legDBL}.)  In a mean-field approximation in which the partons do not interact, the system has five 1D gapless modes corresponding to the five total partially filled bands.  However, in the strong-coupling limit of the full quasi-1D gauge theory (see 
the Supplementary Information for details), two orthonormal linear combinations of the original five modes are rendered massive, leaving an {\it unconventional} Luttinger liquid with $c=3$ gapless modes.

We now provide extensive numerical evidence that this two-leg descendant of the $d$-metal exists as the ground state of the \tJK model over a wide region of the phase diagram.  We summarize these results in Fig.~\ref{fig:phased} by presenting the full phase diagram in the parameters $K/t$ vs.~$J/t$ as obtained by DMRG calculations on length $L_x=24$ and 48 systems at electron density $\rho=1/3$.  For small $K$, we find a conventional one-band (spinful) Luttinger liquid phase which is a two-leg analog of the ``Fermi liquid metal,'' hence the label in Fig.~\ref{fig:phased}.  For moderate $J$ and upon increasing $K$, the system goes into the unconventional ``non-Fermi liquid $d$-wave metal'' phase, which is the main focus of this work.  The phase boundaries in Fig.~\ref{fig:phased}, all of which represent strong first-order transitions, were determined by measuring several standard momentum-space correlation functions in the DMRG (see 
the Supplementary Information for details):  the electron momentum distribution function $\langle c_{\qvec s}^\dagger c_{\qvec s}\rangle$, the density-density structure factor $\langle \delta n_{\qvec} \delta n_{-\qvec}\rangle$, and the spin-spin structure factor $\langle \mathbf{S}_{\qvec}\cdot \mathbf{S}_{-\qvec}\rangle$.

For concreteness, we now focus on the cut along $J/t=2$ in Fig.~\ref{fig:phased} for a $48\times2$ system with $N_e=32$ electrons.  We take one point deep within the conventional one-band metal at $K/t=0.5$ and the other point deep within the exotic $d$-metal at $K/t=1.8$.  First focusing on the former case, we show in Fig.~\ref{fig:FLcorrs} DMRG measurements characteristic of the conventional Luttinger liquid.  The ground state is a spin-singlet with a sharp singularity in the electron momentum distribution function at $q_y=0$ and $q_x=k_F = \pi N_{c\up}/L_x = 8\cdot 2\pi/48$, which is a usual Fermi wavevector determined solely from the electron density.  The density-density and spin-spin structure factors at $q_y=0$ also exhibit familiar features at $q_x=0$ and $q_x=2k_F=16\cdot 2\pi/48$, both characteristic of an ordinary one-band metallic state with gapless charge and spin modes.\cite{Giamarchi03_1D}  We stress that, even with the constraint of no double-occupancy and nonzero $K/t=0.5$ and $J/t=2$, the interacting electronic system is still qualitatively very similar to the two-leg free Fermi gas; analogously, the 2D Fermi liquid is in many ways qualitatively similar to the 2D free Fermi gas.  In both cases, the main differences are basically quantitative and are well-understood.\cite{Giamarchi03_1D, Baym91_FermiLiquids}

\def\phasedCaption
{
\textbf{Phase diagram of the $t$-$J$-$K$ electron ring-exchange model at electron density $\rho=1/3$ on the two-leg ladder.}  In addition to the conventional one-band metal (``Fermi liquid metal'') and exotic ``non-Fermi liquid $d$-wave metal,'' there are two other realized phases.  For small $J$, there is an intermediate phase with fully polarized electrons (region labeled ``FP'').  For large $K$, due to the inherently attractive nature of ring-exchange interactions,\cite{Sheng2008_2legDBL} the system generally phase separates along the ladder (region labeled ``PS'').
}

\ifREVTEX
	\begin{figure}[t]
	\centerline{\includegraphics[width=\columnwidth]{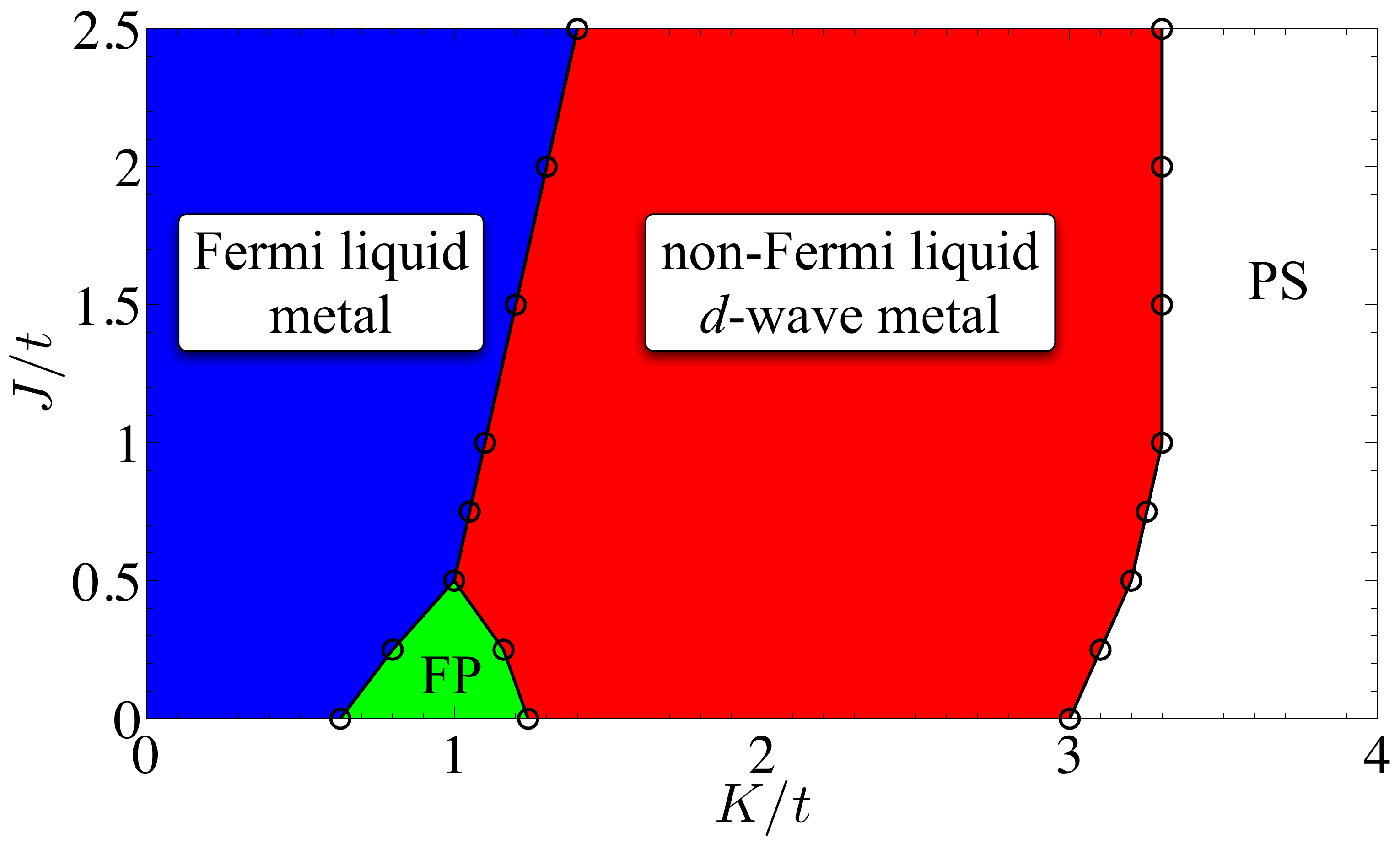}}
	\caption{\phasedCaption}
	\label{fig:phased}
	\end{figure}
\fi

\def\FLcorrsCaption
{
\textbf{DMRG measurements in the conventional Luttinger liquid phase at $J/t = 2$ and $K/t = 0.5$.}
We show (a) the electron momentum distribution function, (b) the density-density structure factor, and (c) the spin-spin structure factor.  The important wavevectors $k_F$ and $2k_F$, as described in the text, are highlighted by vertical dashed-dotted lines.
}

\ifREVTEX
	\begin{figure}
	\centerline{\subfigure{\includegraphics[width=\columnwidth]{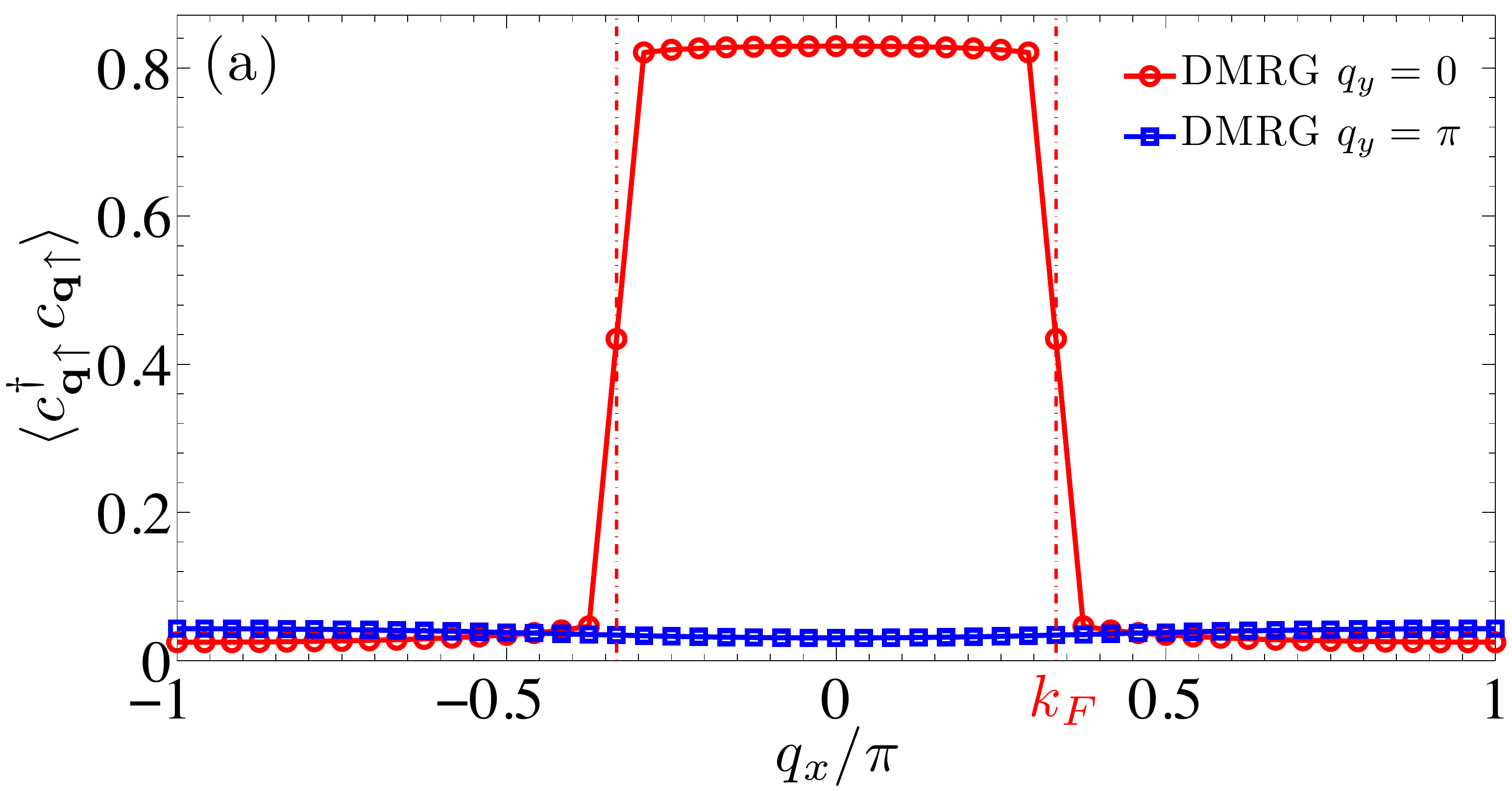}}}
	\centerline{\subfigure{\includegraphics[width=\columnwidth]{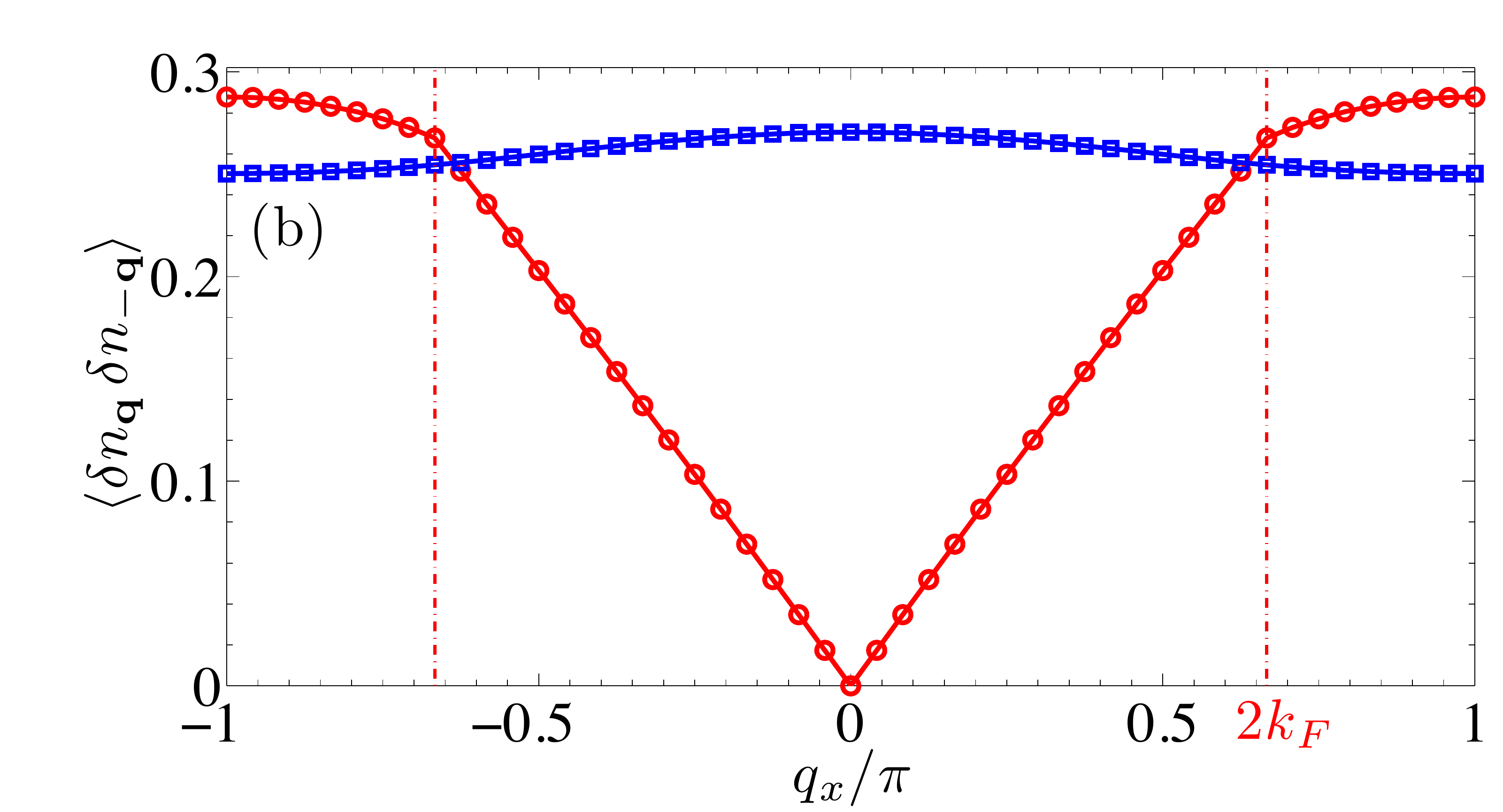}}}
	\centerline{\subfigure{\includegraphics[width=\columnwidth]{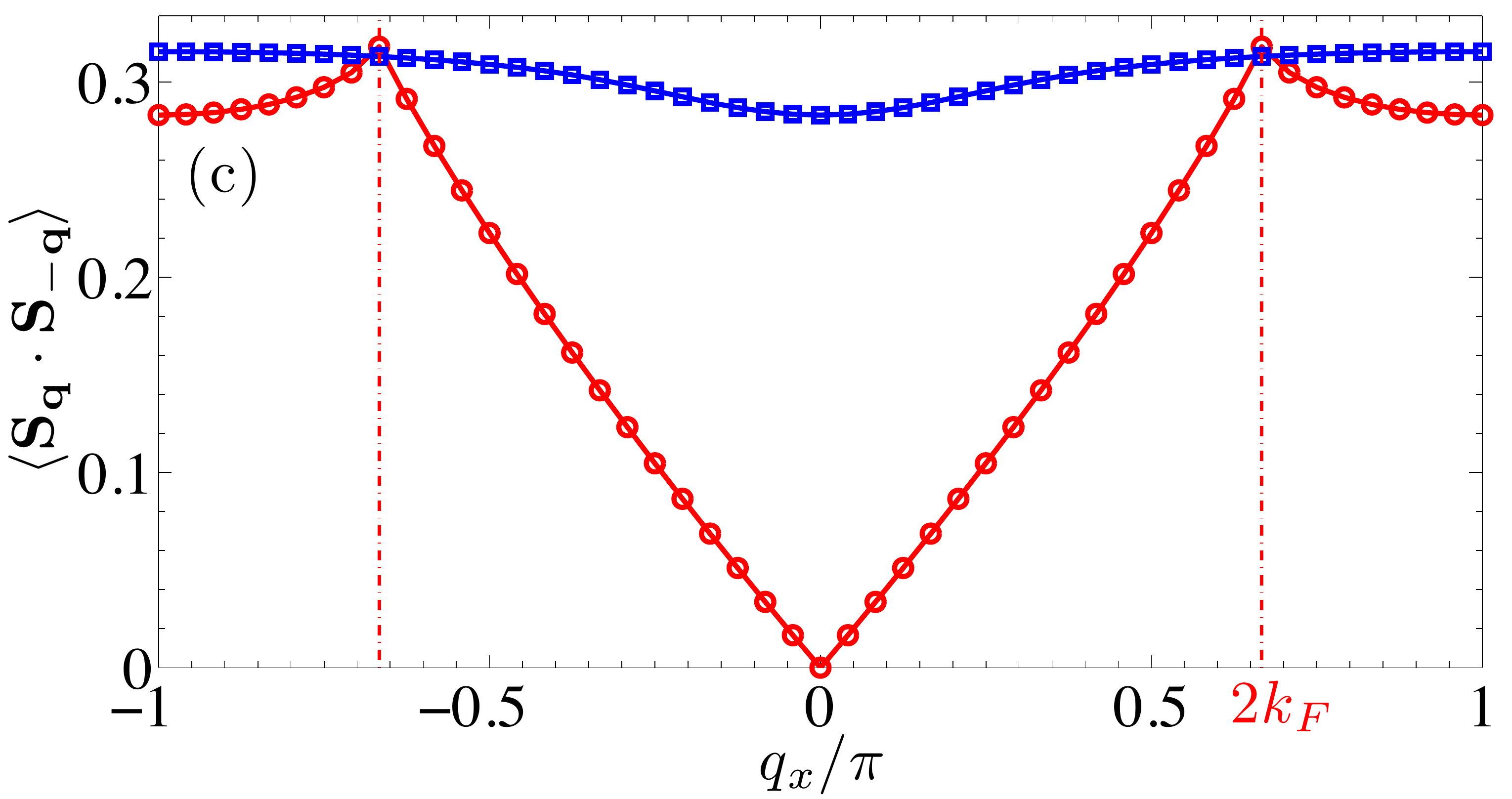}}}
	\caption{\FLcorrsCaption}
	\label{fig:FLcorrs}
	\end{figure}
\fi

\def\dmetcorrsCaption
{
\textbf{DMRG measurements in the unconventional $d$-metal phase at $J/t = 2$ and $K/t = 1.8$.}
We show the same quantities as in Fig.~\ref{fig:FLcorrs}.  Here, we also show the matching VMC measurements using a $d$-metal trial wave function depicted in Fig.~\ref{fig:partonbands}.
}

\ifREVTEX
	\begin{figure}
	\centerline{\subfigure{\includegraphics[width=\columnwidth]{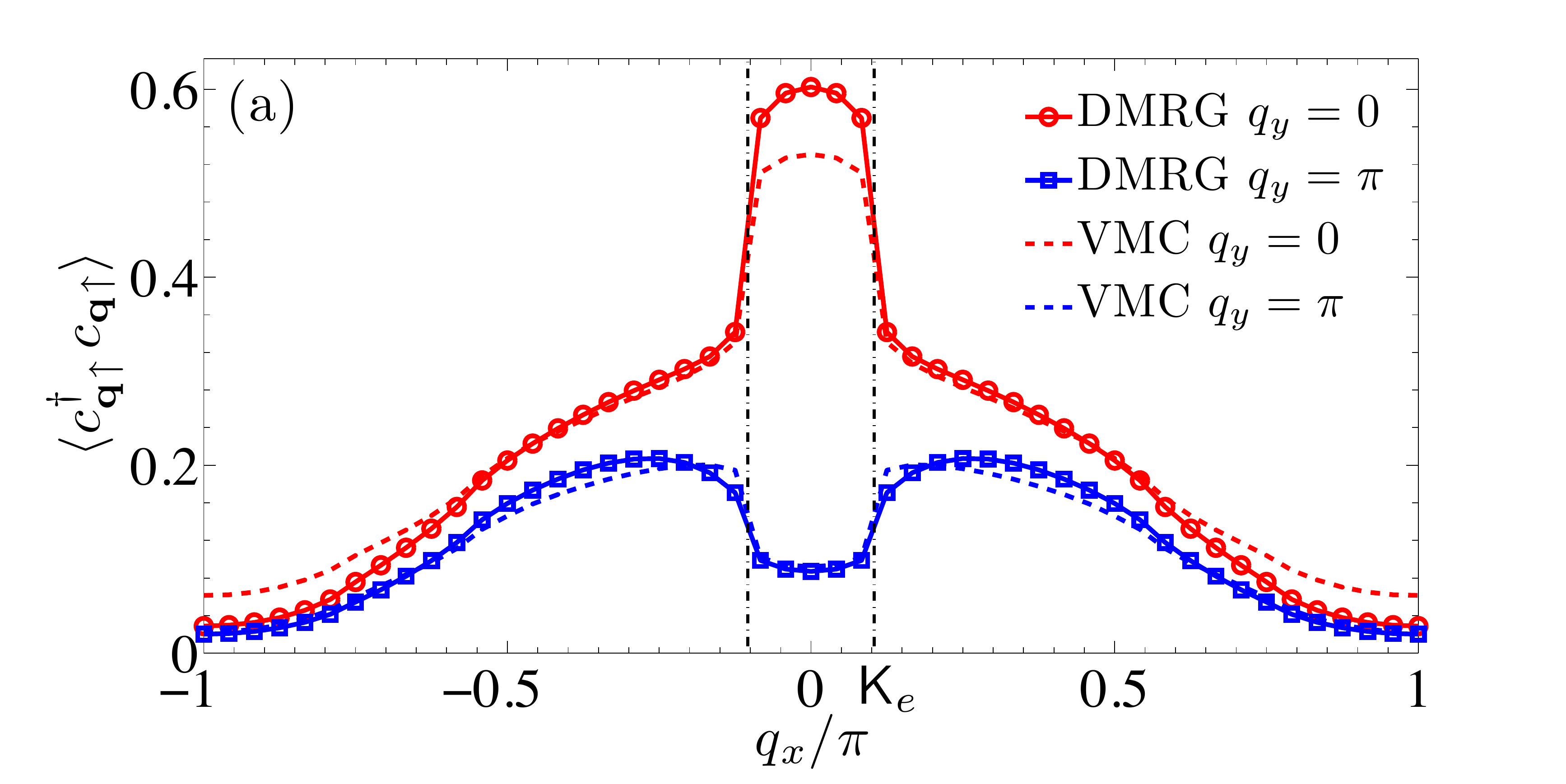}}}
	\centerline{\subfigure{\includegraphics[width=\columnwidth]{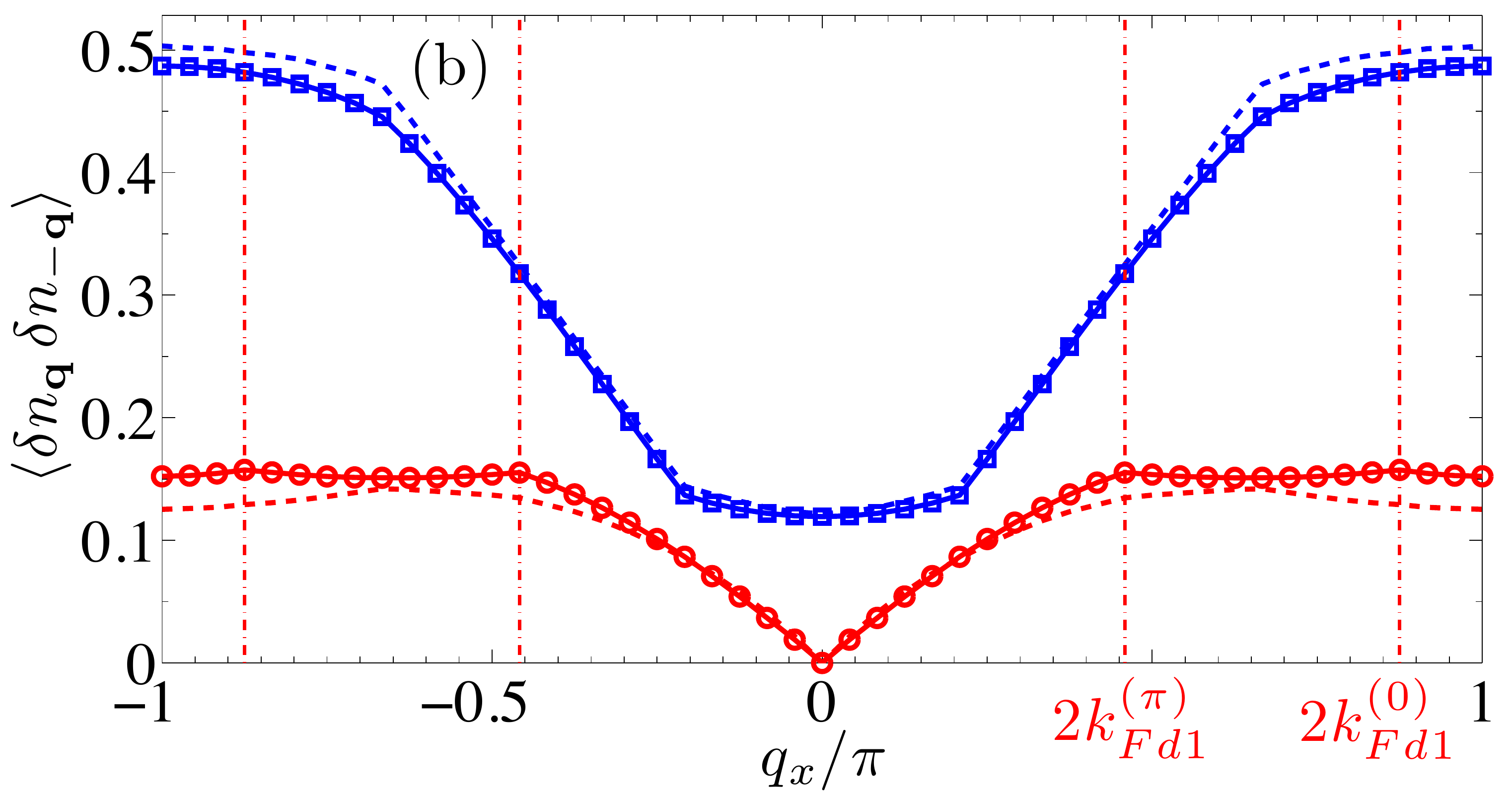}}}
	\centerline{\subfigure{\includegraphics[width=\columnwidth]{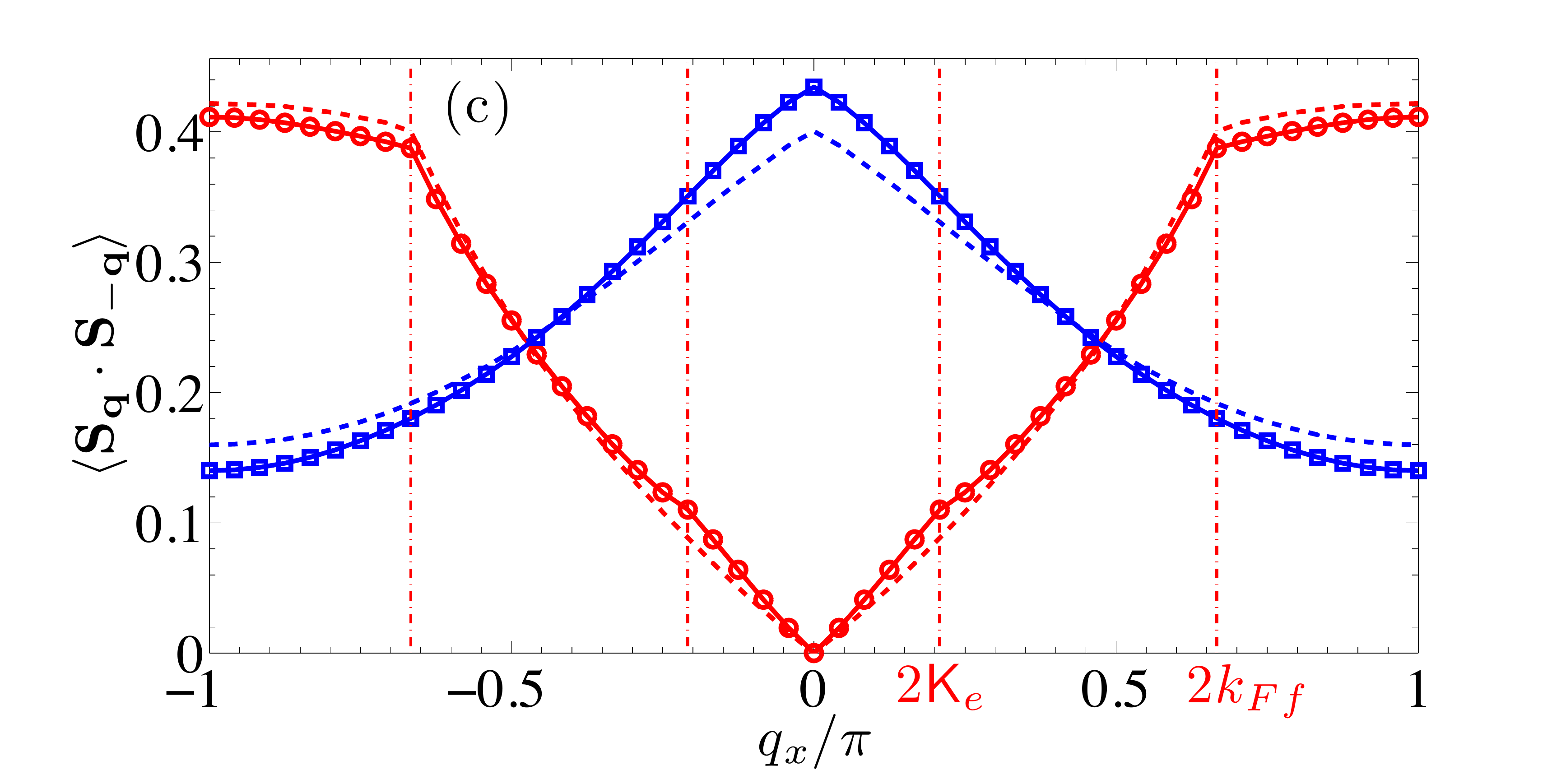}}}
	\caption{\dmetcorrsCaption}
	\label{fig:dmetcorrs}
	\end{figure}
\fi

We turn now to the characteristic point within the $d$-metal phase at $J/t=2$ and $K/t=1.8$.  In Fig.~\ref{fig:dmetcorrs}, we show a set of DMRG measurements at this point, as well as measurements corresponding to a variational wave function chosen such that its singular features best reproduce the DMRG data (see 
the Supplementary Information for details of our VMC methods).  The selected $d$-metal wave function is depicted schematically in Fig.~\ref{fig:partonbands}.  Specifically, we have the following parton Fermi wavevectors:  $2k_{Fd1}^{(0)}=21\cdot2\pi/48$, $2k_{Fd1}^{(\pi)}=11\cdot2\pi/48$, $2k_{Fd2}=32\cdot2\pi/48$, and $2k_{Ff}=16\cdot2\pi/48$.  The overall agreement between the DMRG and VMC measurements is very compelling, and we now summarize our understanding of these results from the perspective of the $d$-metal theory.

In sharp contrast to the conventional Luttinger liquid, the electron momentum distribution function now has singularities
for both $q_y=0$ and $q_y=\pi$ at a wavevector $q_x=\Ke \equiv [k_{Fd1}^{(0)}-k_{Fd1}^{(\pi)}]/2$.  This wavevector corresponds to a composite electron made from a combination of parton fields consisting of a right-moving $d_1$ parton, a left-moving $d_2$ parton, and a right-moving spinon:  $d_{1R}^{(q_y)}d_{2L}f_{\up R}$.  In fact, these ``enhanced electrons'' can be guessed from simple ``Amperian rules''\,\cite{LeeNagaosaWen, Polchinski94, Altshuler94} in our quasi-1D gauge theory as described in detail in 
the Supplementary Information.

The corresponding density-density and spin-spin structure factors, displayed in Fig.~\ref{fig:dmetcorrs}(b)-(c), also show nontrivial behavior.  We expect the density-density structure factor to be sensitive to each parton configuration individually and thus have singular features at various ``$2k_F$'' parton wavevectors (see Refs.~\onlinecite{DBL, Block2011_3legGMI} and 
the Supplementary Information).  In the DMRG measurements, the most noticeable features are at $q_y=0$ and $q_x=2k_{Fd1}^{(0)},2k_{Fd1}^{(\pi)}$, which allow us to directly read off the realized $d_1$ parton configuration (see Fig.~\ref{fig:partonbands}).  The lack of these features in the VMC data, as well as the lack of analogous features at $q_x=2k_{Fd2}$ in the DMRG data, can be understood within our gauge theory framework as presented in 
the Supplementary Information, where we also note that our wave function is only a caricature of the full theory.  Finally, the spin-spin structure factor at $q_y=0$ not only has a familiar, expected feature at $q_x=2k_{Ff}$ coming from the spinon, but also remarkably contains a feature at $q_x=2\Ke$ that can be thought of as a ``$2k_F$'' wavevector from the dominant ``electron'' in Fig.~\ref{fig:dmetcorrs}(a).  All in all, as we detail thoroughly in the Supplementary Information, the DMRG measurements are amazingly consistent, even on a fine quantitative level, with being in a stable non-Fermi liquid $d$-metal phase.

\def\singularQCaption
{
\textbf{Evolution of singular wavevectors in the $d$-metal phase.}
At fixed $J/t=2$ and varying $K/t$, we show the location of the dominant singular wavevector $\Ke$ in the electron momentum distribution function [see Fig.~\ref{fig:dmetcorrs}(a)], as well as the wavevectors identified as $2k_{Fd1}^{(0)}$ and $2k_{Fd1}^{(\pi)}$ in the density-density structure factor [see Fig.~\ref{fig:dmetcorrs}(b)].  These calculations were done with DMRG.
}

\ifREVTEX
	\begin{figure}[t]
	\centerline{\includegraphics[width=\columnwidth]{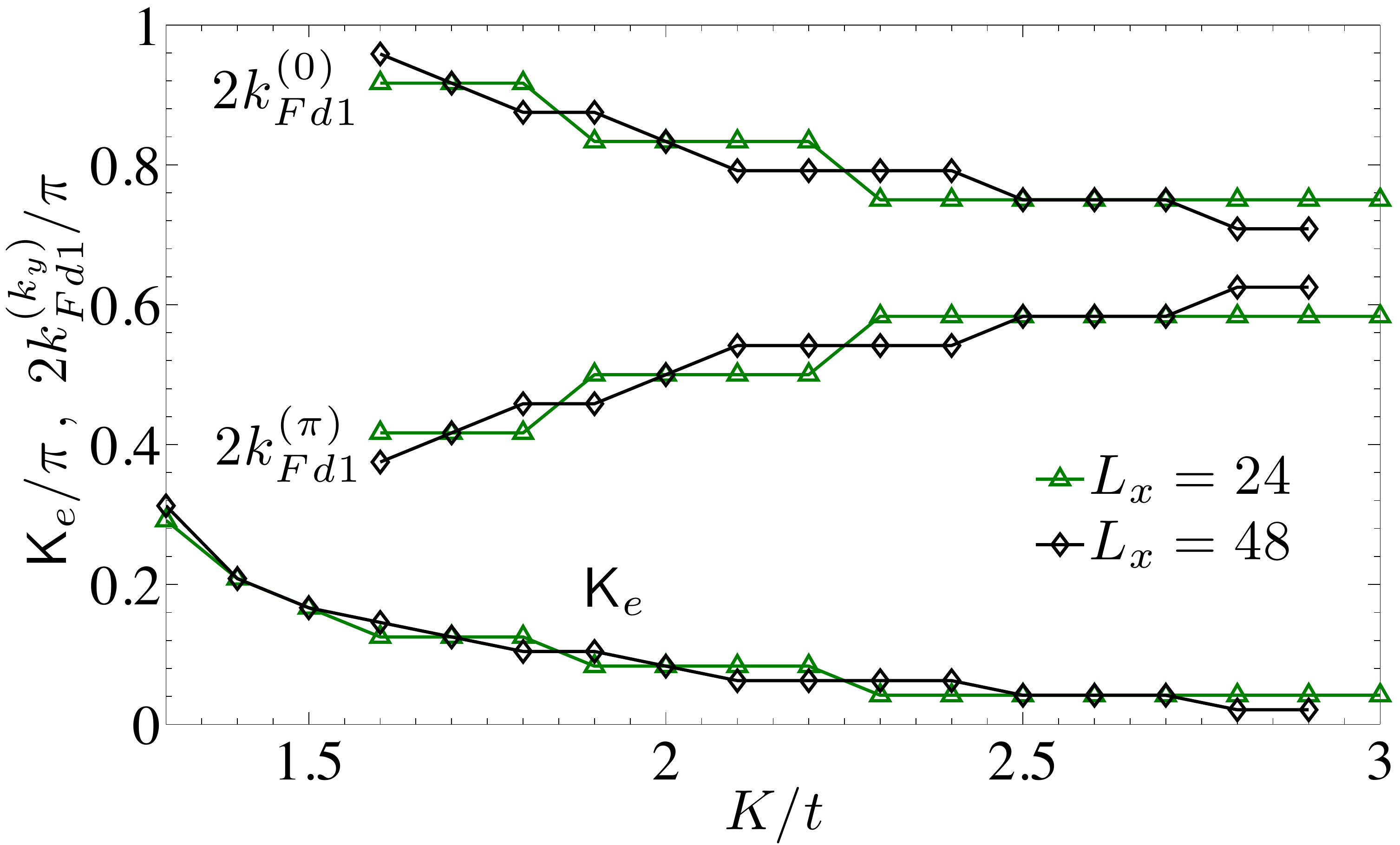}}
	\caption{\singularQCaption}
	\label{fig:singularQ}
	\end{figure}
\fi

It is important to note that the wavevector $\Ke$ depends on the interaction strength $K/t$ since the wavevectors $k_{Fd1}^{(0)}$ and $k_{Fd1}^{(\pi)}$ vary with ring exchange.\cite{Sheng2008_2legDBL}  In Fig.~\ref{fig:singularQ}, we show at $J/t=2$ evolution with $K/t$ of the wavevector $\Ke$, i.e., the location of the sharp steps in the electron momentum distribution function [see Fig.~\ref{fig:dmetcorrs}(a)], as determined by DMRG.  Since the momentum-space ``volume'' enclosed by these singular features depends on the interaction $K/t$ and is not simply determined by the total density of electrons, we may confidently say that the $d$-metal violates Luttinger's volume theorem.\cite{Luttinger60_PhysRev_119_1153}  In fact, the very notion of a single ``Fermi surface'' is actually ambiguous in the $d$-metal phase. We also show in Fig.~\ref{fig:singularQ}, for those values of $K/t$ at which they are discernible, the wavevectors $2k_{Fd1}^{(0)}$ and $2k_{Fd1}^{(\pi)}$ as identified by features in the DMRG-measured density-density structure factor at $q_y=0$ [see Fig.~\ref{fig:dmetcorrs}(b)].  For all points, the locations of the identified features satisfy the nontrivial identity $\Ke=[2k_{Fd1}^{(0)}-2k_{Fd1}^{(\pi)}]/4$, as predicted by our theory.

A remarkable property of the $d$-metal state found in the DMRG is that it has prominent critical $d$-wave Cooper pairs residing on the diagonals, as anticipated earlier from the ring energetics.  We detail these findings in 
the Supplementary Information, while here we only mention that such Cooper pair correlations have the slowest power law decay among all the discussed observables, including the electron Green's function.  This is in stark contrast with the conventional metal and suggests that the $d$-metal phase has some incipient $d$-wave superconductivity in two dimensions.

As a final piece of ``smoking gun'' evidence that the realized DMRG phase is in fact the $d$-metal, we have measured the number of 1D gapless modes, i.e., the effective central charge $c$, via scaling of the bipartite entanglement entropy\cite{Cardy04_JStatMech_P06002,Calabrese10_PRL_104_095701} in the DMRG and VMC\cite{Hastings10_PRL_104_157201,Grover11_PRL_107_067202} wave functions.  As explained above, we expect $c=2$ in the conventional Luttinger liquid and $c=3$ in the $d$-metal.   A detailed comparison of the DMRG and VMC entropy measurements is presented in 
the Supplementary Information, where the DMRG-VMC agreement is just as impressive as it is for the more traditional measurements of Fig.~\ref{fig:dmetcorrs}.  The effective central charge versus $K/t$ at $J/t=2$ as determined by the DMRG is shown in Fig.~\ref{fig:centralCharge}.  Indeed, these measurements indicate that $c\simeq2$ in the conventional one-band metal, while $c\simeq3$ in the exotic $d$-metal.  Since $c=3>2$, our putative $d$-metal phase clearly cannot be understood as an instability out of the conventional one-band metal, but also, since $c=3<4$, the critical bonding and antibonding electrons in Fig.~\ref{fig:dmetcorrs}(a) cannot be reproduced by \emph{any} perturbative treatment starting from free electrons\cite{Balents96_PRB_53_12133} (see also the Supplementary Information).

\def\centralChargeCaption
{
\textbf{Central charge $c$ as a function of interaction $K/t$.}
By measuring the von Neumann entanglement entropy $S_1$ in the DMRG, we calculate the effective central charge $c$ at fixed $J/t=2$ and varying $K/t$.  There is a dramatic jump from $c\simeq2$ to $c\simeq3$ at the transition, as predicted by our theory.  Data for two example points, $K/t=0.8$ and 1.8, is shown in the inset, where $X$ is the number of rungs in each bipartition.  (See also 
the Supplementary Information.)
}

\ifREVTEX
	\begin{figure}[t]
	\centerline{\includegraphics[width=\columnwidth]{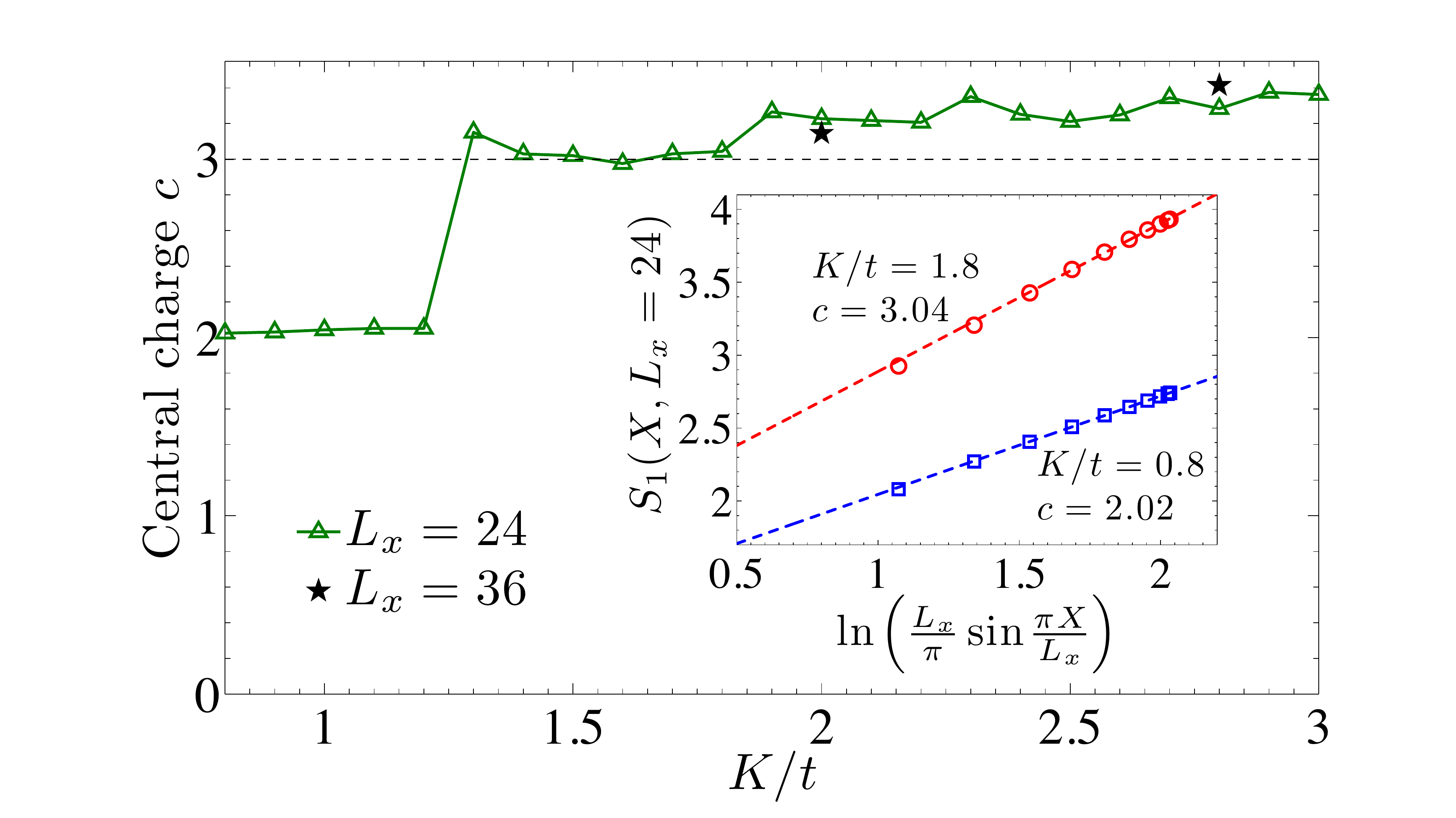}}
	\caption{\centralChargeCaption}
	\label{fig:centralCharge}
	\end{figure}
\fi

\section*{\uppercase{Discussion and outlook}}

In this paper, we have presented exceptionally strong evidence for stability of a two-leg descendant of our exotic strange metal-type phase, the $d$-wave metal, and we conclude here with an outlook on exciting future work.  Firstly, it would be desirable to march towards two dimensions by studying systems with more legs, where our present two-leg $d$-metal treatment is readily extendable.  One of the main purposes of Ref.~\onlinecite{Mishmash2011_4legDBL} was to establish stability of the $d$-wave Bose metal, the main ingredient of the $d$-metal, on three- and four-leg ladders, and this was indeed achieved.  Thus, we do not envision any conceptual obstacles in the way of realizing a similar result for the $d$-metal.  However, we do anticipate that going to more legs will be very challenging numerically for the DMRG due to the large amount of spatial entanglement present in the $d$-metal and Fermi liquid---this is also the current limitation preventing modern 2D tensor network state methods from attacking such problems.\cite{Corboz10_PRB_81_165104}

With the goal of connecting to experiment, it would also be interesting to perform a detailed energetics study of the \tJK model in two dimensions and explore applicability of such models to strongly correlated materials.  By studying 2D variational wave functions based on the $d$-metal, it should be possible to compare physical properties with experimentally observed strange metals, such as that in the cuprates.  This could include various instabilities of the $d$-metal, e.g., spinon pairing as a model of a pseudogap metal or chargon pairing as a model of an ``orthogonal metal'' discussed recently.\cite{Nandkishore12_PRB_86_045128}  It would be particularly exciting to investigate incipient $d$-wave superconductivity of the cuprate variety---this is rather natural given the $d$-wave sign structure already inherent in the nonsuperconducting parent $d$-metal.  Finally, while we have thus far stressed its Luttinger volume violation as a characteristic non-Fermi liquid property of the $d$-metal, we note that the 2D phase will also have no Landau quasiparticle as well as exhibit non-Fermi-liquid-like thermodynamics and transport.  Comparing these predictions with properties of real strange metals would be an interesting endeavor.  In the end, however, we would like to stress the conceptual nature of the present study, and we hope that our ideas may open up new avenues for thinking about non-Fermi liquid electronic fluids.

\ifREVTEX
	\bibliographystyle{prstyNew}
\else
	\bibliographystyle{naturemag}
	\bibliography{bib4ElRings}
\fi

\def\acknowledgementsText
{
We would like to thank T.~Senthil, R.~Kaul, L.~Balents, S.~Sachdev, A.~Vishwanath, and P.~Lee for useful discussions.  This work was supported by the NSF under the KITP grant PHY05-51164 and the MRSEC Program under Award No.~DMR-1121053 (H.C.J.), the NSF under grants DMR-1101912 (M.S.B., R.V.M., J.R.G., and M.P.A.F.), DMR-1056536 (M.S.B.), DMR-0906816 and DMR-1205734 (D.N.S.), DMR-0907145 (O.I.M.), and by the Caltech Institute of Quantum Information and Matter, an NSF Physics Frontiers Center with support of the Gordon and Betty Moore Foundation (O.I.M. and M.P.A.F.).  We also acknowledge support from the Center for Scientific Computing from the CNSI, MRL: an NSF MRSEC (DMR-1121053), and NSF CNS-0960316.
}

\def\authorContributionsText
{
All authors made significant contributions to the research underlying this paper.
}

\def\authorInfoText
{
Reprints and permissions information is available at www.nature.com/reprints. The authors declare no competing financial interests. Correspondence and requests for materials should be addressed to R.V.M. (mishmash@physics.ucsb.edu).
}

\ifREVTEX
	\acknowledgements \acknowledgementsText
\else
	\begin{addendum}
	\item[Acknowledgements] \acknowledgementsText
	\end{addendum}
	
	\begin{addendum}
	\item[Author Contributions] \authorContributionsText
	\end{addendum}
	
	\begin{addendum}
	\item[Author Information] \authorInfoText
	\end{addendum}
	
	\newpage
		
	\begin{figure}
	\centerline{\includegraphics[width=\columnwidth]{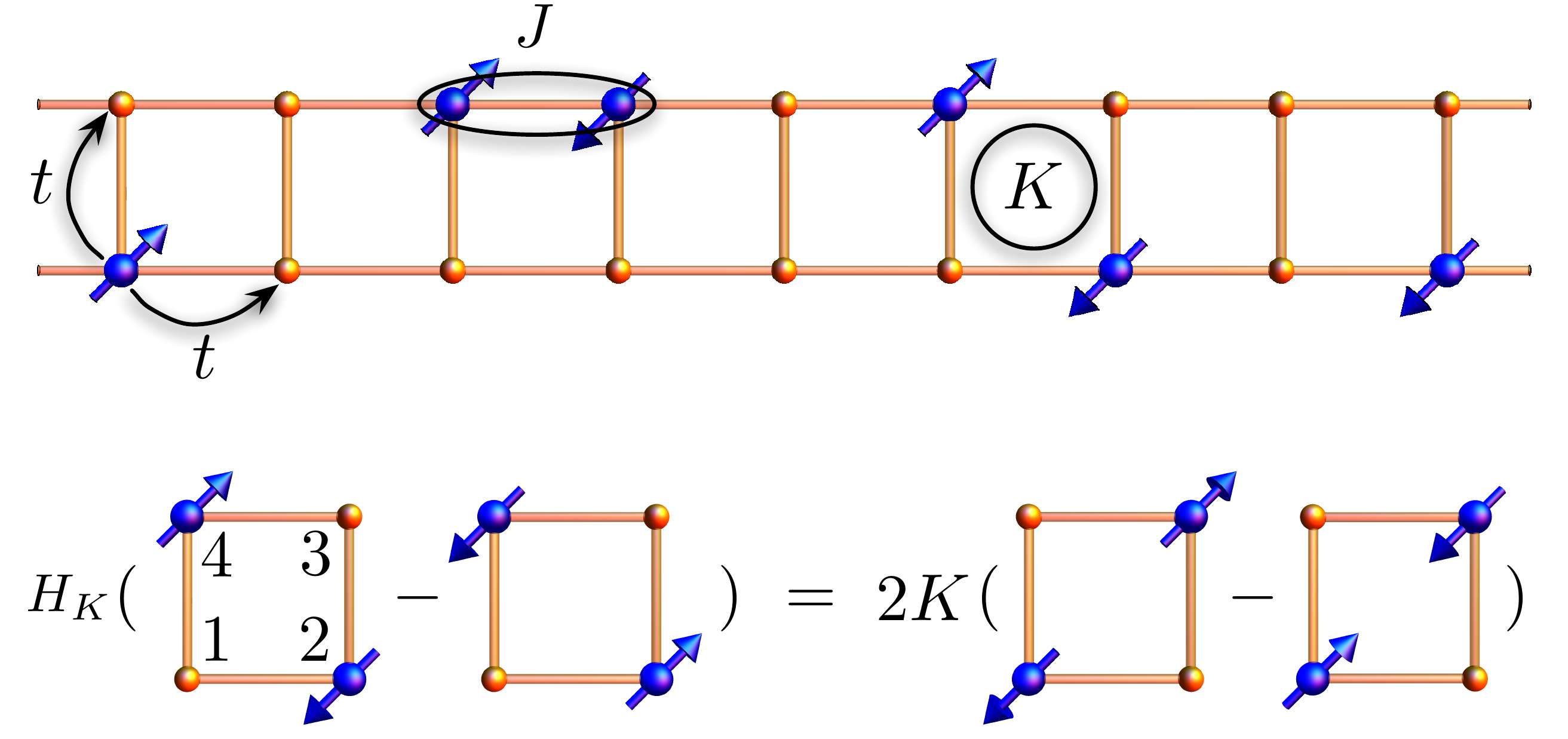}}
	\caption{\modelCaption}
	\label{fig:model}
	\end{figure}
	
	\begin{figure}
	\centerline{\includegraphics[width=\columnwidth]{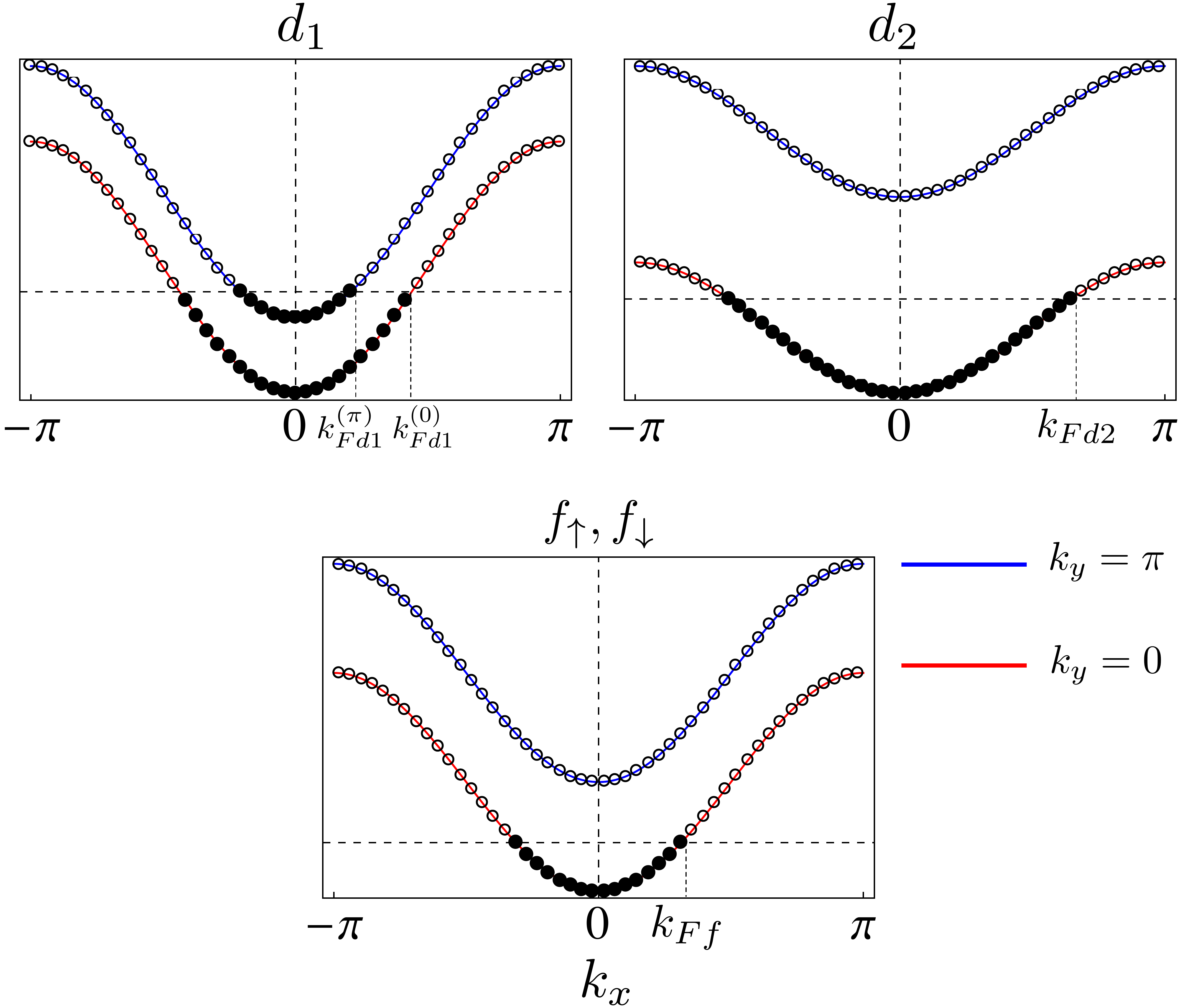}}
	\caption{\partonbandsCaption}
	\label{fig:partonbands}
	\end{figure}
	
	\begin{figure}
	\centerline{\includegraphics[width=\columnwidth]{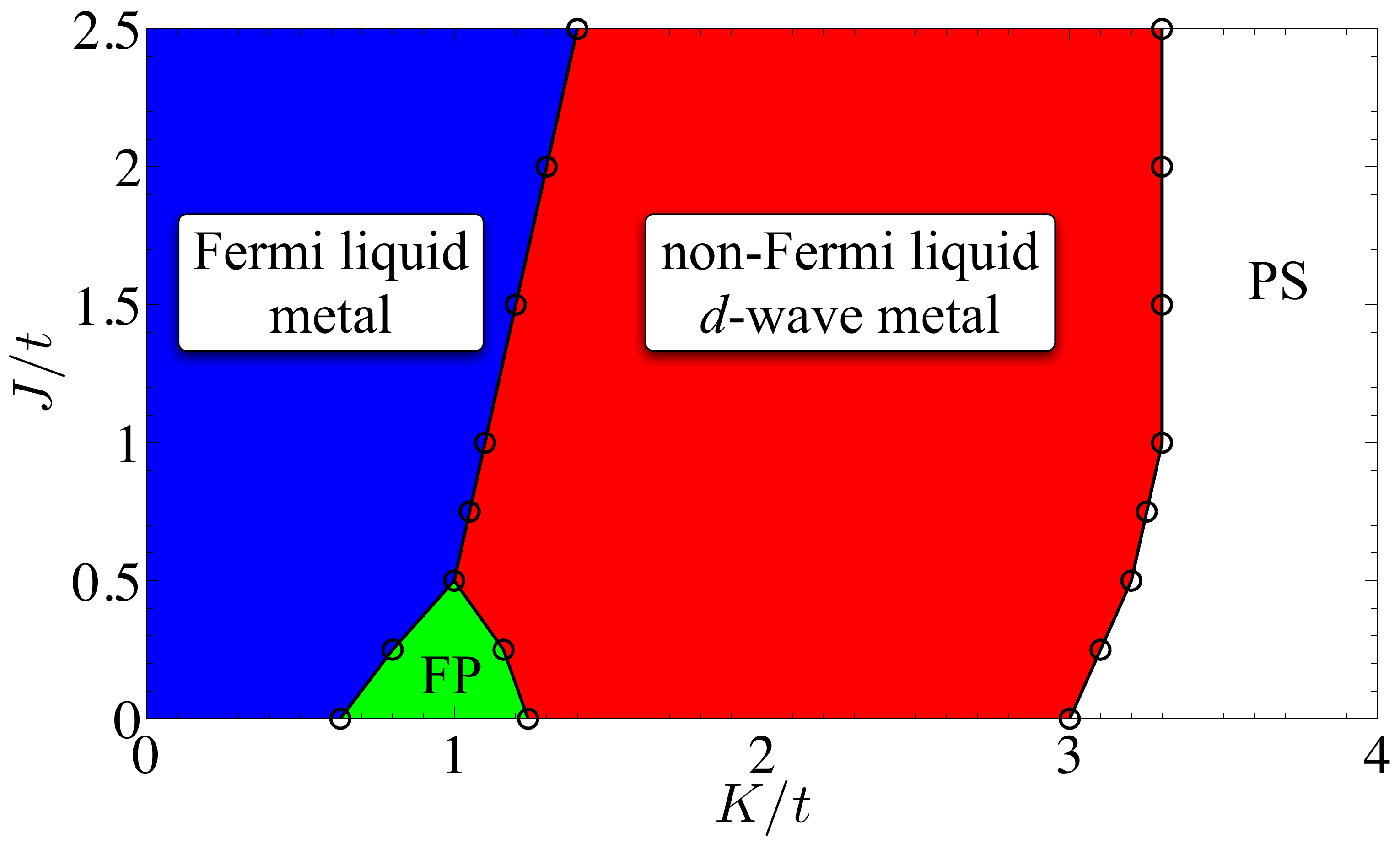}}
	\caption{\phasedCaption}
	\label{fig:phased}
	\end{figure}
	
	\begin{figure}
	\centerline{\subfigure{\includegraphics[width=\corrScaleFac\columnwidth]{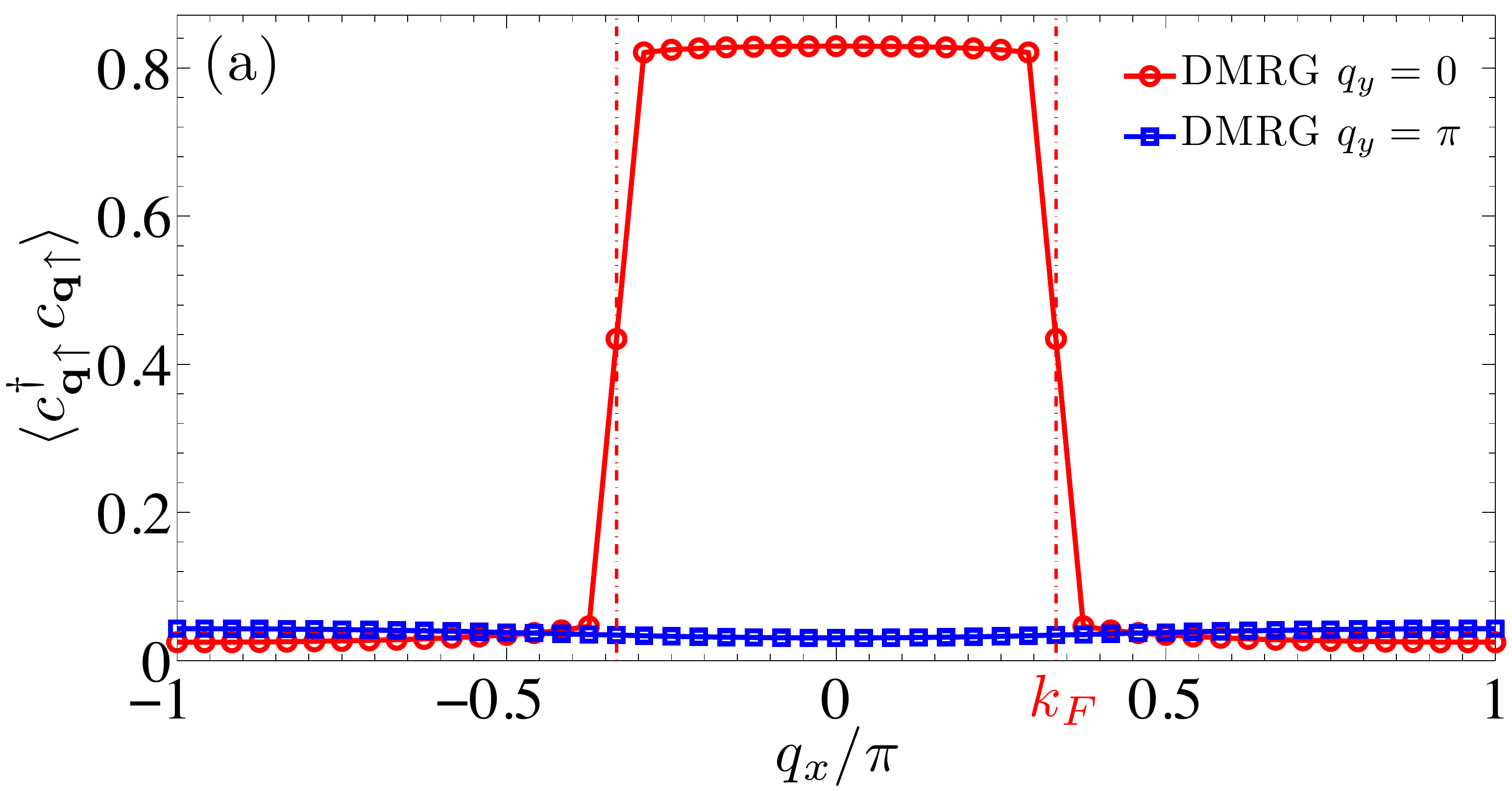}}}
	\centerline{\subfigure{\includegraphics[width=\corrScaleFac\columnwidth]{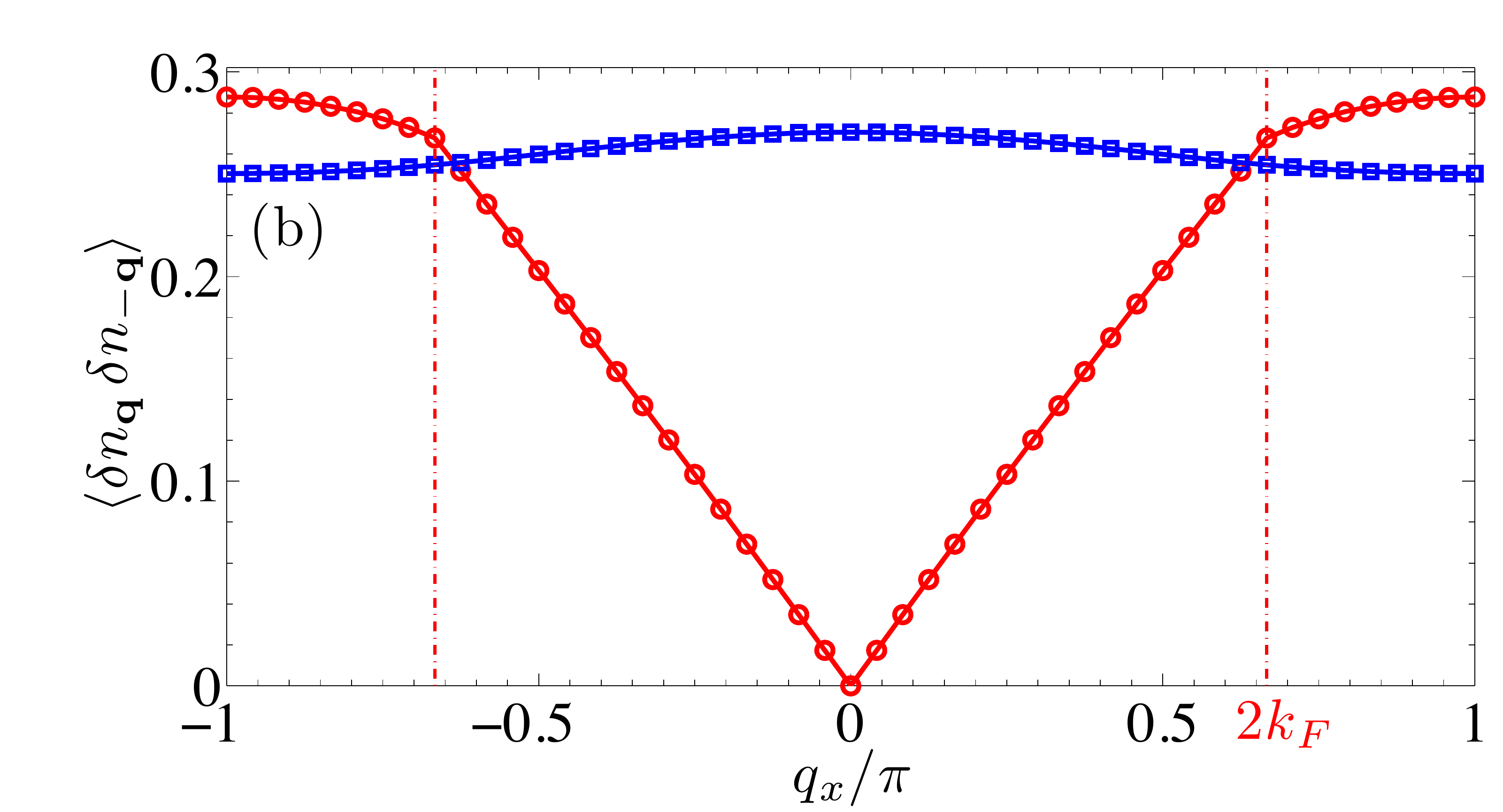}}}
	\centerline{\subfigure{\includegraphics[width=\corrScaleFac\columnwidth]{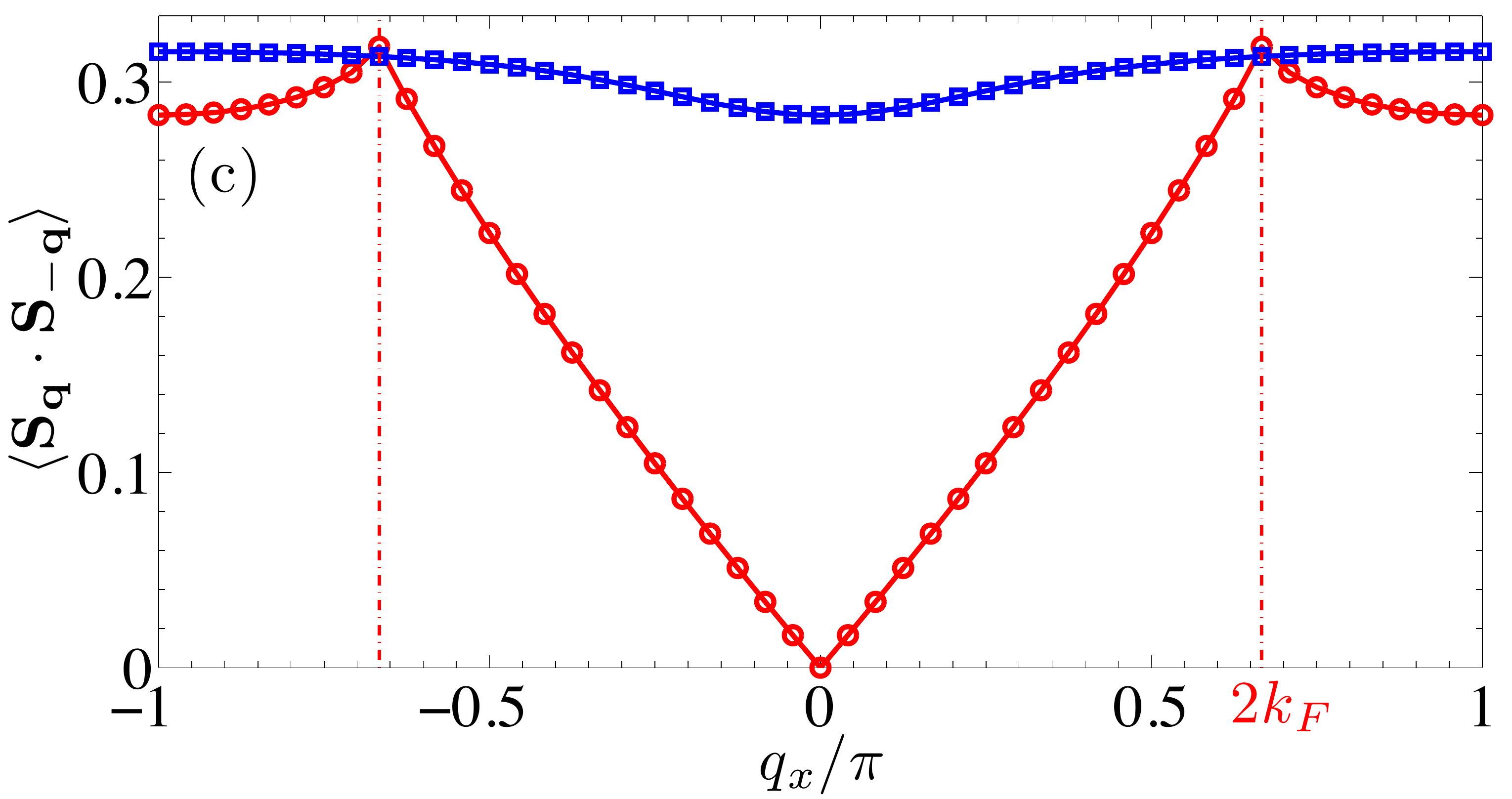}}}
	\caption{\FLcorrsCaption}
	\label{fig:FLcorrs}
	\end{figure}
	
	\begin{figure}
	\centerline{\subfigure{\includegraphics[width=\corrScaleFac\columnwidth]{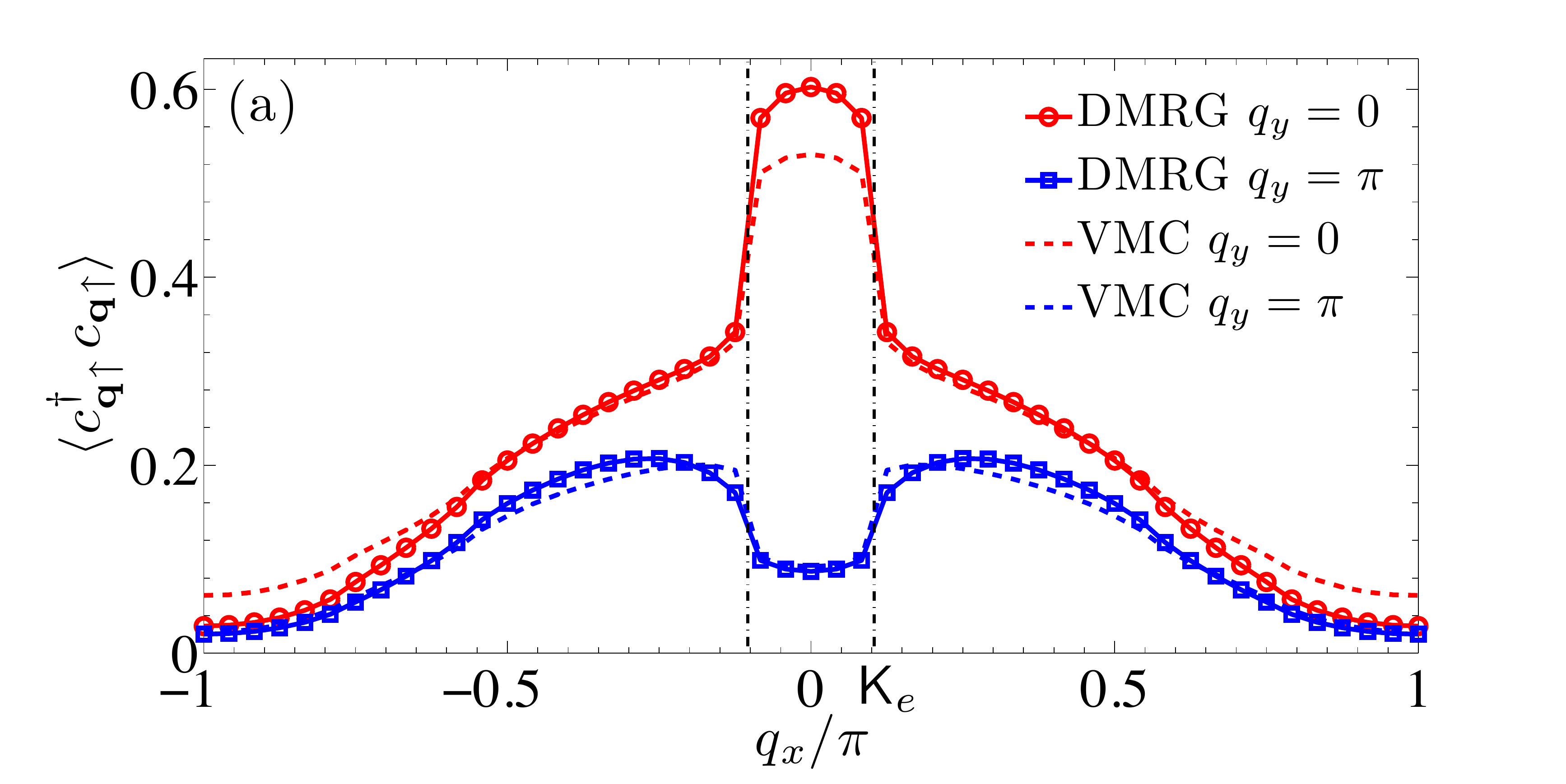}}}
	\centerline{\subfigure{\includegraphics[width=\corrScaleFac\columnwidth]{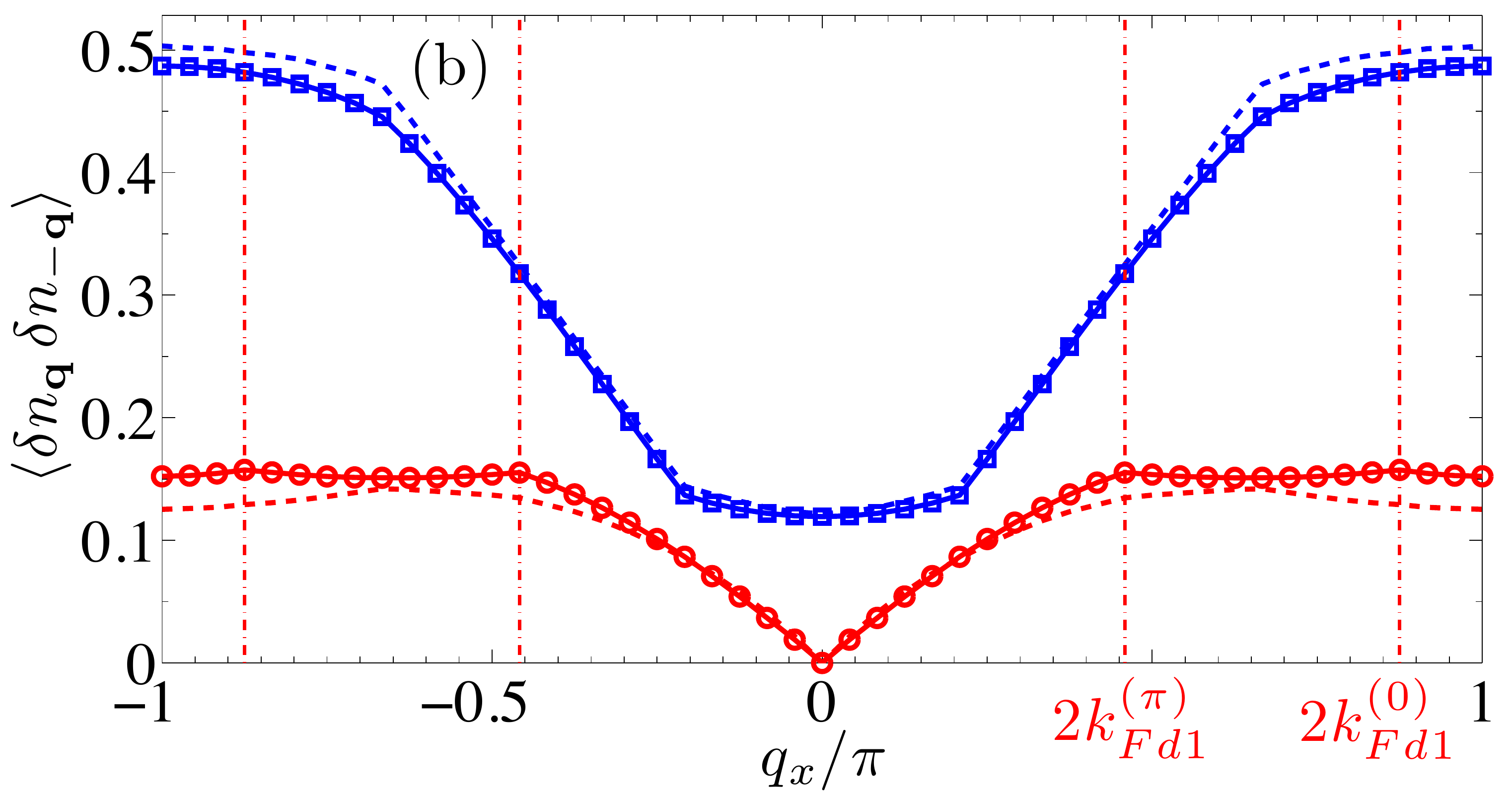}}}
	\centerline{\subfigure{\includegraphics[width=\corrScaleFac\columnwidth]{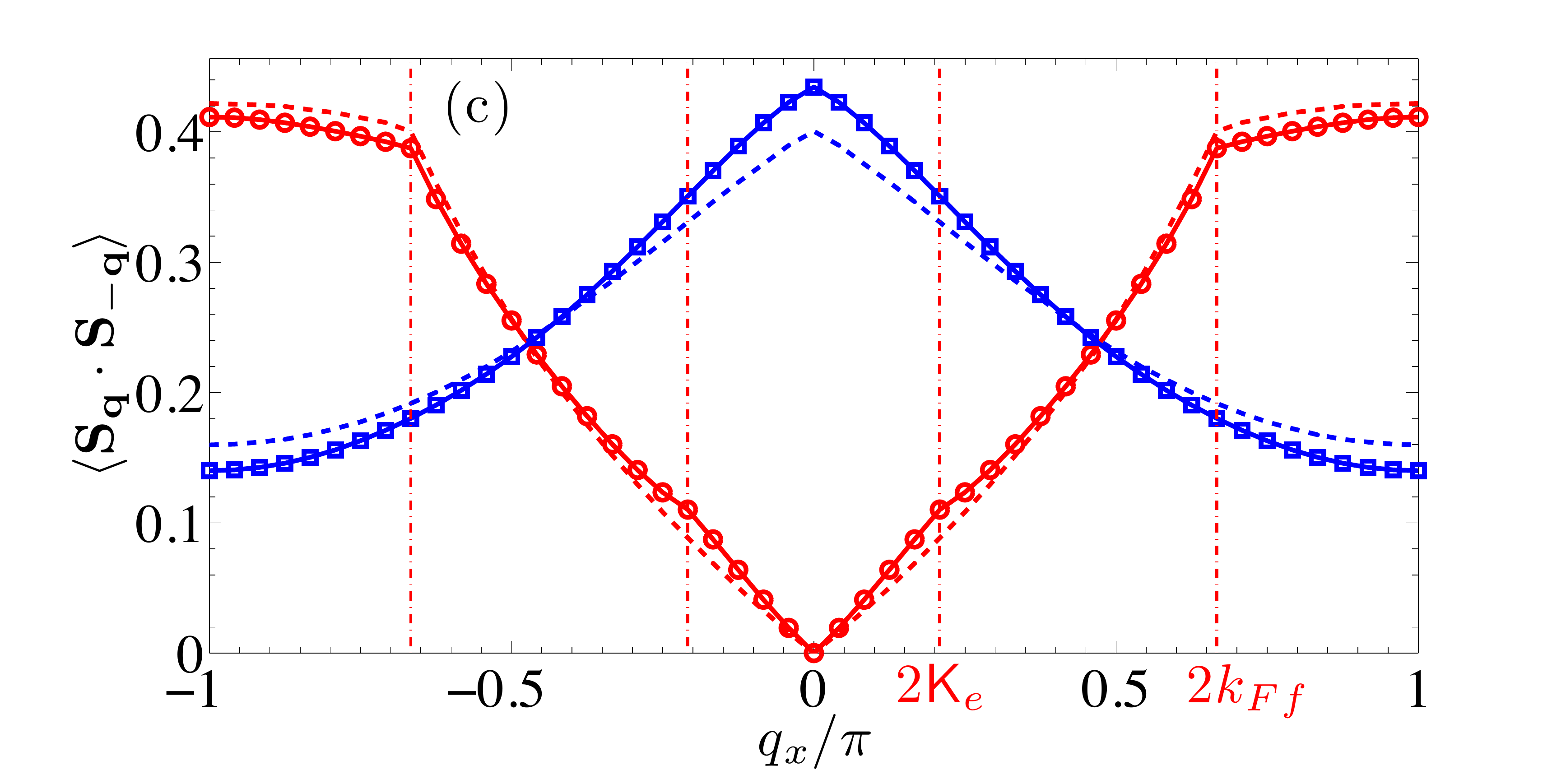}}}
	\caption{\dmetcorrsCaption}
	\label{fig:dmetcorrs}
	\end{figure}
	
	\begin{figure}
	\centerline{\includegraphics[width=\columnwidth]{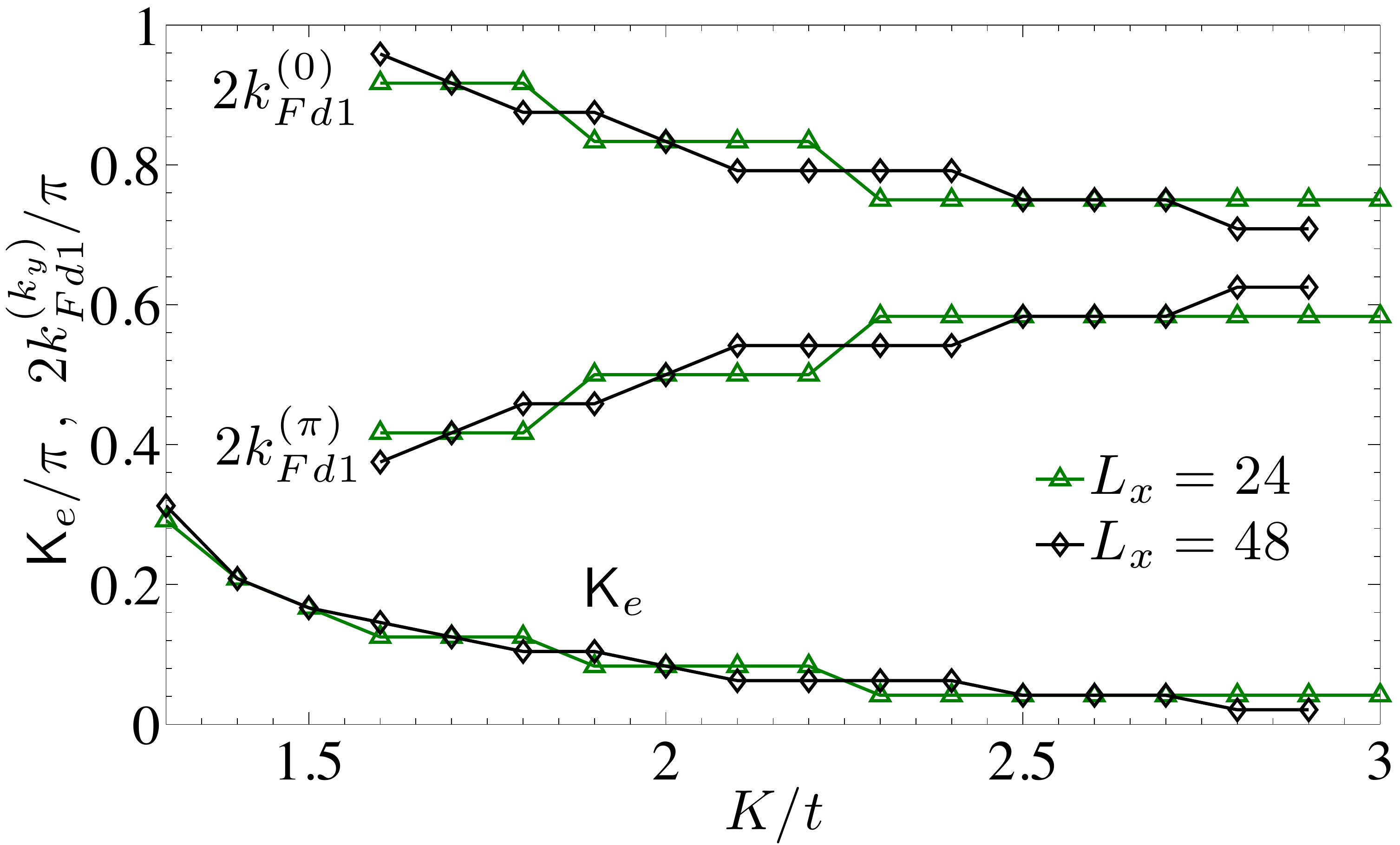}}
	\caption{\singularQCaption}
	\label{fig:singularQ}
	\end{figure}
	
	\begin{figure}
	\centerline{\includegraphics[width=\columnwidth]{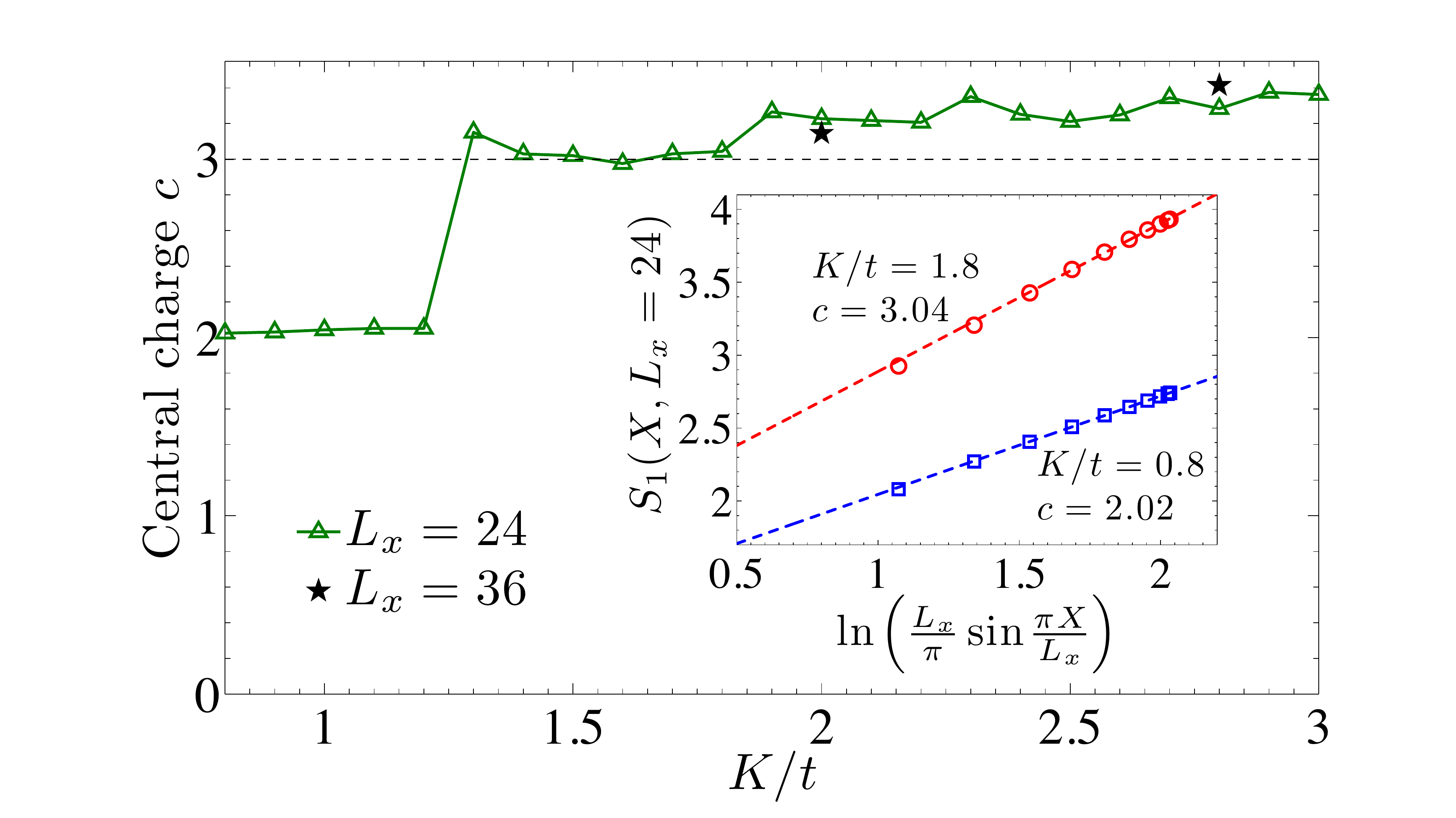}}
	\caption{\centralChargeCaption}
	\label{fig:centralCharge}
	\end{figure}
	\vfill
	
%
%

\fi

\ifREVTEX
	\section*{\underline{\uppercase{Supplementary Information}}}
\else
	\begin{SI}
	\newpage
\fi



\section{Numerical methods and supporting data}
\label{app:num_methods}

In this part, we provide some details of the DMRG and VMC methods and also show comparison of the R\'enyi entropies between the two.  We will present additional data analysis in Sec.~\ref{subsec:fits} after developing a long-wavelength description of the $d$-metal phase.

\subsection{DMRG calculations and observables}

We determine the ground state phase diagram of the $t$-$J$-$K$ model, Eq.~(\ref{eqn:fullmodel}) of the main text, by large-scale DMRG calculations.  We consider square lattice clusters with total number of sites $L_x \times L_y$.  Here, we study the two-leg ladder system, i.e., $L_y=2$, and use periodic boundary conditions along the $\hat{x}$ direction.

Our DMRG calculations generally keep between $m=5$,000 and 20,000 states in each DMRG block.  This is found to give excellent convergence in the measurements such as the ground state energy and various correlation functions defined below, with small errors which can be neglected safely for our sizes up to $L_x = 48$.  The phase boundaries in the ($J/t, K/t$)-parameter space are determined by extensive scans of the derivatives of the ground state energy and by monitoring the correlation functions.  On the other hand, as we will describe later, even with such a large $m$ we can converge the entanglement entropy only for sizes up to $L_x = 36$.

To characterize the ground state properties of the system, as well as properties of the variational wave functions, we calculate the electron Green's function
\begin{eqnarray}
G_e(\rvec_i - \rvec_j) = \langle c_{is}^\dagger c_{js} \rangle,
\end{eqnarray}
where $e=c_s$ with $s =\,\up,\dn$ the electron spin (there is no implied summation over $s$).  For our electron Green's function calculations and analysis, we fix $s$ to one of the two possible flavors of spin, say $s=\, \up$, which in the spin-singlet states considered in this work, gives the same Green's function as the other flavor of spin.  The Fourier transform gives us the electron momentum distribution function
\begin{eqnarray}
\langle c_{\qvec s}^\dagger c_{\qvec s}\rangle = 
\frac{1}{L_x L_y} \sum_{ij} e^{i \qvec \cdot (\rvec_i - \rvec_j)} \langle c_{is}^\dagger c_{js}\rangle ~.
\end{eqnarray}
Similarly, we calculate the electron density-density structure factor in momentum space
\begin{align}
\langle \delta n_{\qvec} \delta n_{-\qvec}\rangle =
\frac{1}{L_x L_y} \sum_{ij} e^{i \qvec \cdot (\rvec_i - \rvec_j)} \langle (n_i - \rho)(n_j - \rho) \rangle,
\end{align}
where $n_i = \sum_s c_{is}^\dagger c_{is}$ is the electron number operator at site $i$ and $\rho$ is the electron density.  
To characterize the magnetic properties of the system, we also study the spin structure factor
\begin{eqnarray}
\langle \mathbf{S}_{\qvec}\cdot \mathbf{S}_{-\qvec}\rangle = 
\frac{1}{L_x L_y} \sum_{ij} e^{i \qvec \cdot (\rvec_i - \rvec_j)} \langle \mathbf{S}_i\cdot \mathbf{S}_j \rangle,
\end{eqnarray}
where the spin operator is defined as $\mathbf{S}_i = \frac{1}{2} \sum_{s,s'} c_{is}^\dagger {\bm \sigma}_{ss'} c_{is'}$.

Finally, we measure Cooper pair correlations
\begin{equation}
G_{\rm Cooper}[\rvec_i, \rvec_j; \rvec_k^\prime, \rvec_l^\prime]
= \langle {\mathsf P}[i, j]^\dagger {\mathsf P}[k, l] \rangle,
\end{equation}
where a Cooper pair operator residing on some nearby sites $i$ and $j$ is defined by
\begin{equation}
\label{CooperPij}
{\mathsf P}[i, j] = \frac{1}{\sqrt{2}}\left(c_{i\up} c_{j\dn} - c_{i\dn} c_{j\up}\right)
= {\mathsf P}[j, i] ~.
\end{equation}
We will specifically be interested in diagonal $d$-wave Cooper pairs as detailed in Sec.~\ref{subsec:Cooper}.  Their correlations can be accessed by considering the combination $G_{\rm Cooper}[(x,1), (x+1,2); (x',1), (x'+1,2)] - G_{\rm Cooper}[(x,1), (x+1,2); (x',2), (x'+1,1)]$; it is this specific quantity which we plot below in Fig.~\ref{fig:DDwaveCooper}.


\subsection{Details of the VMC calculations} \label{app:vmcDetails}
The central task of the VMC analysis is to construct a pool of variational states for which to collect data that can be compared with the DMRG results.  The determinantal VMC with the $d$-metal wave functions is straightforward:  Measurements in the VMC simulation of the energy and correlation functions described above are all accomplished by averaging over relative probability amplitudes corresponding to one or two particle hops.  These are simply ratios of determinants with one or two changed columns, which can be computed efficiently using the previously established methods in Ref.~\onlinecite{Ceperley77_PRB_16_3081}.  The one exception is the density-density correlator, which merely requires averaging over the product of density operators.  These are measured simply by checking for the presence of particles on the relevant sites and therefore involve no relative probability amplitudes.

In principle, one can be quite exhaustive and consider all possible band fillings of the two bands for each of the four partons in the $d$-metal wave function, Eq.~(\ref{eqn:dmetalwf}) of the main text, on the two-leg ladder.  Additionally, there exists the freedom to adjust the boundary conditions for each of the partons so long as the overall conditions are periodic for the electron wave function.  There are four combinations that accomplish this.  Listing the boundary conditions in order of $d_1$, $d_2$, $f_s$, we have: (1) periodic-periodic-periodic, (2) antiperiodic-antiperiodic-periodic, (3) periodic-antiperiodic-antiperiodic, (4) antiperiodic-periodic-antiperiodic.  Finally, we can introduce two other variational parameters in the form of exponents on the magnitudes of the $d_1$ and $d_2$ determinants.\cite{HellbergMele1991, Sheng2008_2legDBL}  That is, we can replace $\psi_{d1}(\{\mathbf{R}_i\}) \rightarrow \lvert\psi_{d1}(\{\mathbf{R}_i\})\rvert^{p_1-1} \psi_{d1}(\{\mathbf{R}_i\})$ and $\psi_{d2}(\{\mathbf{R}_i\}) \rightarrow \lvert\psi_{d2}(\{\mathbf{R}_i\})\rvert^{p_2-1} \psi_{d2}(\{\mathbf{R}_i\})$.

In practice, we do not need to consider all states exhaustively.  From our mean-field understanding of the $d$-metal phase, we can make two assumptions outright.  First, we only need to fill the bands with a single contiguous strip in each band (centered around $k_x=0$ to respect the lattice inversion symmetry).  Second, we only need to fill the $k_y=0$ band for the spinons.  Furthermore, the DMRG results give us three important hints.  First, in the $d$-metal region of the phase diagram, the exact ground states are either spin singlets or have a small spin.  For the cut $J/t=2$, where we focused extensively in our analysis, all $d$-metal states have $S=0$ or $S=1$.  Second, the DMRG states show a strong aversion to nonzero momentum, which, in the VMC language, corresponds to symmetric filling of all the bands.  Finally, by reading off the locations of the singular features in the DMRG data (specifically, the density-density structure factor), we can deduce the corresponding band fillings for the two bands in the $d_1$ parton Fermi sea.  As we increase $K/t$ in the $d$-metal region at fixed $J/t$, the singular features move such that the $d_1$ band fillings evolve accordingly.  For example, for the $48\times 2$ system with $N_e=32$ presented in Fig.~\ref{fig:singularQ} of the main text, we need band fillings ranging from $N_{d1}^{(0)}=23$ and $N_{d1}^{(\pi)}=9$ to $N_{d1}^{(0)}=N_{d1}^{(\pi)}=16$.  To accommodate the zero-momentum constraint, the up and down spinon numbers oscillate between $N_{f\uparrow}=N_{f\downarrow}=16$ for $N_{d1}^{(0)}$ and $N_{d1}^{(\pi)}$ odd and $N_{f\uparrow}=17$, $N_{f\downarrow}=15$ for $N_{d1}^{(0)}$ and $N_{d1}^{(\pi)}$ even.  This corresponds to switching between boundary condition type (3) and type (2) listed above and total spin $S=0$ and $S=1$, respectively.

Amazingly, we actually observe this precise shell-filling effect in the DMRG data itself.  Specifically, in the $2k_{Fd1}^{(k_y)}=N_{d1}^{(k_y)}\cdot2\pi/L_x$ data presented in Fig.~\ref{fig:singularQ} of the main text, we measure a total spin $S=0$ when $N_{d1}^{(k_y)}$ is odd, and $S=1$ when $N_{d1}^{(k_y)}$ is even (since in our calculations $N_e=N_{d1}^{(0)}+N_{d1}^{(\pi)}$ is even, $N_{d1}^{(0)}$ and $N_{d1}^{(\pi)}$ have the same parity). This is very remarkable because the DMRG is producing an unbiased ground state of the \tJK model and, within its inner workings, has no notion of $d$ and $f$ partons. Thus, \emph{a priori} there is no reason to expect such a precise oscillation of the total spin between $S=0$ and $S=1$ as we vary the Hamiltonian parameter $K/t$.  Nonetheless, we do indeed observe this perspicuous effect, exactly as predicted by our $d$-metal theory.


Now going back to the VMC wave functions themselves, once we put all of these considerations together the set of viable band fillings for the four partons is reduced to a very reasonable number.  For example, for the $48\times 2$ system, there are only eight such ``bare" Gutzwiller states across the $J/t=2$ cut.  We use the term ``bare" when the exponents on the $d_1$ and $d_2$ determinants are set to one.  For each of these states, there is then the flexibility to adjust these exponents continuously.  In attempting to match DMRG results, as in Fig.~\ref{fig:dmetcorrs} of the main text, after choosing the correct bare state, we tune the exponents until the features visually match as well as possible.  Admittedly, this process is a bit subjective, especially since gaining agreement of some singular features comes at the expense of poorer agreement of other singular features.  However, an energetics analysis, with respect to the varied exponents but within a fixed bare state, reveals that the chosen states have energies that are very close to the rather robust energetic minima.  Performing energy minimization over all possible orbital fillings usually gives a state that is off by one or two filled orbitals, but again the energies of our chosen states matching the DMRG results are not significantly different from the optimal energy.  In Fig.~\ref{fig:dmetcorrs} of the main text, the chosen exponents are $p_1=0.7$ and $p_2=-0.4$.  Indeed a negative exponent on the $d_2$ determinant is required to achieve a good match with the DMRG.  This corresponds to negating the overly strong Pauli repulsion coming from all determinantal factors of the wave function.  We can gain further understanding of the properties of our VMC wave functions from the $d$-metal gauge theory described in Sec.~\ref{app:dmetTheory} below.


\subsection{Entanglement entropy results}
\label{app:entropy}

The R\'enyi entanglement entropies are defined by
\begin{equation}
  S_\alpha(\rho_A) = \frac{1}{1-\alpha} \ln \left[ \mathrm{Tr} \left( \rho_A^\alpha \right) \right],
\end{equation}
where $\rho_A$ is the reduced density matrix of a subregion $A$ of the lattice.  The limit $\alpha \rightarrow 1$ results in the familiar von Neumann entropy $S_1 = -\mathrm{Tr}\left( \rho_A \ln \rho_A \right)$.

According to conformal field theory, any given R\'enyi entropy is expected to scale as\cite{Cardy04_JStatMech_P06002,Calabrese10_PRL_104_095701}
\begin{equation}
\label{eqn:cardyrenyi}
  S_\alpha^\mathrm{CFT}(X,L_x) = \frac{c}{6}\left( 1 + \frac{1}{\alpha} \right) \ln \left[\frac{L_x}{\pi}\sin\frac{\pi X}{L_x}\right] + c'_\alpha,
\end{equation}
for a quasi-1D gapless system in the ground state with periodic boundary conditions in the $\hat{x}$ direction.  The subsystem length $X$ represents the number of contiguous rungs contained in the subsystem, the central charge $c$ is equal to the number of gapless modes, and $c'_\alpha$ is a non-universal constant.  On the two-leg ladder at $\rho = 1/3$, we expect $c=2$ gapless modes in the conventional Luttinger liquid phase.  In contrast, the bosonization approach to the unconventional $d$-metal phase (see Sec.~\ref{app:dmetTheory}) predicts $c=3$ gapless modes, as there are five partially-filled bands but two are rendered massive in the strong-coupling limit of the gauge theory.  Thus, the central charge as determined by measuring the scaling of the entanglement entropy provides a crucial diagnostic for the presence of the non-Fermi liquid $d$-metal phase.

The DMRG has information about the full entanglement spectrum and hence can calculate any $S_\alpha$ including the von Neumann entropy $S_1$, which is the focus of Fig.~\ref{fig:centralCharge} of the main text.  However, calculations of $S_\alpha$ with DMRG are also very difficult to converge for highly entangled gapless systems such as the $d$-metal state, and while we were able to well converge the von Neumann entropy for $L_x=24$ and 36 length systems, we were unable to obtain convergence for the $L_x=48$ system.  Specifically, for the $L_x=48$ system at $J/t=2$ and $K/t=1.8$ as focused on in the main text, e.g., in Fig.~\ref{fig:dmetcorrs}, fitting to the von Neumann entropy gives $c=2.75$ even when keeping $m=20$,000 states.  We estimate that the entropy data near the middle of the sample, $X\simeq L_x/2$, is still a few percent away from full convergence, and, given the sensitivity of the fit parameter $c$ on such inaccuracies, we believe a result $c\simeq3$ would be obtained in the limit $m\rightarrow\infty$.

\begin{figure}[t]
\centerline{\includegraphics[width=\columnwidth]{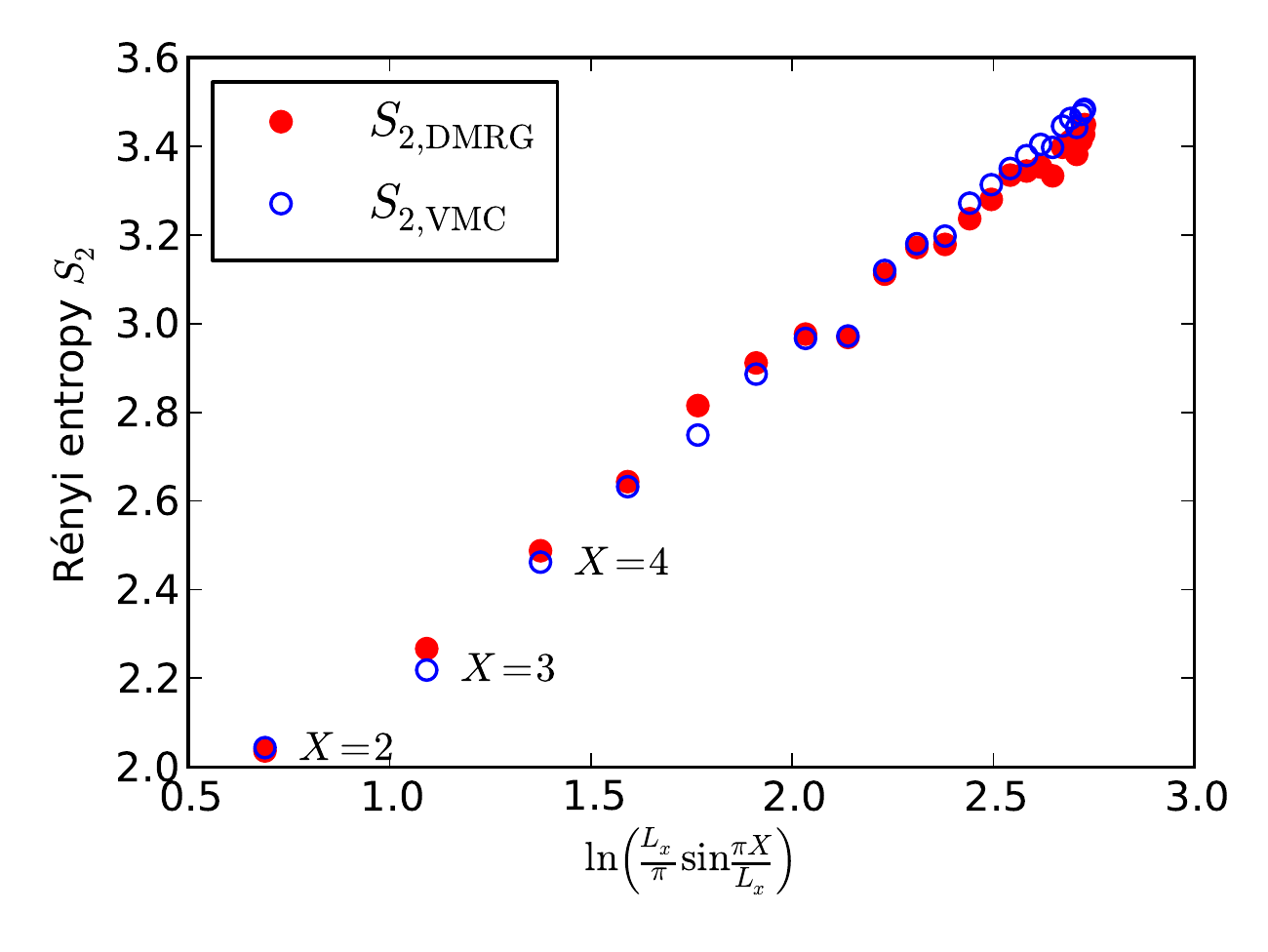}}
\caption{
DMRG and VMC measurements of the scaling of the R\'enyi entropy $S_2$ for $L_x=48$ at the characteristic $d$-metal point of $J/t=2$ and $K/t=1.8$.
}
\label{fig:k18-entropy}
\end{figure}

There are also finite-size uncertainties associated with the range of $X$ values used in the fits to Eq.~(\ref{eqn:cardyrenyi}).  For all data in Fig.~\ref{fig:centralCharge} of the main text, we have restricted the fits to data $X_\mathrm{min}=5 \leq X \leq L_x/2$ in order to focus on the long-distance scaling behavior.  Still, at these system sizes, the extracted values of $c$ do depend on $X_\mathrm{min}$, and this by itself produces a further uncertainty in $c$ of a few percent.

In light of the abovementioned uncertainties associated with extracting $c$ with the DMRG, and with the goal of further bolstering our arguments for the central charge $c=3$ within the $d$-metal on larger systems, we have ventured to compute the entanglement entropy in our variational wave functions and make concrete comparisons to the DMRG data.  Indeed, by measuring the expectation value of a ``swap operator,''\,\cite{Hastings10_PRL_104_157201} VMC can be used to calculate\cite{Grover11_PRL_107_067202} the R\'enyi entropies $S_\alpha$ for integer $\alpha \geq 2$.  As such, we now focus on the scaling of $S_2$ in both the VMC and DMRG for points within the $d$-metal phase.

\begin{figure}[t]
\centerline{\includegraphics[width=\columnwidth]{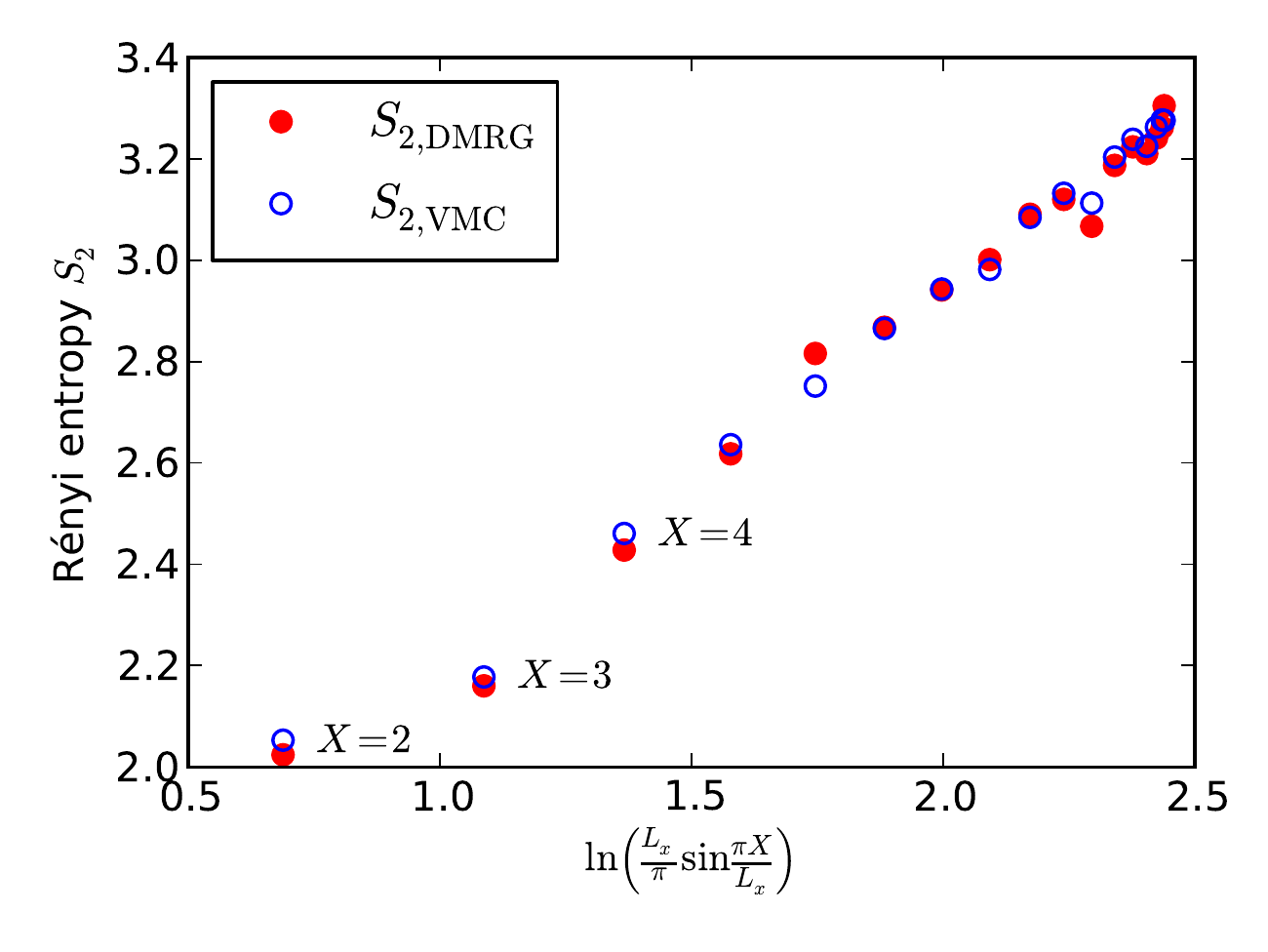}}
\caption{
DMRG and VMC measurements of the scaling of the R\'enyi entropy $S_2$ for $L_x=36$ at $J/t=K/t=2$.
}
\label{fig:k20-entropy}
\end{figure}

Let us consider the characteristic $d$-metal point examined in the main text on the two-leg ladder at $J/t=2$ and $K/t=1.8$, with $L_x=48$.  The scaling of the VMC and DMRG R\'enyi entropies is plotted in Fig.~\ref{fig:k18-entropy}.  There, we observe a striking agreement in both the overall magnitudes and detailed features of the two data sets.  Indeed, the oscillations in $S_2$, which are expected to occur for all but the von Neumann entropy,\cite{Calabrese10_PRL_104_095701} are in sync with each other.  Furthermore, a linear fit gives $c=2.95$ for the VMC and $c=2.75$ for the DMRG, in reasonably close agreement with the expected value of $c=3$ for the $d$-metal phase.  For large $X$, the DMRG entropies lag slightly below the VMC entropies:  As with the von Neumann entropy data discussed above, the DMRG result did not fully converge despite having kept $m=20$,000 states.  (In order to attempt to average over the relatively large oscillations in $S_2$ for small $X$, these quoted values of $c$ used all available data $X_\mathrm{min}=2 \leq X \leq L_x/2$ in the fits.)

On smaller lattices, for instance $L_x=36$ at $J/t=2$ and $K/t=2$ (parameters chosen so that the ground state is spin-singlet), the DMRG result is further converged and the agreement is even better, as seen in Fig.~\ref{fig:k20-entropy}.  Here, a scaled-down version of the $d$-metal VMC state was used, with $N_{d1}^{(0)}=15$ and $N_{d1}^{(\pi)}=9$, exponents $p_1=0.7$ and $p_2=-0.4$, and $N_{f\uparrow}=N_{f\downarrow}=12$.  A linear fit gives $c=2.95$ for the VMC and $c=2.97$ for the DMRG, in excellent agreement with $c=3$.  It is remarkable that the VMC trial wave function is able to reproduce the DMRG data so well and that both give the predicted value of $c$ to such high accuracy.


\section{Long-wavelength description of the $d$-metal phase} \label{app:dmetTheory}

In this part, we develop a long-wavelength theory of the $d$-metal phase and provide detailed characterization of the main observables, including the electron Green's function, density and spin correlations, as well as Cooper pair correlations.  We illustrate such an analysis with the DMRG and VMC data in the $d$-metal phase.  We conclude with a brief discussion of the instabilities of the $d$-metal.

\subsection{Gauge theory description and solution by bosonization}
We start with the decomposition of the electron operator into three fermionic partons as in Eq.~(\ref{d1d2fs}) of the main text.  We assume that the $d_1$ partons hop preferentially along the ladder direction and partially populate both bonding and anti-bonding bands, while the $d_2$ partons hop preferentially in the transverse direction and populate only the bonding band.  The densities of the $d_1$ and $d_2$ partons are the same and equal to the total electron density (i.e., including spin).  We assume that the spinons $f_\up$ and $f_\dn$ populate only the bonding band, and the $\up$ and $\dn$ populations are identical so that the state is spin-singlet.  The parton occupations are illustrated in Fig.~\ref{fig:partonbands} of the main text.  The Fermi wavevectors satisfy
\begin{eqnarray}
k_{Fd1}^{(0)} + k_{Fd1}^{(\pi)} = k_{Fd2} = 2 k_{Ff} = 2\pi\rho ~,
\end{eqnarray}
where $\rho = N_e/(2L_x)$ is the total electron density per site.  When discussing the electron distribution function below, it will be convenient to parametrize the $d_1$ Fermi wavevectors as
\begin{eqnarray}
k_{Fd1}^{(k_y)} = \pi\rho + e^{i k_y} {\mathsf K}_e ~, \quad k_y = 0, \pi ~.
\end{eqnarray}
(The same ${\mathsf K}_e$ was already used in the main body.)

We begin with the mean field Hamiltonian in continuum.  Linearizing near the Fermi points, the kinetic energy density reads
\begin{eqnarray}
h_{\rm kinetic} &=& \sum_{k_y = 0, \pi} \sum_{P = R/L} P v_{d1}^{(k_y)} d_{1P}^{(k_y)\dagger} (-i \partial_x) d_{1P}^{(k_y)} \\
&+& \sum_{P = R/L} P v_{d2} d_{2P}^\dagger (-i \partial_x) d_{2P} \\
&+& \sum_{s = \up, \dn} \sum_{P = R/L} P v_f f_{sP}^\dagger (-i \partial_x) f_{sP} ~.
\end{eqnarray}
Here, superscript $k_y = 0$ or $\pi$ for $d_1$ corresponds to bonding or anti-bonding respectively; $P = R/L = +/-$ refers to right or left moving fields; $v_b$ is the Fermi velocity for fermion ``band'' $b \in \{ d_1^{(0)};~ d_1^{(\pi)};~ d_2;~ f_\up;~ f_\dn \}$.

Going beyond the kinetic energy, we also consider four-fermion interactions, which are required to respect lattice symmetries, time reversal, and spin rotation invariance.  First, we have various density-density interactions in terms of the continuum densities $\rho_{bP}$ for fermion species $b$.  These are strictly marginal (harmonic in the bosonized treatment) and renormalize band velocities as well as Luttinger parameters.  Since the final fixed-point theory will have most general form consistent with the symmetries, we do not spell out the microscopic parton density-density terms.

Next, we have the following four-$d_1$-chargon term:
\begin{eqnarray}
\label{h4d}
h_{{\rm int}, 4d_1} &=& w \left[ d_{1R}^{(0)\dagger} d_{1L}^{(0)\dagger} d_{1L}^{(\pi)} d_{1R}^{(\pi)} + \Hc \right] \\
&=& 2 w \cos[2 \sqrt{2} \varphi_{d1-}] ~.
\end{eqnarray}
The last line shows bosonized expression using fields that will be defined below; as we will describe later, this non-harmonic interaction can destabilize the ``relative charge sector'' ``$d1-$''.

We also have the following four-spinon term:
\begin{eqnarray}
\label{h4f}
&& h_{{\rm int}, 4f} = -u \left( f_R^\dagger \frac{\vec{\sigma}}{2} f_R \right) \cdot \left( f_L^\dagger \frac{\vec{\sigma}}{2} f_L \right) \\
&=& \frac{u^z}{8\pi^2} \left[(\partial_x \varphi_{f\sigma})^2 - (\partial_x \theta_{f\sigma})^2 \right] + u^\perp \cos(2 \sqrt{2} \theta_{f\sigma}) ~.
\nonumber
\end{eqnarray}
Again, the last line shows bosonized expression using fields that will be defined below; this non-harmonic interaction can destabilize the ``spin sector'' ``$f\sigma$''.

The above exhausts residual four-fermion interactions at generic electron densities and generic $k_{Fd1}^{(0/\pi)}$.  On the other hand, at commensurate densities we should also include umklapps.  For example, at density $\rho = 1/3$ studied in the DMRG in the main text, there is a single umklapp term
\begin{eqnarray}
&& h_{\rm int, umkl} = u_{\rm umkl} \left( d_{2R}^\dagger d_{2L} \sum_s f_{sR}^\dagger f_{sL} + \Hc \right) \\
&& = -4 u_{\rm umkl} \cos(\sqrt{2} \theta_{f\sigma}) \cos\left( \frac{3\theta_\rhotot}{\sqrt{2}} + \frac{\theta_A}{\sqrt{6}} - \frac{2\theta_a}{\sqrt{3}} \right) . ~~~~~~
\label{humklapp}
\end{eqnarray}
If this umklapp term is relevant, it leads to a Mott insulator (CDW) of electrons, while it must be irrelevant in the metallic phase.  The absence of any charge ordering in the $d$-metal phase found in the DMRG at $\rho=1/3$ will allow us to crudely bound power laws in some correlations and check the overall consistency of the $d$-metal theory, while the umklapp can be always rendered inoperative if we step away from the commensurate density.

The full theory beyond mean field is a parton-gauge theory.  Unlike the 2D case, in quasi-1D the parton-gauge theory can be solved by bosonization.\cite{Sheng2008_2legDBL, Sheng2009_zigzagSBM, Mishmash2011_4legDBL}  More specifically, treating the gauge field fluctuations on long wavelengths amounts to implementing the constraints Eq.~(\ref{constraints}) of the main text for the coarse-grained densities along the ladder direction.  Since the fixed-point Hamiltonian is harmonic in the bosonized fields and the coarse-grained densities are linear functions of the bosonized fields, implementing the constraints becomes simple.

Our bosonization treatment is as follows.\cite{Shankar_Acta, Lin98, Fjaerestad02}  We bosonize each fermion band species $F_{bR/L}$ ($F_b \in \{ d_1^{(0)};~ d_1^{(\pi)};~ d_2;~ f_\up;~ f_\dn \}$)
\begin{equation}
F_{bP} = \eta_b e^{i (\varphi_b + P\theta_b)} ~,
\label{fbosonize}
\end{equation}
with canonically conjugate boson fields:
\begin{eqnarray}
[\varphi_b(x), \varphi_{b'}(x^\prime)] &=& [\theta_b(x), \theta_{b'}(x^\prime)] = 0 ~, \\ ~ 
[\varphi_b(x), \theta_{b'}(x^\prime)] &=& i \pi \delta_{bb'} \, \Theta(x - x^\prime) ~,
\end{eqnarray}
where $\Theta(x)$ is the Heaviside step function.  Here, we have introduced Klein factors, the Majorana fermions $\{ \eta_b, \eta_{b'} \} = 2 \delta_{bb'}$, which assure that the fermion fields with different flavors anti-commute with one another.  The slowly varying fermionic densities are simply $\rho_{bP} \equiv F_{bP}^\dagger F_{bP} = \partial_x (P\varphi_b + \theta_b)/(2\pi)$, and hence $\rho_b \equiv \rho_{bR} + \rho_{bL} = \partial_x \theta_b/\pi$.

In the mean field, we start with five modes, $\theta_{d1}^{(0/\pi)}$, $\theta_{d2}$, and $\theta_{f\up/\dn}$.  Including gauge fluctuations is simple in the bosonized formulation:\cite{Sheng2008_2legDBL, Sheng2009_zigzagSBM, Mishmash2011_4legDBL} they effectively implement the constraints Eq.~(\ref{constraints}) of the main text for the coarse-grained densities and produce the following pinning of the dual fields (up to fixed constant shifts):
\begin{equation}
\theta_{d1}^{(0)} + \theta_{d1}^{(\pi)} = \theta_{d2} = \theta_{f\up} + \theta_{f\dn} ~.
\label{thtpin}
\end{equation}
Formally, we perform an orthonormal transformation on the $\theta$ fields:
\begin{eqnarray}
\theta_{f\sigma} &=& \frac{1}{\sqrt{2}} \left( \theta_{f\up} - \theta_{f\dn} \right), \\
\theta_{d1-} &=& \frac{1}{\sqrt{2}} \left( \theta_{d1}^{(0)} - \theta_{d1}^{(\pi)} \right), \\
\theta_\rhotot &=& \frac{1}{2\sqrt{2}} \left( \theta_{f\up} + \theta_{f\dn} + \theta_{d1}^{(0)} + \theta_{d1}^{(\pi)} + 2\theta_{d2} \right), ~~~\\
\theta_a &=& \frac{1}{\sqrt{3}} \left( \theta_{d1}^{(0)} + \theta_{d1}^{(\pi)} - \theta_{d2} \right), \\
\theta_A &=& \frac{\sqrt{3}}{2\sqrt{2}} \left(\theta_{f\up} + \theta_{f\dn} - \frac{\theta_{d1}^{(0)} + \theta_{d1}^{(\pi)} + 2\theta_{d2}}{3} \right). ~~~
\end{eqnarray}
We perform the same transformation on the $\varphi$ fields, so the new $\theta$ and $\varphi$ fields are again canonically conjugate.  Gauge field fluctuations render the combinations $\theta_a$ and $\theta_A$ in the last two lines massive, effectively leading to the pinnings in Eq.~(\ref{thtpin}), and only the $\theta_{f\sigma}$, $\theta_{d1-}$, and $\theta_\rhotot$ modes remain.  Note that the definition of $\theta_\rhotot$ is fixed once we define the convenient fields $\theta_{f\sigma}$ and $\theta_{d1-}$ and require orthogonality among these modes and orthogonality to the linear space of equations~(\ref{thtpin}).  On the other hand, the definitions of $\theta_a$ and $\theta_A$ are somewhat arbitrary (roughly, massiveness of $\theta_a$ corresponds to gluing $d_1$ and $d_2$ partons to form the bosonic chargon $b$, while massiveness of $\theta_A$ then glues the $b$ and $f$); we can take any orthonormal linear combination of $\theta_a$ and $\theta_A$ and such a choice does not affect the $\theta_{f\sigma}$, $\theta_{d1-}$, and $\theta_\rhotot$ content of any physical operator below.

In the fixed-point theory, the ``$f\sigma$'' (spin) sector decouples from the ``$d1-$'' and ``$\rhotot$'' (charge) sectors:
\begin{eqnarray}
&& {\cal L} = {\cal L}_1[\theta_{f\sigma}; v_{f\sigma}; g_{f\sigma}] + {\cal L}[\theta_{d1-}, \theta_\rhotot] ~,\\
&& {\cal L}_1[\theta; v; g] = \frac{1}{2\pi g} \left[v (\partial_x \theta)^2 + \frac{1}{v} (\partial_\tau \theta)^2 \right] ~.
\end{eqnarray}
The last line is a generic Lagrangian for a single mode with velocity $v$ and Luttinger parameter $g$ written in Euclidean space-time.  In the spin sector, $g_{f\sigma} = 1$ is fixed by the condition of SU(2) spin invariance.  The residual interaction Eq.~(\ref{h4f}) is marginally irrelevant if $u > 0$.

On the other hand, in the charge sectors, we have most general quadratic Lagrangian
\begin{equation*}
{\cal L}[\theta_{d1-}, \theta_\rhotot] = \frac{1}{2\pi}
\left[\partial_x \bm{\Theta}^T \cdot {\bf A} \cdot \partial_x \bm{\Theta}
      + \partial_\tau \bm{\Theta}^T \cdot {\bf B} \cdot \partial_\tau \bm{\Theta}
\right],
\end{equation*}
where we defined $\bm{\Theta}^T \equiv (\theta_{d1-}, \theta_\rhotot)$.  In general, ${\bf A}$ and ${\bf B}$ can be arbitrary positive-definite symmetric matrices.

To get some feel for the charge sectors, we can consider the case where the only interactions between different parton species are due to gauge fluctuations enforcing the constraints Eq.~(\ref{thtpin}).  For more generality, we allow density-density interactions within each parton type and encode these in the corresponding renormalized velocities $v_b$ and Luttinger parameters $g_b$.  The matrices ${\bf A}$ and ${\bf B}$ are readily evaluated to be
\begin{eqnarray}
\label{A11}
A_{11} &\!=\!& \frac{1}{2} \left(\frac{v_{d1}^{(0)}}{g_{d1}^{(0)}} + \frac{v_{d1}^{(\pi)}}{g_{d1}^{(\pi)}} \right) , \\
A_{22} &\!=\!& \frac{1}{8} \left(\frac{v_{d1}^{(0)}}{g_{d1}^{(0)}} + \frac{v_{d1}^{(\pi)}}{g_{d1}^{(\pi)}} + 4 \frac{v_{d2}}{g_{d2}} + 2 \frac{v_f}{g_f} \right) , \\
A_{12} &\!=\!& A_{21} = \frac{1}{4} \left(\frac{v_{d1}^{(0)}}{g_{d1}^{(0)}} - \frac{v_{d1}^{(\pi)}}{g_{d1}^{(\pi)}} \right) , \\
B_{11} &\!=\!& \frac{1}{2} \left(\frac{1}{g_{d1}^{(0)} v_{d1}^{(0)}} + \frac{1}{g_{d1}^{(\pi)} v_{d1}^{(\pi)}} \right) , \\
B_{22} &\!=\!& \frac{1}{8} \left(\frac{1}{g_{d1}^{(0)} v_{d1}^{(0)}} + \frac{1}{g_{d1}^{(\pi)} v_{d1}^{(\pi)}} + \frac{4}{g_{d2} v_{d2}} + \frac{2}{g_f v_f} \right) , ~~~~\\
B_{12} &\!=\!& B_{21} = \frac{1}{4} \left(\frac{1}{g_{d1}^{(0)} v_{d1}^{(0)}} - \frac{1}{g_{d1}^{(\pi)} v_{d1}^{(\pi)}} \right) .
\label{B12}
\end{eqnarray}
As an example, if we take all $g_b = 1$ but allow general velocities $v_b$, we can check that the scaling dimension of the interaction Eq.~(\ref{h4d}) is greater than 2 and hence this interaction is irrelevant.  At generic incommensurate electron density, this is the only allowed non-linear interaction other than the spin interaction Eq.~(\ref{h4f}); the latter can be marginally irrelevant, and hence the $d$-metal phase can be stable as a matter of principle.

For illustrations later in the text, we will consider a schematic model where the ``$d1-$'' and ``$\rhotot$'' modes also decouple, so
\begin{eqnarray}
{\cal L} &=& {\cal L}_1[\theta_{f\sigma}; v_{f\sigma}; g_{f\sigma}]
+ {\cal L}_1[\theta_{d1-}; v_{d1-}; g_{d1-}] \\
&+& {\cal L}_1[\theta_\rhotot; v_\rhotot; g_\rhotot] ~.
\end{eqnarray}
This situation arises, e.g., in the above case with gauge-only interactions, Eqs.~(\ref{A11})-(\ref{B12}), if we further require $v_{d1}^{(0)} = v_{d1}^{(\pi)} \equiv v_{d1}$ and $g_{d1}^{(0)} = g_{d1}^{(\pi)} \equiv g_{d1}$, where we find
\begin{eqnarray}
g_{d1-} &=& g_{d1}, \label{gd1-_decoupled} \\
g_\rhotot &=& 4 \left( \frac{v_{d1}}{g_{d1}} + \frac{2 v_{d2}}{g_{d2}} + \frac{v_f}{g_f} \right)^{-1/2} \times  \label{grhotot_decoupled} \\
&& \times \left(\frac{1}{g_{d1} v_{d1}} + \frac{2}{g_{d2} v_{d2}} + \frac{1}{g_f v_f} \right)^{-1/2}  ~.
\end{eqnarray}
The schematic model holds approximately when the $d_1$ bonding and anti-bonding populations are approximately equal.  Furthermore, we think it is valid for the ``bare Gutzwiller'' wave function, cf.~Sec.~\ref{app:vmcDetails}, since the projection does not know about the intra-species interactions and band velocities; in this case, it is natural to set all bare Luttinger parameters equal to $g_b = 1$ and all velocities equal, and we obtain $g_{d1-} = g_\rhotot = 1$.  If we require all bare $g_b = 1$ but allow general velocities, Eq.~(\ref{grhotot_decoupled}) would give $g_\rhotot \leq 1$.

Here we note that the $d$-metal phase found in the DMRG is roughly consistent with the approximation of decoupled ``$d1-$'' and ``$\rhotot$'' modes with $g_\rhotot$ significantly larger than $1$ (see our discussion of observables below).  Also, as described in Sec.~\ref{app:vmcDetails}, our optimal VMC wave functions have powers on the determinants and are significantly away from the bare Gutzwiller states.  We can crudely model the effect of adding power on the $d_2$ determinant, $\psi_{d2}(\{\mathbf{R}_i\}) \rightarrow \lvert\psi_{d2}(\{\mathbf{R}_i\})\rvert^{p_2-1} \psi_{d2}(\{\mathbf{R}_i\})$, by including the Luttinger parameter $g_{d2}$ in the bare Lagrangian for the $\theta_{d2}$ mode.  The Slater determinant $\psi_{d2}$ fills only one band, and a crude guess from studies\cite{HellbergMele1991} with such 1D Jastrow-Luttinger wave functions is that $g_{d2} = 1/p_2$ for $p_2 > 0$.  In particular, $g_{d2} \to \infty$ for $p_2 \to 0$.  In this limit, $g_\rhotot$ in Eq.~(\ref{grhotot_decoupled}) can be as large as $2$ even when $g_{d1} = g_f = 1$.  Below, we will use values $g_{d1-} = 1$ and $g_\rhotot = 2$ to illustrate such a ``dressed Gutzwiller'' wave function.  Our VMC measurements with the bare and dressed Gutzwiller wave functions support the above conjectures for the effective Luttinger parameters in both cases.

Related to the above discussion, we remark that the DMRG does not find any translational symmetry breaking that could be driven by the umklapp Eq.~(\ref{humklapp}) at density $\rho = 1/3$.  In the approximation of decoupled ``$d1-$'' and ``$\rhotot$'' modes, the irrelevance of the umklapp requires $g_\rhotot > 4/3$.  This is consistent with the large effective $g_\rhotot$ deduced from the DMRG correlations and from the optimal VMC wave functions, see Sec.~\ref{subsec:fits} below.

Having discussed the general structure of the $d$-metal fixed point and some approximate models, we now turn to the characterization of the phase in terms of electron Green's function, density and spin correlations, as well as Cooper pair correlations that are measured in the DMRG.


\begin{table*}[t!]
\ifREVTEX
\else
	{\scriptsize
\fi
\begin{tabular}{| c | c | c || c | c | c | c | c | c |}
\hline
  & & & & $\theta_\rhotot$ coeff. & $\Delta_{\alpha\beta\gamma}^{(k_y)}$, Eq.~(\ref{Delta_exact_bound}) & $\Delta_{\alpha\beta\gamma}^{(k_y)}$, Eq.~(\ref{Delta_crude_model}), crude model & $\Delta_{\alpha\beta\gamma}^{(k_y)}$, bare Gutzwiller & $\Delta_{\alpha\beta\gamma}^{(k_y)}$, dressed Gutzwiller \\
  $\alpha$ & $\beta$ & $\gamma$ & $Q_{\alpha\beta\gamma}^{(k_y)}$, Eq.~(\ref{Q_abg}) & $\frac{\alpha + 2\beta + \gamma}{2\sqrt{2}}$ & exact bound & decoupled ``$d1-$'' and ``$\rhotot$'' & $g_{d1-} = 1$, $g_\rhotot = 1$ & $g_{d1-} = 1$, $g_\rhotot = 2$ \\
\hline
\hline
  + & - & + & $e^{ik_y} {\mathsf K}_e$             & 0                     & $\geq 1/2$ & $\Delta_{+-+}$ & 1 & 3/4 \\
\hline
  + & - & - & $-2\pi\rho + e^{ik_y} {\mathsf K}_e$ & $-\frac{1}{\sqrt{2}}$ & $\geq 1/2$ & $\Delta_{+-+} + g_\rhotot/8$ & 9/8 & 1 \\
\hline
  + & + & - & $2\pi\rho + e^{ik_y} {\mathsf K}_e$  & $\frac{1}{\sqrt{2}}$  & $\geq 1$   & $\Delta_{+-+} + g_\rhotot/8$ & 9/8 & 1 \\
\hline
  + & + & + & $4\pi\rho + e^{ik_y} {\mathsf K}_e$  & $\sqrt{2}$            & $\geq 3/2$ & $\Delta_{+-+} + g_\rhotot/2$ & 3/2 & 7/4 \\
\hline
\end{tabular}
\ifREVTEX
\else
	}
\fi
\caption{
Analysis of different contributions, Eq.~(\ref{c_abg}), to the electron operator.  We list the wavevector $Q_{\alpha\beta\gamma}^{(k_y)}$; the coefficient of $\theta_\rhotot$ in Eq.~(\ref{Theta_abg}), which strongly affects the scaling dimension $\Delta_{\alpha\beta\gamma}^{(k_y)}$; exact lower bound on $\Delta_{\alpha\beta\gamma}^{(k_y)}$; the scaling dimension $\Delta_{\alpha\beta\gamma}^{(k_y)}$ in the approximation of decoupled ``$d1-$'' and ``$\rhotot$'' modes, with $\Delta_{+-+}$ given in Eq.~(\ref{Delta_+-+}); specialization to $g_{d1-} = g_\rhotot = 1$ appropriate for the bare Gutzwiller wave function; and specialization to $g_{d1-} = 1$, $g_\rhotot = 2$ appropriate for the dressed Gutzwiller wave function that roughly captures the DMRG structure factors (see text for details).  Note that in general the scaling dimensions also depend on $k_y$.
}
\label{tab:Ge}
\end{table*}

\subsection{Electron Green's function}
We begin with the electron Green's function $G_e(x,y)$, $e = c_{s = \up/\dn}$.  In the mean field,
\begin{eqnarray*}
G_e^{\rm mf}(x, y) &=& G_{d1}(x, y) ~ G_{d2}(x) ~ G_f(x) ~, \\
G_e^{\rm mf}(x, k_y) &\sim& \frac{\sin(k_{Fd1}^{(k_y)} x) ~ \sin(k_{Fd2} x) ~ \sin(k_{Ff} x)}{x^3} ~.
\end{eqnarray*}
In the first line, we used the fact that the $d_2$ and $f_s$ partons populate only the bonding band, hence the $y$-dependence comes only from the $d_1$ parton.  In the second line, it is convenient to work with bonding or anti-bonding components, $k_y = 0$ or $\pi$.  The general structure of the oscillating contributions is obtained by expanding the sines in the last equation,
\begin{eqnarray}
G_e(x, k_y) &\approx& \sum_{\alpha,\beta,\gamma = \pm}
\frac{ i A_{\alpha\beta\gamma}^{(k_y)} ~~ e^{i Q_{\alpha\beta\gamma}^{(k_y)} x} }{ {\rm sign}(x) |x|^{2 \Delta_{\alpha\beta\gamma}^{(k_y)}} } ~, \\
Q_{\alpha\beta\gamma}^{(k_y)} &\equiv& \alpha k_{Fd1}^{(k_y)} + \beta k_{Fd2} + \gamma k_{Ff} \\
&=& \alpha e^{i k_y} {\mathsf K}_e + (\alpha + 2\beta + \gamma) \pi \rho ~,
\label{Q_abg}
\end{eqnarray}
where we allowed general amplitudes $A_{\alpha\beta\gamma}^{(k_y)}$ and scaling dimensions $\Delta_{\alpha\beta\gamma}^{(k_y)}$.  Contribution at the wavevector $Q_{\alpha\beta\gamma}^{(k_y)}$ comes from the following combination of long-wavelength parton fields,
\begin{eqnarray}
d_{1\alpha}^{(k_y)} d_{2\beta} f_{s\gamma} \sim e^{i (\Phi_{\alpha\beta\gamma, s}^{(k_y)} + \Theta_{\alpha\beta\gamma, s}^{(k_y)})} ~,
\label{c_abg}
\end{eqnarray}
where $\alpha,\beta,\gamma = R/L = +/-$; $s = \up/\dn$; and
\begin{eqnarray}
\Phi_{\alpha\beta\gamma, s}^{(k_y)} &=& \varphi_{d1}^{(k_y)} + \varphi_{d2} + \varphi_{fs} \\
&=& \frac{e^{ik_y}}{\sqrt{2}} \varphi_{d1-} + \sqrt{2}\varphi_\rhotot + \frac{s}{\sqrt{2}} \varphi_{f\sigma} ~, \\
\Theta_{\alpha\beta\gamma, s}^{(k_y)} &=& \alpha \theta_{d1}^{(k_y)} + \beta \theta_{d2} + \gamma \theta_{fs} \label{Theta_abg} \\
&=& \frac{\alpha e^{ik_y}}{\sqrt{2}} \theta_{d1-} + \frac{\alpha + 2\beta + \gamma}{2\sqrt{2}} \theta_\rhotot + \frac{\gamma s}{\sqrt{2}} \theta_{f\sigma} ~~~~~~~~~ \\
&+& \frac{\alpha - \beta}{\sqrt{3}}\theta_a + \frac{-\alpha - 2\beta + 3\gamma}{2\sqrt{6}} \theta_A ~.
\end{eqnarray}
The last line contains $\theta_a$ and $\theta_A$ that are pinned upon including gauge fluctuations, and from here on we often drop these fields when they only modify overall complex phases of the operator expressions.  For ease of reference, Table~\ref{tab:Ge} lists different cases where we fix $\alpha = +$.  We will consider these cases and entries in the table after some more general discussion.  We are primarily interested in the scaling dimensions $\Delta_{\alpha\beta\gamma}^{(k_y)}$.

First, we note that the spin sector ``$f\sigma$'' is decoupled from the rest and has $g_{f\sigma} = 1$; the corresponding fields contribute $1/4$ to the scaling dimension of the electron correlation for any wavevector $Q_{\alpha\beta\gamma}^{(k_y)}$.  Thus, the scaling dimension depends only on the ``$d1-$'' and ``$\rhotot$'' content of both $\Phi_{\alpha\beta\gamma, s}^{(k_y)}$ and $\Theta_{\alpha\beta\gamma, s}^{(k_y)}$ fields.  We expect that the main difference among the cases comes from the $\theta_\rhotot$ content, since its coefficient $(\alpha + 2\beta + \gamma)/(2\sqrt{2})$ can vary in magnitude from $0$ for $\alpha = \gamma = -\beta$ to $\sqrt{2}$ for $\alpha = \gamma = \beta$; this content is listed in one of the columns in Table~\ref{tab:Ge}.  Note, however, that in general the relative signs of the ``$d1-$'' and ``$\rhotot$'' components are also important and affect the scaling dimensions, thus making the scaling dimensions also depend on $k_y$; this is indeed what we find from fitting the DMRG data (see Sec.~\ref{subsec:fits}).  It is only in the case when the ``$d1-$'' and ``$\rhotot$'' decouple that the scaling dimensions do not depend on the signs of the coefficients.

Second, by using the general result
$\Delta[e^{i \sum_j (a_j \varphi_j + b_j \theta_j)}] \geq \frac{1}{2}|\sum_j a_j b_j|$
valid for any canonically conjugate set $\varphi_j, \theta_j$ and any coefficients $a_j, b_j$, we can obtain an exact general bound
\begin{equation}
\Delta_{\alpha\beta\gamma}^{(k_y)} \geq \frac{1}{4} + \frac{|2\alpha + 2\beta + \gamma|}{4} ~,
\label{Delta_exact_bound}
\end{equation}
which is also listed in Table~\ref{tab:Ge}.

Third, we can calculate the scaling dimensions in the approximation that the ``$d1-$'' and ``$\rhotot$'' modes decouple:
\begin{eqnarray}
\Delta_{\alpha\beta\gamma}^{(k_y)} &=& \Delta_{+-+} + \frac{(\alpha + 2\beta + \gamma)^2}{32} g_\rhotot ~,  \label{Delta_crude_model} \\
\Delta_{+-+} &=& \frac{1}{4} + \frac{1}{8}\left(\frac{1}{g_{d1-}} + g_{d1-} \right) + \frac{1}{2g_\rhotot} ~.  \label{Delta_+-+}
\end{eqnarray}
This illustrates the preceding discussion of the dependence of the scaling dimensions on the $\theta_\rhotot$ content.  By setting $g_{d1-} = g_\rhotot = 1$, we obtain scaling dimensions in the bare Gutzwiller wave function also listed in the table.  On the other hand, by setting $g_{d1-} = 1$ and $g_\rhotot = 2$, we obtain scaling dimensions in the dressed Gutzwiller wave function (tentatively corresponding to powers on the determinants $p_1 = 1$ and $p_2 = 0$); these dimensions are listed in the last column in Table~\ref{tab:Ge}.  Power law fits for the electron Green's function measured in the VMC are consistent with these predictions.

Let us now consider different entries in Table~\ref{tab:Ge}.  We can roughly understand the trends in the scaling dimensions by appealing to so-called ``Amperean'' rules for interactions mediated by gauge fields:\cite{LeeNagaosaWen, Polchinski94, Altshuler94}  Parallel currents attract and hence processes containing such currents are enhanced, while anti-parallel currents repel and hence such processes are suppressed.  The splitting of the electron into three partons leads to two gauge fields.  We again use the picture where we first break the electron into the spinon $f$ and chargon $b$, and then break the chargon $b$ into the partons $d_1$ and $d_2$.  We can then think of the two gauge fields as follows.  The first gauge field works to glue the $d_1$ and $d_2$ together to form the chargon; the two partons carry opposite gauge charges with respect to this gauge field, hence processes that contain the $d_1$ and $d_2$ moving in opposite (same) directions are enhanced (suppressed).  The enhanced combinations produce the main features in the boson distribution function in the so-called DBL[2,1] phase of Ref.~\onlinecite{Sheng2008_2legDBL} at wavevectors $(k_{Fd2} - k_{Fd1}^{(k_y)}, k_y)$ marked in the top panel in Fig.~\ref{fig:convolution}.  The second gauge field works to glue the chargons and spinon to form the electron; the $d_{1/2}$ and $f$ carry opposite gauge charges with respect to this gauge field, and hence processes that contain $d_{1/2}$ and $f$ moving in opposite (same) directions are enhanced (suppressed).

The precise mathematics in $(1+1)D$ is that the gauge field fluctuations pin the fields $\theta_a$ and $\theta_A$ and hence can reduce the fluctuating content in Eq.~(\ref{Theta_abg}), particularly for the Amperean-enhanced combinations.
The first two rows in Table~\ref{tab:Ge} correspond to oppositely moving $d_1$ and $d_2$ partons.  From the exact lower bound, we see that these entries can be potentially more enhanced than the other two rows.  In the first row, the spinon moves parallel to the $d_1$ and anti-parallel to the $d_2$, while in the second row the situation is interchanged; both situations produce the same exact lower bound, but the first one is apparently more enhanced in the approximate model with decoupled ``$d1-$'' and ``$\rhotot$'' modes.
The third row in Table~\ref{tab:Ge} has the $d_1$ and $d_2$ partons moving in the same direction, but the corresponding Amperean suppression is somewhat compensated by the spinon moving in the opposite direction to the two; in the approximate model with decoupled ``$d1-$'' and ``$\rhotot$'', the second and third row have the same scaling dimension.
The last row in Table~\ref{tab:Ge} has all partons moving in the same direction; this combination is suppressed by all gauge field fluctuations, and we see that it has the largest scaling dimension, which is always larger than the mean field value of $3/2$.

\begin{figure}[t]
\centerline{\includegraphics[width=\convScaleFac\columnwidth]{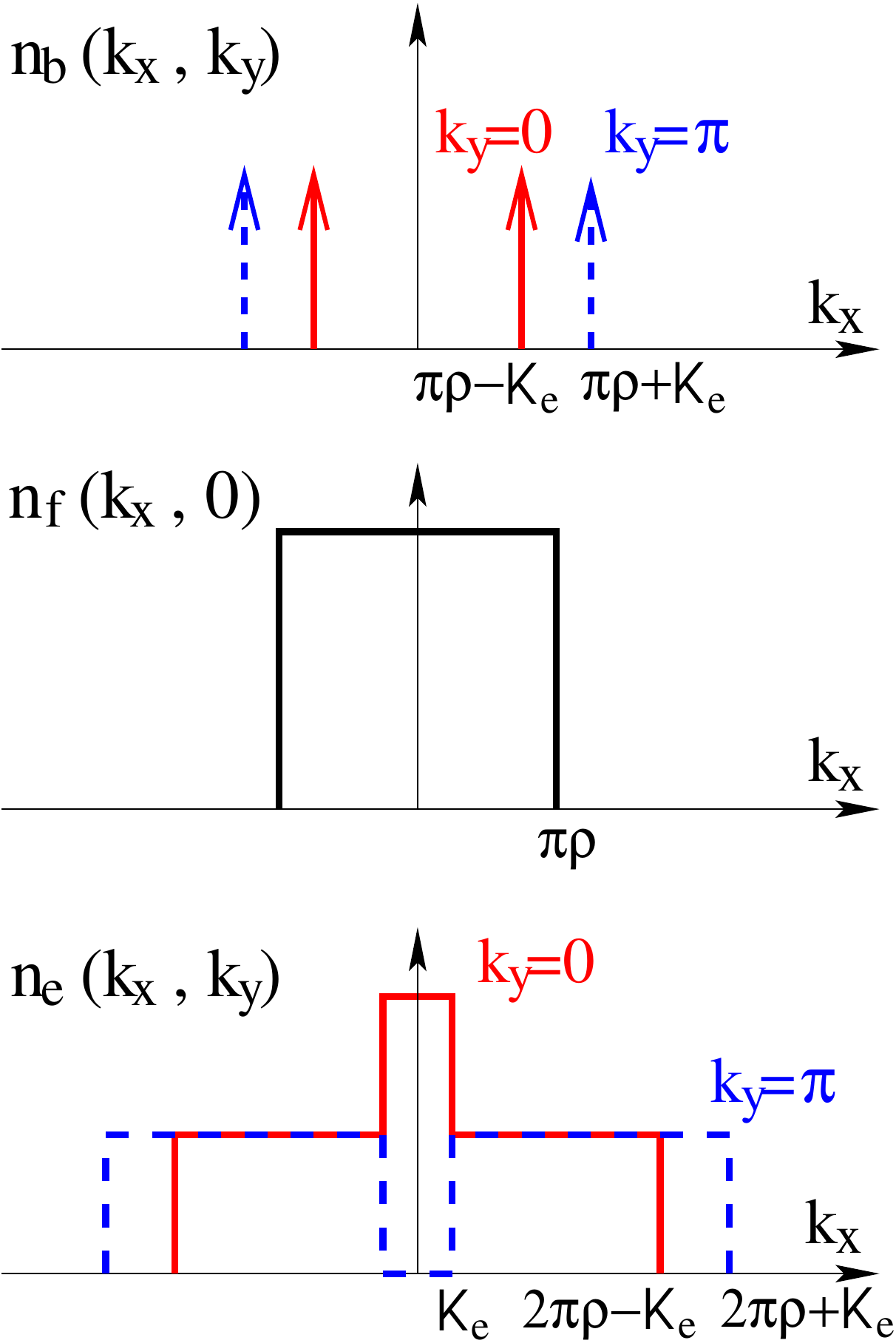}}
\caption{
The electronic $d$-metal is obtained by putting the chargons into the DBL[2,1] (Bose-metal) state of Ref.~\onlinecite{Sheng2008_2legDBL} and spinons into the Fermi sea (spin liquid) state.  Top panel:  Crude approximation to the chargon distribution function where we replace the peaks in Fig.~8(a) of Ref.~\onlinecite{Sheng2008_2legDBL} with delta-functions; the wavevectors are $k_{Fd2} - k_{Fd1}^{(k_y)} = \pi\rho - e^{ik_y} {\mathsf K}_e$ and arise from the oppositely moving $d_1$ and $d_2$ partons.  Middle panel: Spinon distribution function.  Bottom panel: Electron distribution function obtained as a convolution of the chargon and spinon distribution functions.  The wavevectors of the step singularities correspond to the first two rows in Table~\ref{tab:Ge}.  Note that this sketch does not contain potentially important wavevectors listed in the third row in Table~\ref{tab:Ge}.
}
\label{fig:convolution}
\end{figure}

We conclude with a simple understanding of the main landscapes in the electron distribution function, as measured in the DMRG and VMC in Fig.~\ref{fig:dmetcorrs}(a) of the main text.  This discussion also re-iterates the origin of the wavevectors in the first two rows in Table~\ref{tab:Ge}.  We can take the chargon distribution function from the earlier DBL[2,1] study.\cite{Sheng2008_2legDBL}  This has dominant features at wavevectors obtained by combining oppositely moving $d_1$ and $d_2$ partons.  For simplicity, we approximate the dominant features by $\delta$-functions as shown in the top panel in Fig.~\ref{fig:convolution}.  For the spinons $f$, we take a simple step distribution shown in the middle panel in Fig.~\ref{fig:convolution}.  The electron distribution function is a convolution of the chargon and spinon distribution functions.  The result is shown in the bottom panel in Fig.~\ref{fig:convolution}.  The inner steps arise from combinations where $f$ is moving parallel to the $d_1$ and anti-parallel to the $d_2$ and correspond precisely to the first row in Table~\ref{tab:Ge}, while the outer steps arise from combinations where $f$ is moving anti-parallel to the $d_1$ and parallel to the $d_2$ and correspond to the second row in Table~\ref{tab:Ge}.  We can see from Table~\ref{tab:Ge} that the inner singularities are likely stronger than the outer ones, particularly if $g_\rhotot$ is large, and this is consistent with the DMRG findings in the main text.

\subsection{Density and spin correlations}
We now discuss the electron density and spin correlations.  From the various ways of writing the electron number in terms of partons, cf.\ Eq.~(\ref{constraints}) of the main text, we can immediately obtain fermionic bilinear contributions to the electron number operator that have the right symmetry properties:
\begin{eqnarray}
&& n_{(2k_{Fd2}, 0)} \equiv d_{2L}^\dagger d_{2R}
\sim i e^{i \sqrt{2}\theta_\rhotot} ~, \\
&& n_{(2k_{Fd1}^{(k_y)}, 0)} \equiv d_{1L}^{(k_y)\dagger} d_{1R}^{(k_y)}
\sim i e^{i (\pm \sqrt{2}\theta_{d1-} + \frac{\theta_\rhotot}{\sqrt{2}})} ~,~~~~ \\
&& n_{(k_{Fd1}^{(0)} + k_{Fd1}^{(\pi)}, \pi)} \equiv d_{1L}^{(0)\dagger} d_{1R}^{(\pi)} + d_{1L}^{(\pi)\dagger} d_{1R}^{(0)} \label{nk0+pi} \\
&& ~~~~~~~~~~~~~~~~~~~~~ \sim -i \eta_1^{(0)} \eta_1^{(\pi)} \sin(\sqrt{2} \varphi_{d1-}) e^{i \frac{\theta_\rhotot}{\sqrt{2}}} ~, ~~\\
&& n_{(k_{Fd1}^{(0)} - k_{Fd1}^{(\pi)}, \pi)} \equiv d_{1L}^{(0)\dagger} d_{1L}^{(\pi)} + d_{1R}^{(\pi)\dagger} d_{1R}^{(0)} \label{nk0-pi}\\
&& ~~~~~~~~~~~~~~~~~ \sim -i \eta_1^{(0)} \eta_1^{(\pi)} \sin(\sqrt{2} \varphi_{d1-}) e^{i \sqrt{2}\theta_{d1-}} ~, ~~ \\
&& n_{(2k_{Ff}, 0)} \equiv \sum_s f_{sL}^\dagger f_{sR}
\sim i \cos(\sqrt{2}\theta_{f\sigma}) e^{i \frac{\theta_\rhotot}{\sqrt{2}}} ~. ~~~~~~~~~
\end{eqnarray}
In the second line, $\pm$ refers to $e^{i k_y}$.  Note also that at $\rho = 1/3$ we have $2k_{Fd2} = -2k_{Ff}$, and the product $n_{(2k_{Fd2}, 0)} n_{(2k_{Ff}, 0)}$ gives the umklapp term Eq.~(\ref{humklapp}); however, at any other density these are distinct wavevectors.  Combinations similar to those in Eqs.~(\ref{nk0+pi}) and (\ref{nk0-pi}) but with minus sign between the terms contribute to a current operator rather than the density fluctuation and are not spelled out here.

We can similarly obtain fermionic bilinear contributions to the electron spin operator, e.g.,
\begin{eqnarray}
\label{eqn:spinon2kF}
S^z_{(2k_{Ff}, 0)} &\equiv& \sum_s s f_{sL}^\dagger f_{sR}
\sim -\sin(\sqrt{2}\theta_{f\sigma}) e^{i \frac{\theta_\rhotot}{\sqrt{2}}} ~. ~~~~~~
\end{eqnarray}
In the bosonized expressions above, we have omitted $\theta_a$ and $\theta_A$ as these are pinned by the gauge field fluctuations; the discussion of the power law correlations in the $d$-metal phase only depends on the displayed fluctuating field content.

Finally, we have bilinears carrying zero momentum contributing to both charge and spin, e.g.,
\begin{eqnarray}
S^z_{(0, 0)} &\equiv& \sum_{s,P} s f_{sP}^\dagger f_{sP}
= \frac{\sqrt{2}}{\pi} \partial_x \theta_\sigma ~;
\end{eqnarray}
such contributions have scaling dimension $1$.

Let us first note some general relations among scaling dimensions for various density and spin correlations:
\begin{eqnarray}
\Delta[n_{(2k_{Ff}, 0)}] = \Delta[\vec{S}_{(2k_{Ff}, 0)}] &=& \frac{1}{2} + \frac{\Delta[n_{(2k_{Fd2}, 0)}]}{4} ~, ~~~~ \\
\Delta[n_{(k_{Fd1}^{(0)} - k_{Fd1}^{(\pi)}, \pi)}] &\geq& 1 ~.
\label{Delta_bound}
\end{eqnarray}
Thus, the scaling dimension of $n_{(k_{Fd1}^{(0)} - k_{Fd1}^{(\pi)}, \pi)}$ is always larger than the mean field value of $1$.

Next, we can get some quantitative feel by using the approximation of decoupled ``$d1-$'' and ``$\rhotot$'' modes; the scaling dimensions in this approximation are:
\begin{eqnarray}
&& \Delta[n_{(2k_{Fd2}, 0)}] = \frac{g_\rhotot}{2} ~,\\
&& \Delta[n_{(2k_{Fd1}^{(k_y)}, 0)}] = \frac{g_{d1-}}{2} + \frac{g_\rhotot}{8} ~, \\
&& \Delta[n_{(k_{Fd1}^{(0)} + k_{Fd1}^{(\pi)}, \pi)}] = \frac{1}{2 g_{d1-}} + \frac{g_\rhotot}{8} ~,\\
&& \Delta[n_{(k_{Fd1}^{(0)} - k_{Fd1}^{(\pi)}, \pi)}] = \frac{1}{2}\left(\frac{1}{g_{d1-}} + g_{d1-} \right) ~, \\
&& \Delta[n_{(2k_{Ff}, 0)}] = \Delta[\vec{S}_{(2k_{Ff}, 0)}] = \frac{1}{2} + \frac{g_\rhotot}{8} ~.
\end{eqnarray}

As before, we obtain results for the bare Gutzwiller wave function by setting $g_{d1-} = g_\rhotot = 1$.  In this case, the smallest scaling dimension is $\Delta[n_{(2k_{Fd2}, 0)}] = \frac{1}{2}$ corresponding to slow $|x|^{-1}$ power law.  We also have
$\Delta[n_{(2k_{Fd1}^{(k_y)}, 0)}] = \Delta[n_{(k_{Fd1}^{(0)} + k_{Fd1}^{(\pi)}, \pi)}] = \Delta[n_{(2k_{Ff}, 0)}] = \Delta[\vec{S}_{(2k_{Ff}, 0)}] = \frac{5}{8}$ corresponding to $|x|^{-5/4}$ power law, which is also enhanced over the mean field $|x|^{-2}$.  The enhancement of the correlations comes from the reduction of the fluctuating $\theta$ content, which is the (1+1)D realization of the ``Amperean'' enhancement\cite{LeeNagaosaWen, Polchinski94, Altshuler94} of correlations for processes that contain parallel gauge charge currents.  On the other hand, $\Delta[n_{(k_{Fd1}^{(0)} - k_{Fd1}^{(\pi)}, \pi)}] = 1$ and is not enhanced over the mean field, since the particle and hole partons are on the same side of the Fermi sea and create anti-parallel currents.  We performed measurements in sample bare Gutzwiller wave functions and verified the dominant $|x|^{-1}$ and $|x|^{-5/4}$ power laws for different oscillating components in the density structure factor.

We remark that the $d$-metal realized in the DMRG appears to have larger $g_\rhotot > 1$ and hence $n_{(2k_{Fd2}, 0)}$ can be suppressed compared to $n_{(2k_{Fd1}^{(k_y)}, 0)}$.  In fact, from our earlier discussion of the umklapp term Eq.~(\ref{humklapp}) at density $1/3$ and absence of charge order in the $d$-metal phase found in the DMRG, we can conclude that $\Delta[h_{\rm umkl}] > 2$, and hence have an exact bound:
\begin{eqnarray}
\Delta[n_{(2k_{Fd2}, 0)}] = \frac{4}{9} \Delta[h_{\rm umkl}] - \frac{2}{9} \geq \frac{2}{3} ~.
\end{eqnarray}
In the approximation of decoupled ``$d1-$'' and ``$\rhotot$'' modes, this corresponds to $g_\rhotot > 4/3$, and in this case $n_{(2k_{Fd2}, 0)}$ already has larger scaling dimension than $n_{(2k_{Fd1}^{(k_y)}, 0)}$ (if we assume $g_{d1-} = 1$).  Let us also quote the numbers for our earlier dressed Gutzwiller example where we set $g_{d1-} = 1$ and $g_\rhotot = 2$:  We get $\Delta[n_{(2k_{Fd2}, 0)}] = 1$ that is larger than $\Delta[n_{(2k_{Fd1}^{(k_y)}, 0)}] = \Delta[n_{(k_{Fd1}^{(0)} + k_{Fd1}^{(\pi)}, \pi)}] = \Delta[n_{(2k_{Ff}, 0)}] = \Delta[\vec{S}_{(2k_{Ff}, 0)}] = 3/4$, so the latter singularities are more strong.
The DMRG indeed finds visible features in the density structure factors at wavevectors $(2k_{Fd1}^{(k_y)}, 0)$, which we track to identify the $d_1$ bonding/antibonding orbital populations.  On the other hand, at density 1/3, $2k_{Ff} = -2k_{Fd2} = 2\pi/3$ and these wavevectors are not visible in the density structure factors in much of the data.  In the above approximation, we can make the $(2k_{Fd1}^{(k_y)}, 0)$ to be dominant if we also assume $g_{d1-} < 1$.  This is consistent with the condition that the interaction $h_{{\rm int}, 4d_1}$ in Eq.~(\ref{h4d}) is irrelevant.

Furthermore, inspired by the DMRG observation of a singularity in the spin structure factor at wavevector $(2{\mathsf K}_e, 0)$, we have also considered four-fermion contributions to the spin operator, since, in the strongly coupled theory, these can be comparable in prominence to the fermionic bilinears.  We have identified two interesting terms, which can be constructed by combining the already exhibited bilinears,
\begin{eqnarray}
S^z_{(2{\mathsf K}_e, 0)} &\!\sim\!& A S^z_{(-2k_{Ff}, 0)} n_{(2k_{Fd1}^{(0)}, 0)} + B S^z_{(2k_{Ff}, 0)} n_{(-2k_{Fd1}^{(\pi)}, 0)} \nonumber \\
&\!\sim\!& \sin(\sqrt{2}\theta_{f\sigma}) e^{i \sqrt{2}\theta_{d1-}} ~,\\
S^z_{(0, \pi)} &\!\sim\!& S^z_{(-2k_{Ff}, 0)} n_{(k_{Fd1}^{(0)} + k_{Fd1}^{(\pi)}, \pi)} + \Hc \nonumber \\
&\!\sim\!& i \eta_1^{(0)} \eta_1^{(\pi)} \cos\left[\frac{2 (\sqrt{2}\theta_A - \theta_a)}{\sqrt{3}} \right]
\label{thtA_thta_comb}\\
&& \times \sin(\sqrt{2}\theta_{f\sigma}) \sin(\sqrt{2} \varphi_{d1-}) ~.
\label{Sz0pi}
\end{eqnarray}
Note that the first term can be also constructed by combining $c^{(k_y)\dagger}_{-{\mathsf K}_e}$ and $c^{(k_y)}_{{\mathsf K}_e}$ and is a kind of ``electron $2k_F$'' from the dominant feature observed in the electron distribution function.  Similarly, the second term can be constructed by combining, e.g., $c^{(0) \dagger}_{{\mathsf K}_e}$ and $c^{(\pi)}_{{\mathsf K}_e}$.  Writing out such electron bilinears gives six-parton terms, which turn out to be equivalent to the above four-parton terms upon considering the pinning of $\theta_A$ and $\theta_a$.  In the last equation, we have also carefully kept track of these pinned fields, since the very presence or absence of this contribution can depend on the pinning values.  Both the DMRG and VMC have a visible feature in the spin structure factor at the wavevector $(0, \pi)$, cf.~Fig.~\ref{fig:dmetcorrs}(c) of the main text, and hence we conjecture that the pinning of the fields $\theta_{A}$ and $\theta_a$ is such as to give nonzero cosine for their combination exhibited above.  This knowledge will be useful when discussing Cooper pair correlations in the next section (note that the precise pinning values of the $\theta_{A}$ and $\theta_a$ fields were not important in the observables discussed earlier).

In the same spirit, we can also construct four-parton contributions to the electron density, $n_{(2{\mathsf K}_e, 0)}$ and $n_{(0, \pi)}$.  We expect the former to be present generically with properties similar to $S^z_{(2{\mathsf K}_e, 0)}$.  On the other hand, the expression for the latter contains sine of the $\theta_{A}$ and $\theta_a$ combination in Eq.~(\ref{thtA_thta_comb}), and we conjecture that the pinning is such that the $n_{(0, \pi)}$ term vanishes.  This is consistent with the absence of any feature in the DMRG and VMC density structure factor at $(0,\pi)$, compare Figs.~\ref{fig:dmetcorrs}(b) and (c) of the main text.

From the bosonized expressions, we can see that the scaling dimension of $S^z_{(0, \pi)}$ is related to that of the allowed interaction $h_{{\rm int}, 4d_1}$, Eq.~(\ref{h4d}):
\begin{eqnarray}
\Delta[S^z_{(0, \pi)}] = \frac{1}{2} + \frac{\Delta[h_{{\rm int}, 4d_1}]}{4} ~.
\label{DeltaSz0pi}
\end{eqnarray}
Since the stability of the $d$-metal requires $\Delta[h_{{\rm int}, 4d_1}] > 2$, we conclude that $\Delta[S^z_{(0, \pi)}] > 1$.  This is indeed what we found from the DMRG measurements.  In the approximation with decoupled ``$d1-$'' and ``$\rhotot$'' modes, we find
\begin{eqnarray}
\Delta[S^z_{(2{\mathsf K}_e, 0)}] &=& \frac{1}{2} + \frac{g_{d1-}}{2} ~,\\
\Delta[S^z_{(0, \pi)}] &=& \frac{1}{2} + \frac{1}{2 g_{d1-}} ~.
\end{eqnarray}

To summarize, many such qualitative and semi-quantitative considerations of the observables, relations among exponents, and stability to perturbations are internally consistent, giving us more confidence that the phase found in the DMRG is indeed the electronic $d$-metal.

\subsection{Cooper pair correlation}
\label{subsec:Cooper}
To complete the discussion of observables, we also consider electron Cooper pair operators.  Besides being prominent observables themselves, these are needed, e.g., in Sec.~\ref{subsec:instabilities} for the detailed description of the phases proximate to the $d$-metal when the electron operator gets gapped.  We consider only spin singlets and can start with microscopic Cooper operators defined as in Eq.~(\ref{CooperPij}) on some nearby pair of sites.  We can write each electron operator in terms of the partons and obtain contributions containing six continuum parton fields (two for each species $d_1$, $d_2$, and $f$).  We examined all such six-fermion terms, and below we present most interesting ones bearing in mind the application to proximate phases.  Rather than doing direct expansion, we can use symmetry arguments to identify which microscopic Cooper pairs receive particular contributions.

The most important such six-fermion terms are
\begin{eqnarray}
{\mathcal P}_{(0,\pi)}^{(e)} &\equiv& d_{2R} d_{2L} d_{1R}^{(0)} d_{1R}^{(\pi)} f_{\up L} f_{\dn L} + (R \leftrightarrow L) \\
&\sim& \eta_1^{(0)} \eta_1^{(\pi)} \eta_\up \eta_\dn \sin\left[\frac{2 (\sqrt{2}\theta_A - \theta_a)}{\sqrt{3}} \right] e^{i 2\sqrt{2} \phi_\rhotot} ~, \nonumber \label{P-0pi-e} \\
{\mathcal P}_{(0,\pi)}^{(o)} &\equiv& -i \left[d_{2R} d_{2L} d_{1R}^{(0)} d_{1R}^{(\pi)} f_{\up L} f_{\dn L} - (R \leftrightarrow L) \right] \\
&\sim& \eta_1^{(0)} \eta_1^{(\pi)} \eta_\up \eta_\dn \cos\left[\frac{2 (\sqrt{2}\theta_A - \theta_a)}{\sqrt{3}} \right] e^{i 2\sqrt{2} \phi_\rhotot} ~. \nonumber \label{P-0pi-o} 
\end{eqnarray}
These terms carry momentum $(0,\pi)$, i.e., they are translationally invariant along the ladder and are odd under the leg interchange.  They are defined to be invariant under time reversal, as is appropriate for the singlet Cooper pairs.  The ``$(e)$''~[``$(o)$''] combination is even~[odd] under mirror $(x,y) \to (-x,y)$; the numerical constant is sine~[cosine] of the particular combination of the pinned fields $\theta_A$ and $\theta_a$.  We expect that one or the other combination is nonzero, but which one depends on the detailed pinning specifying the $d$-metal phase (see below).
From the identified symmetry properties, we can see that the combination ${\mathcal P}_{(0,\pi)}^{(e)}$ contributes to leg-bond Cooper pairs that are anti-symmetric under the leg interchange:
\begin{equation}
{\mathsf P}[(x,1), (x+1,1)] - {\mathsf P}[(x,2), (x+1,2)]
\sim {\mathcal P}_{(0,\pi)}^{(e)}(x) + \dots ~.
\label{P-leg-dwave}
\end{equation}
On the other hand, the combination ${\mathcal P}_{(0,\pi)}^{(o)}$ contributes to diagonal Cooper pairs anti-symmetric under the leg interchange:
\begin{equation}
{\mathsf P}[(x,1), (x+1,2)] - {\mathsf P}[(x,2), (x+1,1)]
\sim {\mathcal P}_{(0,\pi)}^{(o)}(x) + \dots ~.
\label{P-diag-dwave}
\end{equation}
The ${\mathcal P}_{(0,\pi)}$ terms have the smallest content of fluctuating fields in the $d$-metal --- only the $\phi_\rhotot$ part that is necessary to encode the electrical charge of the Cooper pairs.  In the approximation of decoupled ``$d1-$'' and ``$\rhotot$'' modes, the scaling dimension is $2/g_\rhotot$ (equal to $2$ in the bare Gutzwiller and $1$ in the dressed Gutzwiller wave functions).  By measuring whether the leg-bond or diagonal Cooper pair shows power law behavior, we can constrain the appropriate pinning of the fields $\theta_A$ and $\theta_a$.  As we will present in Sec.~\ref{subsec:fits}, both the DMRG and VMC find that it is the diagonal Cooper pairs that have dominant power law correlations in real space.  Hence we conclude that the pinning is such that ${\mathcal P}_{(0,\pi)}^{(o)}$ is nonzero while ${\mathcal P}_{(0,\pi)}^{(e)}$ is zero, which is also consistent with the presence of $\vec{S}_{(0,\pi)}$, Eq.~(\ref{Sz0pi}), and the absence of $n_{(0,\pi)}$ features in the DMRG and VMC.  The specific diagonal Cooper pairs Eq.~(\ref{P-diag-dwave}) are natural in the model with electronic ring exchange: e.g., they arise when solving the ring Hamiltonian for two electrons on a single placket.  They can be viewed as having a $d$-wave character, which is one of the motivations for naming our non-Fermi-liquid phase as ``$d$-wave metal.''

We also mention the following six-fermion terms
\begin{eqnarray}
\label{C00}
{\mathcal P}_{(0,0)} &\equiv& d_{2R} d_{2L} d_{1R}^{(k_y)} d_{1L}^{(k_y)} \left( f_{\up R} f_{\dn L} - f_{\dn R} f_{\up L} \right) \\
&\sim& -\eta_\up \eta_\dn e^{i (2\sqrt{2} \phi_\rhotot \pm \sqrt{2}\phi_{d1-})} \cos(\sqrt{2}\theta_{f\sigma}) ~,
\end{eqnarray}
where $\pm$ refers to $e^{ik_y}$.
Such terms carry zero momentum, i.e., they are translationally invariant along the ladder and are even under the leg interchange.  They are also even under the mirror $(x,y) \to (-x,y)$ and contribute, e.g., to a rung Cooper pair ${\mathsf P}[(x,1), (x,2)]$ (as well as leg-bond or diagonal Cooper pairs symmetric under the leg interchange).   In some loose sense, they can be viewed as ``$s$-wave'' Cooper pairs, while the ${\mathcal P}_{(0,\pi)}$ ones are ``$d$-wave'', the precise distinction lying in the transformation properties under the discrete ladder symmetries.  Because of the additional fluctuating field content, we expect the $s$-wave ones to have larger scaling dimension [equal to $2/g_\rhotot + 1/(2 g_{d1-}) + 1/2$ in the decoupled ``$d1-$'' and ``$\rhotot$'' approximation] and to be less visible in the $d$-metal than the $d$-wave ones.  However, the $s$-wave can become comparably prominent in a spin gap phase discussed in Sec.~\ref{subsec:instabilities}.

\subsection{Sample of the DMRG power law fits}
\label{subsec:fits}
Guided by the long-wavelength description of the $d$-metal, we performed detailed fits for the power laws in various correlations at dominant wavevectors.  Here we highlight some results at the DMRG point $J/t = 2$, $K/t = 1.8$, presented in Fig.~\ref{fig:dmetcorrs} in the main text.  To remind the readers, the best $d$-metal candidate for this $48\times 2$ ladder with $32$ electrons has $2 k_{Fd1}^{(0)} = 21 \cdot 2\pi/48$, $2 k_{Fd1}^{(\pi)} = 11 \cdot 2\pi/48$, cf.~Fig.~\ref{fig:partonbands} of the main text.

The main features in the electron distribution function occur at the wavevectors listed in the first three rows of Table~\ref{tab:Ge}: ${\mathsf K}_e = 5\pi/48$, $2\pi\rho + {\mathsf K}_e = 37\pi/48$, and $2\pi\rho - {\mathsf K}_e = 27\pi/48$, for either $k_y=0$ or $\pi$.  We fit the electron Green's function to an expression
\begin{equation}
G_e(x, k_y) = \sum_a \frac{C_a^{(k_y)} \sin(Q_a x)}{\left[(L_x/\pi) \sin(\pi x/L_x) \right]^{2\Delta_a^{(k_y)}}} ~.
\label{Gfit}
\end{equation}
The dominant oscillation is at the wavevector ${\mathsf K}_e$, which is readily recognized in the electron momentum distribution function in Fig.~\ref{fig:dmetcorrs}(a) of the main text.  As illustrated in Fig.~\ref{fig:Gfits}, including only this wavevector already captures the overall real-space dependence and gives $2\Delta_{+-+}^{(0)} \approx 1.3$ and $2\Delta_{+-+}^{(\pi)} \approx 1.5$ for the bonding and antibonding electrons respectively.  Fits including the other two wavevectors (not shown) capture also finer features without affecting much the estimates of $\Delta_{+-+}^{(0/\pi)}$.  We thus confirm the general expectation that the exponents can be different in the bonding and antibonding electron distribution functions.  Figure~\ref{fig:Gfits} also shows the electron Green's function measured in the VMC wave function.  Similar fits in this case (not shown) give roughly equal values $2\Delta_{+-+}^{(0)} \approx 2\Delta_{+-+}^{(\pi)} \approx 1.25$, which is also expected in the approximate gauge theory treatment with decoupled ``$d1-$'' and ``$\rhotot$''.  Our trial wave functions and the approximate treatment of the gauge theory are not qualitatively accurate in this respect.  The structure of the gauge theory is qualitatively accurate---it is only that we do not know numerical values of the couplings in the full theory with coupled ``$d1-$'' and ``$\rhotot$'' modes.  Nevertheless, we see that the approximate treatment provides a reasonable semi-analytical guide.

\begin{figure}[t]
\centerline{\includegraphics[width=\columnwidth]{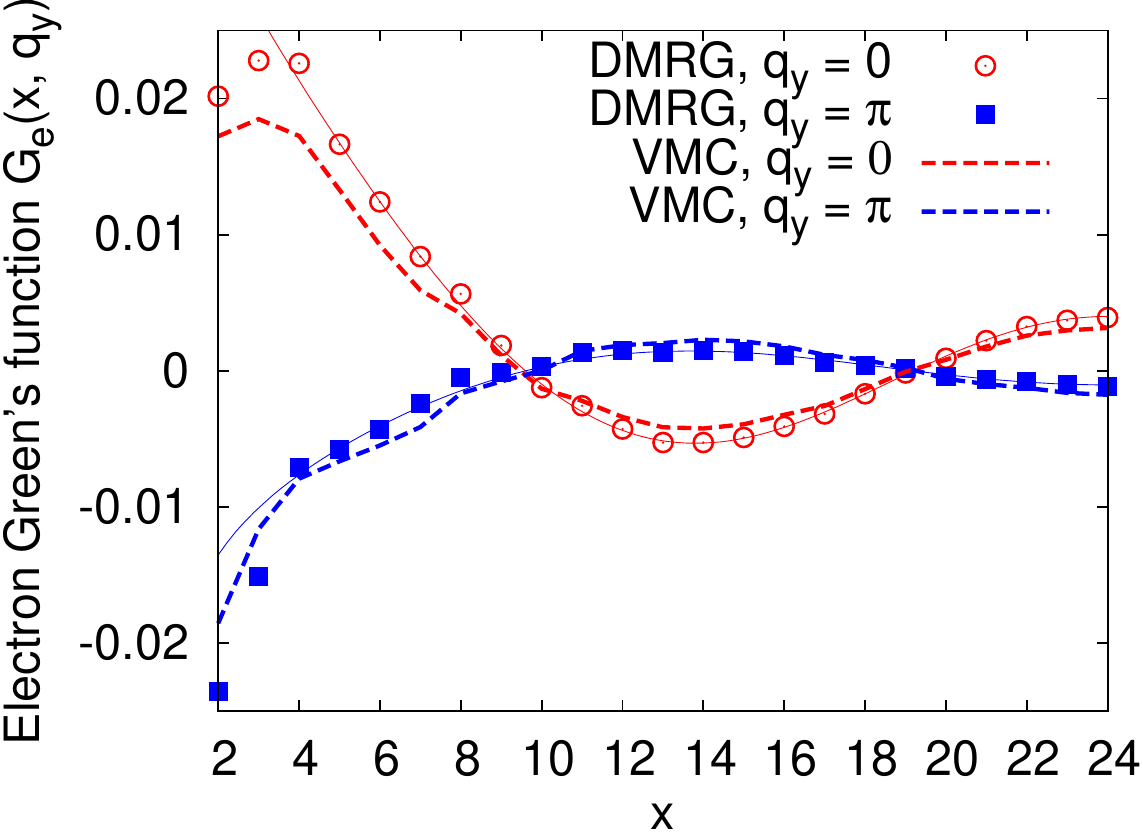}}
\caption{
Real-space DMRG and VMC electron Green's functions $G_e(x)$ in the $L_x=48$ ladder at $J/t=2$, $K/t=1.8$, corresponding to the momentum distribution functions shown in Fig.~\ref{fig:dmetcorrs} of the main text.  Thin solid lines show power law fits of the DMRG data to Eq.~(\ref{Gfit}) restricted to a single oscilating component with wavevector ${\mathsf K}_e$.  We show data for all distances $2\leq x \leq L_x/2$ to bring out the oscillation more clearly, while the fit itself is done over $4<x<L_x/2$ to pick up the long-distance behavior.
}
\label{fig:Gfits}
\end{figure}

Turning to the spin and density correlations, we fit these to an expression similar to Eq.~(\ref{Gfit}), but with cosines instead of sines.  It is simple to fit the spin correlations at $k_y = \pi$, since there is only one feature, Eq.~(\ref{Sz0pi}).  We estimate $2 \Delta[S^z_{(0, \pi)}] \approx 2.2$, which satisfies the $d$-metal stability requirement discussed after Eq.~(\ref{DeltaSz0pi}) and gives us an estimate $g_{d1-} \approx 0.85$ in the approximation with decoupled ``$d1-$'' and ``$\rhotot$''.
On the other hand, the spin correlations at $k_y = 0$ have three important wavevectors: $0$, $2\pi/3$, and $2{\mathsf K}_e = 10\pi/48$.  The power law for the zero-momentum component is fixed at $x^{-2}$.  Fitting power law decays at the other two wavevectors allows us to estimate $g_\rhotot \approx 3.5 - 4$ and $g_{d1-}$ consistent with the previous estimate.  The large value of $g_\rhotot$ explains the weakness of the feature at $2\pi/3$.  As discussed in Sec.~\ref{app:vmcDetails}, our matching VMC state has small negative power $p_2 = -0.4$ on the $d_2$ determinant, which can indeed reproduce such large $g_\rhotot$.  On the other hand, the value of $g_{d1-} < 1$ implies that the component at $2{\mathsf K}_e$ has power law decay slower than $x^{-2}$ in real space and singularity in the structure factor that is stronger than slope change; the singularity at this wavevector can be noticed already in the DMRG data in Fig.~\ref{fig:dmetcorrs}(c) of the main text, while the VMC has harder time reproducing this.

It is also simple to fit the density correlations at $k_y = \pi$, where we have two wavevectors $k_{Fd1}^{(0)} + k_{Fd1}^{(\pi)} = 2\pi/3$ and $k_{Fd1}^{(0)} - k_{Fd1}^{(\pi)} = 10\pi/48$.  The power law fits give $2 \Delta[n_{(k_{Fd1}^{(0)} + k_{Fd1}^{(\pi)}, \pi)}] \approx 2.3$ and $2 \Delta[n_{(k_{Fd1}^{(0)} - k_{Fd1}^{(\pi)}, \pi)}] \approx 2.1$.  The latter is consistent with the general bound in Eq.~(\ref{Delta_bound}) and with $g_{d1-}$ value close to $1$, while the former is consistent with $g_\rhotot \approx 4$.
On the other hand, the density correlations at $k_y = 0$ have many wavevectors: $0$, $2k_{Fd1}^{(0/\pi)}$, $2\pi/3$, and $2{\mathsf K}_e$.  Based on the previous estimates, we expect that all these components have similar scaling dimension of order $1$.  We can indeed get nice fits, but because of the many parameters it is difficult to give accurate individual exponents; nevertheless, all are consistent with the previous estimates of $g_{d1-}$ and $g_\rhotot$.

Finally, we also measured the diagonal $d$-wave Cooper pair correlations in the DMRG and VMC.  For the $48 \times 2$ sample above, we found it difficult to fully converge the pair correlations in the DMRG (the electron Green's function shown earlier converged more readily).  Nevertheless, we observed very clearly a slow power law decay in this Cooper channel, fitting roughly $\sim 1/x^{1.15}$ in the DMRG and $\sim 1/x^{0.85}$ in the VMC.  For illustration, in Fig.~\ref{fig:DDwaveCooper} we show the measurements in the smaller $36 \times 2$ sample at $J/t=2$ and $K/t=2$, where the DMRG data is better converged, with the remaining uncertainty less than 5\% (the same sample was used in the R\'enyi entropy comparisons in Fig.~\ref{fig:k20-entropy}).  We see good match between the DMRG and VMC results, with power law fits giving roughly $\sim 1/x^{1.1}$ in the DMRG and $\sim 1/x^{0.9}$ in the VMC.  Comparing with the predictions in Sec.~\ref{subsec:Cooper},
the exponents are again consistent with $g_\rhotot$ value close to $4$.
All other Cooper pair correlators, including also the leg-bond $d$-wave Eq.~(\ref{P-leg-dwave}), decay much faster.  This observation constrains the pinning values of the $\theta_A$ and $\theta_a$ fields in our fixed-point theory of the $d$-metal, as discussed with regards to which of the two Cooper pairs Eq.~(\ref{P-0pi-e}) or Eq.~(\ref{P-0pi-o}) are present.  
Furthermore, note that the measured power law decay is comparable or even slower than that in the electron Green's function, so the diagonal $d$-wave Cooper pair operator is one of the most prominent observables in the $d$-metal phase on the two-leg ladder.  It is this observation that leads us to speculate about incipient $d$-wave superconductivity in this strange metal phase.

\begin{figure}[t]
\centerline{\includegraphics[width=\columnwidth]{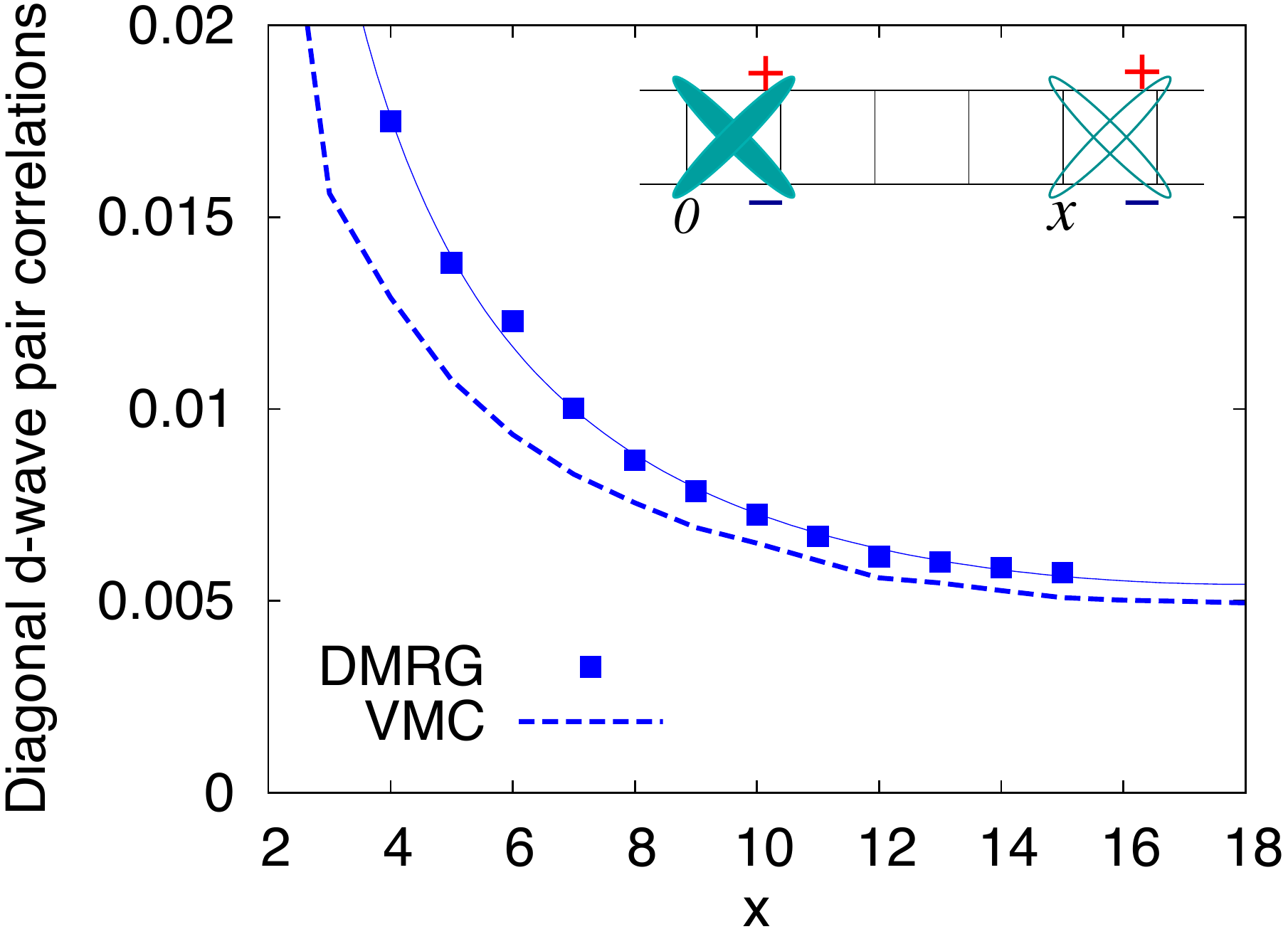}}
\caption{
DMRG and VMC diagonal $d$-wave Cooper pair correlations [see expression below Eq.~(\ref{CooperPij})] in the system of length $L_x=36$ at $J/t=2$ and $K/t=2$ (same as in Fig.~\ref{fig:k20-entropy}).  We show the DMRG data for $x \leq 15$, where the convergence error is estimated to be less than 5\% having kept $m=8$,500 states in the DMRG.  Thin line shows fit of the DMRG data to the functional form $\sim \left[(L_x/\pi) \sin(\pi x/L_x) \right]^{-p}$, with $p \approx 1.1$.  In the inset, we illustrate the computed correlation function by showing two diagonal $d$-wave ($d_{xy}$) Cooper pairs, one being created and the other being destroyed, separated by a distance $x$.
} 
\label{fig:DDwaveCooper}
\end{figure}

To conclude, many detailed properties of the various measured correlations match well with those expected in the $d$-metal theory.  This provides strong support for the identification of the phase found in the DMRG as the $d$-metal.

\subsection{Instabilities of the $d$-metal}
\label{subsec:instabilities}
With all observables at hand, we can now consider phases that would be obtained out of the $d$-metal when one or both of the interactions $h_{{\rm int}, 4d_1}$ and $h_{{\rm int}, 4f}$, Eqs.~(\ref{h4d}) and (\ref{h4f}) respectively, become relevant.  Since the electron operator gets gapped out, a detailed description requires consideration of the Cooper pair operators.  In the $d$-metal phase and in all phases below, the Cooper pair ${\mathcal P}_{(0,\pi)} \sim e^{i 2\sqrt{2} \phi_\rhotot}$ discussed in Sec.~\ref{subsec:Cooper} is always prominent, and only one more Cooper pair will come to prominence in one of the phases.  On the other hand, clear distinctions between possible phases are already apparent with the density and spin observables.  There are three cases:

1) If the interaction $h_{{\rm int}, 4f}$ is relevant, it pins $\theta_{f\sigma}$ and gaps the spin.  Assuming $h_{{\rm int}, 4d_1}$ remains irrelevant, we have two gapless modes ``$d1-$'' and ``$\rhotot$'' and the full theory is similar to the DBL[2,1] theory in Ref.~\onlinecite{Sheng2008_2legDBL}.  A prominent density observable is $n_{(2k_{Ff}, 0)} \sim e^{i \theta_\rhotot/\sqrt{2}}$, whose wavevector $(2\pi\rho, 0)$ is determined by the electron density.  Loosely speaking, this phase can be thought of as a two-leg analog of a ``pseudogap,'' i.e., a nonsuperconducting quantum fluid with a spin gap.

2) If the interaction $h_{{\rm int}, 4d_1}$ is relevant, it pins $\varphi_{d1-}$.  Assuming $h_{{\rm int}, 4f}$ remains irrelevant, we have two gapless modes ``$f\sigma$'' and ``$\rhotot$'', with one Luttinger parameter in the latter sector.  In this case, the spin correlations remain gapless.  Depending on the precise pinning of $\varphi_{d1-}$ determined by the sign of $w$ in Eq.~(\ref{h4d}), we get dominant density correlations $n_{(k_{Fd1}^{(0)} + k_{Fd1}^{(\pi)}, \pi)} \sim \sin(\sqrt{2} \varphi_{d1-}) e^{i \theta_\rhotot/\sqrt{2}}$ or current correlations $j_{(k_{Fd1}^{(0)} + k_{Fd1}^{(\pi)}, \pi)} \sim \cos(\sqrt{2} \varphi_{d1-}) e^{i \theta_\rhotot/\sqrt{2}}$; the wavevector is $(2\pi\rho, \pi)$ and is distinct from the case 1).
This phase can be viewed as a phase where the bosonic chargons form a paired-boson state,\cite{Sheng2008_2legDBL} while the spin remains gapless, and is akin to a non-Fermi liquid termed ``orthogonal metal'' discussed recently in Ref.~\onlinecite{Nandkishore12_PRB_86_045128}.

3) Finally, if both $h_{{\rm int}, 4d_1}$ and $h_{{\rm int}, 4f}$ are relevant, we have pinning of the fields $\theta_{f\sigma}$ and $\varphi_{d1-}$ as in the cases 1) and 2) above.  Only one gapless mode ``$\rhotot$'' remains.  We have presence of both $(2\pi\rho, 0)$ and $(2\pi\rho, \pi)$ wavevectors in the density (or appropriate current depending on the sign of $w$) correlation with the same power law.  Furthermore, the Cooper pair ${\mathcal P}_{(0,0)}$, Eq.~(\ref{C00}), at wavevector $(0,0)$ now becomes as prominent as the ${\mathcal P}_{(0,\pi)}$ at $(0,\pi)$.  This phase is the two-leg analog of a superconductor.

Since the DMRG did not observe strong density correlations at the wavevector $2\pi\rho$ along the ladder or any other signatures of instability, we conclude that our $d$-metal phase is stable in the $t$-$J$-$K$ model.  In future work, it would be interesting to modify the model to further explore the above proximate phases.

\section{Eliminating conventional Luttinger liquid scenarios in favor of the $d$-metal}
\label{app:convLL}

In light of the remarkable success to date of describing 1D and quasi-1D interacting quantum systems with conventional Luttinger liquid theory,\cite{Giamarchi03_1D} it is natural to ask whether the results in our putative $d$-metal phase can be reproduced with such a conventional weak-coupling approach.\cite{Balents96_PRB_53_12133}  Clearly, since the number of gapless modes in the putative $d$-metal ($c=3$) is larger than in the conventional one-band metal ($c=2$), the former cannot be understood as an instability of the latter.  However, there are more complicated scenarios that one may envision that involve strong Fermi surface renormalization, as well as electron pairing, but that still lie within the \emph{conventional} Luttinger framework and still assume a free electron starting point.

For instance, one could first imagine the $K$ term renormalizing the free electron band structure such that the antibonding band eventually gets populated (as if $K$ had the effect of simply renormalizing the interchain hopping $t_\perp$ towards zero---this is admittedly somewhat natural given that $H_K$ conserves the number of electrons in each chain).  If we denote a conventional Luttinger liquid with $\alpha$ gapless charge modes and $\beta$ gapless spin modes as C$\alpha$S$\beta$ (see Ref.~\onlinecite{Balents96_PRB_53_12133}), then this free electron state would be some C2S2 metal with $c=4$ gapless modes, say a charge ($\rho$) and spin ($\sigma$) mode for each band ($0/\pi$): $\theta_{\rho/\sigma}^{(0/\pi)}$.  In principle, a spin gap in the antibonding band could be opened through relevance of a term involving only a cosine of the $\theta_\sigma^{(\pi)}$ field,\cite{Balents96_PRB_53_12133} giving a C2S1 metal with $c=3$ gapless modes.  However, this possibility can be immediately ruled out in our putative $d$-metal region by noting that the DMRG state unambiguously has a critical Green's function for the antibonding electrons:  See the sharp step in $\langle c_{\qvec s}^\dagger c_{\qvec s}\rangle$ at $\qvec=(q_x,q_y)=(\Ke,\pi)$ in Fig.~\ref{fig:dmetcorrs}(a) of the main text, as well as the discussion in Sec.~\ref{subsec:fits} of the slow power law decay of the Green's function in real space.  Pinning of the $\theta_\sigma^{(\pi)}$ field, on the other hand, directly implies that the electron Green's function would decay \emph{exponentially} at $q_y=\pi$, in clear contradiction with our DMRG data.

Other aspects of the DMRG data are also markedly inconsistent with this C2S1 scenario.  For example, throughout the $d$-metal phase at, say, $J/t=2$ (to avoid small polarization observed at smaller $J/t$), we observe an enhanced feature in the spin-spin structure factor $\langle \mathbf{S}_{\qvec}\cdot \mathbf{S}_{-\qvec}\rangle$ at $\qvec=(2\pi\rho,0)$ for all $K/t$.  The location of this feature is fixed by the electron density and is readily explainable by our $d$-metal theory [see Eq.~(\ref{eqn:spinon2kF})].  In the C2S1 state discussed above, however, the only singularity in $\langle \mathbf{S}_{\qvec}\cdot \mathbf{S}_{-\qvec}\rangle$ at $q_y=0$ would be at $q_x=2k_F^{(0)}$, where $\pm k_F^{(0)}$ denotes the Fermi points for the bonding electrons assumed to be gapless.  This wavevector is not fixed by the electron density and is more akin to our observed feature at $q_x=2\Ke$.  Hence, presence of the feature at $\qvec=(2\pi\rho,0)$ in $\langle \mathbf{S}_{\qvec}\cdot \mathbf{S}_{-\qvec}\rangle$ [see Fig.~\ref{fig:dmetcorrs}(c) in the main text] is not consistent with a conventional C2S1.  The C2S1 state also fails to explain things like why $\langle c_{\qvec s}^\dagger c_{\qvec s}\rangle$ has substantial weight outside the main ``step'' for the bonding electrons, whereas this observation is very naturally explained by the $d$-metal theory as shown explicitly by the VMC calculations in Fig.~\ref{fig:dmetcorrs}(a) of the main text as well as heuristically according to the convolution argument in Fig.~\ref{fig:convolution}.  There are yet more features in our data that are clearly inconsistent with this scenario which we do not mention further.

Having now eliminated the possibility that our putative $c=3$ $d$-metal could instead be some conventional C2S1 obtained through band renormalization and pairing, we now consider an even more complicated ``standard'' Luttinger scenario.  Rather than focusing on the fact that we measure central charge $c=3$ in the putative $d$-metal region, we can instead try to directly accommodate the obtained electron momentum distribution function $\langle c_{\qvec s}^\dagger c_{\qvec s}\rangle$ of Fig.~\ref{fig:dmetcorrs}(a) in the main text by postulating a band structure in which the antibonding electrons fill two disconnected Fermi segments, while the bonding electrons fill a single segment centered about $q_x=0$.  That is, consider a C3S3 metal with a Fermi sea $\left[-k_F^{(0)},+k_F^{(0)}\right]$ in the bonding band and Fermi seas $\left[-k_{F2}^{(\pi)},-k_{F1}^{(\pi)}\right]$, $\left[+k_{F1}^{(\pi)},+k_{F2}^{(\pi)}\right]$ in the antibonding band $\left(k_{F2}^{(\pi)}>k_{F1}^{(\pi)}>0\right)$.  Such a band structure could arise, for example, if the $K$ term renormalizes the antibonding electrons to have substantial next-nearest-neighbor hopping.  Furthermore, to be consistent with the DMRG data, we must take $k_F^{(0)}=k_{F1}^{(\pi)}\equiv\Ke$, which would of course not generally be true for such a band structure, but such a fine-tuned state could at least exist in principle.

There exist many possible phases obtainable by gapping out various modes in the $c=6$ C3S3 state.  However, to reproduce the DMRG data within this framework, we must retain gapless electron fields at $\pm k_F^{(0)}$ and $\pm k_{F1}^{(\pi)}$ which directly implies $c\geq4$.  Thus, due to our measurement of $c=3<4$ with the DMRG [see Fig.~\ref{fig:centralCharge} of the main text and Sec.~\ref{app:entropy}] we can eliminate \emph{on very general grounds} any weak-coupling scenario that ends with critical electrons in both the bonding and antibonding bands.

All in all, the above weak-coupling scenarios clearly have severe difficulty describing the DMRG data obtained in the putative $d$-metal region of the phase diagram.  Of course, we cannot rule out every single weak-coupling scenario, including even more complicated and contrived ones, but the above two possibilities would be the most natural in our view, and they are clearly not working.  On the other hand, we stress that our $d$-metal framework can basically describe the \emph{entire} DMRG data set in a very natural, unified fashion, giving us a high degree of confidence that our novel \emph{non-perturbative} $d$-metal theory is indeed correct.  In addition, as discussed in the next section, the structure of the $d$-metal gauge theory itself gives us reason to at least anticipate that the \tJK model Hamiltonian may harbor the non-Fermi liquid $d$-metal phase.

\section{Motivation for the $t$-$J$-$K$ model to realize electronic $d$-metal} \label{app:ringMot}

The purpose of this section is to give some analytical intuition why the electronic ring Hamiltonian likes the particular $d$-metal phase.  Of course, we need direct numerical studies to determine what actually happens in the particular Hamiltonian, as done in the main text.

We can gain some feeling for the energetics by doing a slave particle mean field calculation.  For convenience, we write the electronic ring term as the following electron pair-hopping,
\begin{equation}
\hat{R}_{1234} = \left(c_{1\alpha}^\dagger c_{2\alpha} \right) \left(c_{3\beta}^\dagger c_{4\beta} \right) + \left(c_{1\alpha}^\dagger c_{4\alpha} \right) \left(c_{3\beta}^\dagger c_{2\beta} \right) ~.
\label{R1234}
\end{equation}
Here and below, summation over repeated spin indices is implied.
Assuming decoupled $d_1$, $d_2$, and $f$ partons with translationally invariant hopping mean field, we can calculate the expectation value of the ring term on an elementary square as
\begin{eqnarray}
&& \la \hat{R}_{r, r+\hat{x}, r+\hat{x}+\hat{y}, r+\hat{y}} \ra_{\rm mf} = 2 \left( |\chi_{f,x}|^2 + |\chi_{f,y}|^2 \right) \times \nonumber \\
&& ~~~~~ \times \left( |\chi_{d1,x}|^2 - |\chi_{d1,y}|^2 \right) \left( |\chi_{d2,x}|^2 - |\chi_{d2,y}|^2 \right) ~, ~~~~
\end{eqnarray}
where $\chi_{d_a,\mu} \equiv \la d_a^\dagger(r) d_a(r+\hat{\mu}) \ra_{\rm mf}$, and $\chi_{f,\mu}$ is evaluated similarly for one spinon species (we assume identical hopping for the $\up$ and $\dn$ spinons).  The plus sign between $|\chi_x|^2$ and $|\chi_y|^2$ for the $f$-spinons comes from adding the two terms in Eq.~(\ref{R1234}) and summing over the spin labels, while we have the minus sign for the $d$-partons from the Fermi statistics.  If $d_1$ ($d_2$) partons hop preferentially in the $\hat{x}$ (respectively $\hat{y}$) direction, then $|\chi_{d1,x}|^2 > |\chi_{d1,y}|^2$ (respectively $|\chi_{d2,x}|^2 < |\chi_{d2,y}|^2$).  In this case, the expectation value of the ring term is negative, so for the positive $K$ in our model we obtain low ring energy; increasing the anisotropy in the $d_1$ and $d_2$ hoppings lowers the ring energy, while the optimal anisotropy is determined from the competition with the electronic kinetic and spin exchange energies.  This is our crude mean field argument why the ring terms prefer the particular fractionalized state and why the $d_1$ and $d_2$ partons develop strong anisotropies for large $K$.

We can also provide a more constructive motivation for the electronic ring model as a candidate for realizing the $d$-metal phase.  This ``reverse engineering'' argument starts with an effective lattice gauge theory for the $d$-metal phase with partons $d_1$, $d_2$, and $f$.  To simplify further, let us go back to splitting the electron $c_\alpha = b f_\alpha$ into bosonic chargon $b$ and fermionic spinon $f_\alpha$ and seek a system where the chargons are in a ``$d$-wave Bose metal' (DBM) phase introduced in Ref.~\onlinecite{DBL} while the spinons are in a spinon Fermi sea state.  The latter can be obtained by postulating simple hopping for the spinons.  On the other hand, to put the chargons into the DBM state, we follow prior works proposing such a phase in frustrated boson models with ring exchanges and realizing it in two-leg, three-leg, and four-leg ladder studies.\cite{Sheng2008_2legDBL, Block2011_3legGMI, Mishmash2011_4legDBL}  Note that this ``mean field Hamiltonian'' for the chargons is actually strongly coupled; it was motivated in Ref.~\onlinecite{DBL} as a candidate that would produce the DBM phase where we further split $b$ into $d_1$ and $d_2$ partons with anisotropic hopping, precisely as we want in the $d$-metal of electrons.  Here by starting with the ring Hamiltonian for the $b$'s we avoid duplicating efforts in Refs.~\onlinecite{DBL, Sheng2008_2legDBL, Block2011_3legGMI, Mishmash2011_4legDBL}.

Going beyond the ``mean field,'' we need to include a compact U(1) gauge field that attempts to glue the $b$ and $f_\alpha$ partons to produce the physical electron.\cite{LeeNagaosaWen}  Denoting this gauge field on the (oriented) lattice links $\la rr' \ra$ as $A_{rr'}$ and the conjugate integer-valued electric field as $E_{rr'}$, the lattice gauge theory Hamiltonian for the $d$-metal phase has the structure
\ifREVTEX
	\begin{widetext}
\fi
\begin{eqnarray}
H & = & h \sum_{\la rr' \ra} E_{rr'}^2 - \kappa \sum_\square \cos(\nabla \times A) -\sum_{\la rr' \ra} \left(t_{rr'}^{(f)} e^{-iA_{rr'}} f_{r\alpha}^\dagger f_{r'\alpha} + \Hc \right)
-\sum_{\la rr' \ra} \left(t_{rr'}^{(b)} e^{iA_{rr'}} b_r^\dagger b_{r'} + \Hc \right) \nonumber \\
& + & K^{(b)} \sum_\square \left[\left(e^{i A_{r, r+\hat{x}}} e^{i A_{r+\hat{x}+\hat{y}, r+\hat{y}}} + e^{i A_{r, r+\hat{y}}} e^{i A_{r+\hat{x}+\hat{y}, r+\hat{x}}} \right) b_r^\dagger b_{r+\hat{x}} b_{r+\hat{x}+\hat{y}}^\dagger b_{r+\hat{y}} + \Hc \right] ~.
\end{eqnarray}
\ifREVTEX
	\end{widetext}
\fi
The Hamiltonian is supplemented by the Gauss law constraint
\begin{equation}
\left(\vec{\nabla} \cdot \vec{E} \right)(r) = b_r^\dagger b_r - f_{r\alpha}^\dagger f_{r\alpha} ~.
\end{equation}
In the gauge theory Hamiltonian, $t_{rr'}^{(f)}$ and $t_{rr'}^{(b)}$ are hopping amplitudes for the spinon and chargon respectively.  The $b$ and $f$ partons are coupled to the gauge field $A$ with opposite gauge charges, so that the physical electron operator is gauge neutral.  We have also generalized the boson ring term with coupling $K^{(b)}$ to a gauge-invariant form that also respects the square lattice symmetries.  The $h$ and $\kappa$ terms are standard for the lattice gauge field dynamics.

For large $h$, the electric fields are pinned at $E_{rr'}=0$ and the partons are confined.  We can identify the sector $E_{rr'}=0$ with the physical Hilbert space of electrons.  Starting with this limit and working perturbatively in $t^{(f)}$, $t^{(b)}$, and $K^{(b)}$, we obtain an effective Hamiltonian for electrons that contains the following terms:
\ifREVTEX
\begin{widetext}
\begin{eqnarray}
H_{\rm el} = -\sum_{\la rr' \ra} \left( \frac{2 t_{rr'}^{(f)} t_{rr'}^{(b)}}{h} c_{r\alpha}^\dagger c_{r'\alpha} + \Hc \right)
+ \sum_{\la rr' \ra} \frac{4 |t_{rr'}^{(f)}|^2}{h} \vec{S}_r \cdot \vec{S}_{r'}
+ \frac{4 |t^{(f)}|^2 K^{(b)}}{h^2} \sum_\square \left[\hat{R}_{r, r+\hat{x}, r+\hat{x}+\hat{y}, r+\hat{y}} + \Hc \right] + \dots ~.
\end{eqnarray}
\end{widetext}
\else
\begin{eqnarray}
H_{\rm el} & = & -\sum_{\la rr' \ra} \left( \frac{2 t_{rr'}^{(f)} t_{rr'}^{(b)}}{h} c_{r\alpha}^\dagger c_{r'\alpha} + \Hc \right)
+ \sum_{\la rr' \ra} \frac{4 |t_{rr'}^{(f)}|^2}{h} \vec{S}_r \cdot \vec{S}_{r'} \nonumber \\
& + & \frac{4 |t^{(f)}|^2 K^{(b)}}{h^2} \sum_\square \left[\hat{R}_{r, r+\hat{x}, r+\hat{x}+\hat{y}, r+\hat{y}} + \Hc \right] + \dots ~.
\end{eqnarray}
\fi
The electron hopping is obtained from second-order processes hopping both chargon and spinon, while the spin-spin interaction is obtained from second-order processes exchanging only spinons.  The electron ring term is obtained from third-order processes involving hopping spinons on opposite edges of a square and chargon ring exchange on the square [for simplicity, we have used isotropic spinon hopping $t_{r,r+\hat{x}}^{(f)} = t_{r,r+\hat{y}}^{(f)} = t^{(f)}$].  Some other terms are also generated at the same orders in $1/h$ but are only indicated with dots; these include terms that correlate the electron densities and similarly modify the already present terms, but are not qualitatively new.

The terms exhibited above give precisely the electronic ring model studied in the main text, where we allow ourselves to vary independently the electron hopping and ring-exchange amplitudes, as well as the antiferromagnetic exchange coupling.  In the end, it is the detailed numerical study that establishes the phase diagram of this Hamiltonian.

We conclude by mentioning that the $d$-wave correlations present in our two-leg $d$-wave metal are of the $d_{xy}$ variety.  These are built in by taking the ring term to operate on \emph{elementary} plaquettes of the square lattice:  $(r,r+\hat{x},r+\hat{x}+\hat{y},r+\hat{y})$.  Looking forward, both when going to more legs with the DMRG and when studying two dimensions directly with the VMC, it will be interesting to consider a $d$-wave metal of the $d_{x^2-y^2}$ variety, a phase which is potentially more relevant to the cuprates.  A promising model to realize this phase would include, instead of that considered in this work, a ring-exchange term operating on all plaquettes $(r,r+\hat{x}+\hat{y},r+2\hat{y},r-\hat{x}+\hat{y})$.  Addressing the applicability of such models to real cuprate materials, as well investigating incipient $d_{x^2-y^2}$-wave superconductivity out of the putative $d_{x^2-y^2}$-wave metal, are very exciting future problems.

\section{Possible microscopic origin of the electronic ring terms by projecting the Coulomb interaction into a tight-binding band}
\label{}

Here we show how the electronic ring terms can appear by projecting the Coulomb interaction into a tight-binding band.  Our goal is to emphasize that such terms are rather simple and natural, even if they do not look familiar.  This analysis can also motivate more realistic estimates of the ring couplings from ab-initio calculations, which would be interesting to pursue for the cuprates and other strongly correlated materials.

We start with a tight-binding model for electrons moving in a periodic ionic potential,
\begin{equation}
H_{\rm kin} = -\sum_{\la ij \ra} \left( t_{ij} c_{i\alpha}^\dagger c_{j\alpha} + \Hc \right) ~,
\end{equation}
where $c_{i\alpha}^\dagger$ creates an electron with spin $\alpha$ in a Wannier orbital $w_i(\rvec)$ localized near an ion $\Rvec_i$ (here and below, summation over repeated spin indices is implied).  In the context of a particular material, the Wannier orbitals might be obtained, e.g., from the Bloch states of a narrow band in an LDA-type band structure that crosses the Fermi energy.\cite{Imada2010}  Next we want to include the electron-electron interaction.  We assume spin-independent pair-wise interaction such as screened Coulomb repulsion $u_{\rm Coul}(\rvec, \rvec')$, which could be obtained from the bare Coulomb within an RPA approach, integrating out the filled and empty LDA bands.\cite{Imada2010}  Projecting the interaction into the tight-binding band, we obtain
\begin{eqnarray}
H_{\rm int} = \frac{1}{2} \sum_{ijk\ell} \la ij |\hat{U}| k\ell \ra
c_{i\alpha}^\dagger c_{j\beta}^\dagger c_{\ell\beta} c_{k\alpha} ~,
\label{projUCoul}
\end{eqnarray}
with
\begin{eqnarray*}
\la ij |\hat{U}| k\ell \ra \equiv \int_{\rvec, \rvec'}
w_i^*(\rvec) w_j^*(\rvec') u_{\rm Coul}(\rvec, \rvec') w_k(\rvec) w_\ell(\rvec') ~.
\end{eqnarray*}

At this stage, it is customary to focus on terms that do not change the electron number on each site:
\begin{eqnarray*}
H_{\rm int}^{(0)} &=& 
\frac{1}{2} \sum_i \la ii |\hat{U}| ii \ra n_i (n_i - 1)
+ \frac{1}{2} \sum_{i \neq j} \la ij |\hat{U}| ij \ra n_i n_j \\
&-& \frac{1}{2} \sum_{i \neq j} \la ij |\hat{U}| ji \ra \left(2 \mathbf{S}_i \cdot \mathbf{S}_j + \frac{1}{2} \right) ~.
\end{eqnarray*}
The first and second terms are the familiar on-site (Hubbard) and inter-site repulsion.  The last term is the inter-site spin interaction, where we used $(c_{i\alpha}^\dagger c_{i\beta}) (c_{j\beta}^\dagger c_{j\alpha}) = 2 \mathbf{S}_i \cdot \mathbf{S}_j + 1/2$.  Note that in such a derivation, the spin exchanges can come out as ferromagnetic.  On the other hand, in materials with Mott physics, effective antiferromagnetic exchanges arise from a further interplay of the hopping $t$ and Hubbard repulsion $U$ and dominate over the bare ferromagnetic couplings in $H_{\rm int}^{(0)}$.  This is reminiscent of the Hartree-Fock over-estimation of the ferromagnetic tendencies in an electron gas due to neglect of the electronic correlations.  While we will not belabor this concern further, it is good to keep in mind that one needs to look at $H_{\rm int}$ as a whole, a warning that applies equally to our schematic discussion below.

We now want to stress that keeping only the terms in $H_{\rm int}^{(0)}$ is not complete.  There are more four-fermion terms that arise at the same level of treatment but that do not preserve the number of electrons on each site.  Let us assume for simplicity that our electronic orbitals are peaked on sites of a square lattice and respect the symmetries of the lattice.  Such a microscopic model is not applicable to the cuprates but can serve as a good illustration.  
Consider a square placket formed by sites 1, 2, 3, and 4 referring to ions $\Rvec_1$, $\Rvec_2 = \Rvec_1 + \hat{\mathbf{x}}$, $\Rvec_3 = \Rvec_1 + \hat{\mathbf{x}} + \hat{\mathbf{y}}$, and $\Rvec_4 = \Rvec_1 + \hat{\mathbf{y}}$ respectively.  Among various terms in $H_{\rm int}$, we find
\begin{equation}
\la 1,3 |\hat{U}| 2,4 \ra \left(
c_{1\alpha}^\dagger c_{3\beta}^\dagger c_{4\beta} c_{2\alpha} 
+ c_{1\alpha}^\dagger c_{3\beta}^\dagger c_{2\beta} c_{4\alpha} \right) + \Hc~, 
\end{equation}
where we used $\la 1,3 |\hat{U}| 4,2 \ra = \la 1,3 |\hat{U}| 2,4 \ra$ from the assumed orbital symmetries.  The pair-hopping term in the brackets is precisely the ring term in Eq.~(\ref{R1234}), with the coupling
\begin{eqnarray}
K &=& \la 1,3 |\hat{U}| 2,4 \ra \\
&=& \int_{\rvec, \rvec'} 
w_1^*(\rvec) w_3^*(\rvec') u_{\rm Coul}(\rvec, \rvec') w_2(\rvec) w_4(\rvec') ~.
\end{eqnarray}
The $w_{1,2,3,4}$ orbitals are peaked on ions $\Rvec_{1,2,3,4}$ respectively.  Assuming fairly localized orbitals, the main contribution in the above integral will come from configurations where $\rvec$ is somewhere between sites $\Rvec_1$ and $\Rvec_2$, while $\rvec'$ is somewhere between sites $\Rvec_3$ and $\Rvec_4$.  Then we could approximate
$u_{\rm Coul}(\rvec, \rvec') \approx u_{\rm Coul}(a) f_{12}(\rvec) f_{34}(\rvec')$, 
$u_{\rm Coul}(a) \approx u_{\rm Coul}[(\Rvec_1 + \Rvec_2)/2, (\Rvec_3 + \Rvec_4)/2]$, and obtain:
\begin{eqnarray*}
K &\approx& u_{\rm Coul}(a) 
\int_\rvec w_1^*(\rvec) w_2(\rvec) f_{12}(\rvec)  
\int_{\rvec'} w_3^*(\rvec') w_4(\rvec') f_{34}(\rvec') \\
&=& u_{\rm Coul}(a) \left| \int_\rvec w_1^*(\rvec) w_2(\rvec) f_{12}(\rvec) \right|^2 ~.
\end{eqnarray*}
Here $f_{12}(\rvec)$ is an $O(1)$ function peaked between the sites $\Rvec_1$ and $\Rvec_2$, and similarly for $f_{34}(\rvec')$.  We introduced these functions ad-hoc to implement an observation that $O(1)$ variations of $u_{\rm Coul}(\rvec, \rvec')$ with $\rvec$ or $\rvec'$ will eliminate cancellations that lead to orthogonality of the Wannier orbitals; instead, we anticipate obtaining factors like $\int_\rvec w_1^*(\rvec) w_2(\rvec) f_{12}(\rvec)$, which is on the order of the overlap of the non-orthogonal atomic orbitals.  It would clearly be desirable to perform such calculations more accurately in realistic contexts.  Here we want to point out that the above estimate gives a positive value for $K$, a key assumption we have made in our electronic ring model and presumably necessary for the $d$-metal.  More realistic calculations would hopefully give a reliable estimate of the sign of $K$ as well as its magnitude.

Because of the overlap integrals, the ring term will be significantly smaller than the on-site Hubbard repulsion.  However, we treat the latter by prohibiting double occupancy, and then the relevant energy scales to compare with are some effective hopping and antiferromagnetic spin exchange couplings.  Note that the hopping amplitudes themselves are set by overlaps between atomic orbitals times typical ionic potentials, so it is not inconceivable to estimate the ring terms as $K \sim t^2/u_{\rm Coul}$, which can be comparable to the spin exchange couplings.

It is important to note that the electron ring terms in our work are different from four-spin ring exchange terms that arise at order $t^4/U^3$ in effective spin models for so-called weak Mott insulators.\cite{Sheng2009_zigzagSBM, Block11_PRL_106_157202}  Our electron ring terms are also four-site terms but move two charges from one diagonal of a square to the other previously unoccupied diagonal.  Thus they are four-fermion rather than four-spin (eight-fermion) terms and can arise more directly from the Coulomb interaction.

Finally, as noted in the previous section, to search for seeds of $d_{x^2 - y^2}$ pair correlations in the context of the cuprates, we would want to consider electron ring terms that act on four sites 
$\Rvec_1$, $\Rvec_1 + \hat{\mathbf{x}} + \hat{\mathbf{y}}$, $\Rvec_1 + 2\hat{\mathbf{y}}$ and $\Rvec_1 - \hat{\mathbf{x}} + \hat{\mathbf{y}}$.  More reliable estimates of such terms would be highly desirable, as well as estimates of all other terms in the projected Coulomb interaction Eq.~(\ref{projUCoul}).

\ifREVTEX
	\bibliography{bib4ElRings}
\else
	\end{SI}
\fi

\end{document}